\documentclass[aps,prl,reprint,twocolumn,superscriptaddress,
               amsmath,amssymb,floatfix, nofootinbib]{revtex4-2}

\usepackage{graphicx}
\usepackage{dcolumn}
\usepackage{bm}
\usepackage{multirow}
\usepackage{booktabs}
\usepackage{braket}
\usepackage{xcolor}
\usepackage{textcomp}
\usepackage[colorlinks=true,linkcolor=blue,citecolor=blue,urlcolor=blue]{hyperref}
\usepackage{sibib}
\usepackage{pifont}

\graphicspath{{figures/}}

\newcommand{\Ba}{$^{138}$Ba$^{+}$}

\newcommand{\SiV}{SiV$^{-}$}
\providecommand{\ket}[1]{\left|#1\right\rangle}

\newcommand{\Fid}{\mathcal{F}}
\newcommand{\Rate}{R_{\mathrm{ent}}}
\newcommand{\etaion}{\eta_{\mathrm{ion}}}
\newcommand{\etasig}{\eta_{\mathrm{sig}}}
\newcommand{\etamem}{\eta_{\mathrm{mem}}}
\newcommand{\etaprot}{\eta_{\mathrm{prot}}}

\newcommand{\eqcontrib}{\thanks{These authors contributed equally to this work, and are listed in a random order.}}
\newcommand{\IonQ}{IonQ, Inc., Boston, Massachusetts 02135, USA}
\newcommand{\Duke}{Duke Quantum Center, Departments of Electrical and Computer Engineering and Physics, Duke University, Durham, North Carolina 27708, USA}

\newcommand{\circnum}[1]{\ding{\numexpr201+#1\relax}}

\begin{document}

\makeatletter
\@ifundefined{frontmatter@footnote@produce@footnote}{}{%
  \let\frontmatter@footnote@produce\frontmatter@footnote@produce@footnote}
\makeatother

\title{Quantum Networking at the Speed of Quantum Computation: \texorpdfstring{\\}{} Kilohertz Entanglement in a Heterogeneous Quantum System}

\author{Lukas Hartung}\eqcontrib\affiliation{\IonQ}
\author{Pierre Barral}\eqcontrib\affiliation{\IonQ}
\author{Noah Glachman}\eqcontrib\affiliation{\IonQ}
\author{George Toh}\eqcontrib\affiliation{\IonQ}
\author{Jacob H. Davidson}\eqcontrib\affiliation{\IonQ}
\author{Sagnik Saha}\eqcontrib\affiliation{\IonQ}
\author{Alexander Chuang}\eqcontrib\affiliation{\IonQ}
\author{Matthew C. Cambria}\eqcontrib\affiliation{\IonQ}
\author{Madison Sutula}\eqcontrib\affiliation{\IonQ}

\author{Alexander Abulnaga}\affiliation{\IonQ}
\author{Julia Brevoord}\affiliation{\IonQ}
\author{John Chiaverini}\affiliation{\IonQ}
\author{Ian Counts}\affiliation{\IonQ}
\author{Aos Dabbagh}\affiliation{\IonQ}
\author{Christian Dangel}\affiliation{\IonQ}
\author{Skylar Deckoff-Jones}\affiliation{\IonQ}
\author{Chawina De-Eknamkul}\affiliation{\IonQ}
\author{Jonathan Dietz}\affiliation{\IonQ}
\author{Riley Forst}\affiliation{\IonQ}
\author{Prithvi Gundlapalli}\affiliation{\IonQ}
\author{Jeonghoon Ha}\affiliation{\IonQ}
\author{Michael Haas}\affiliation{\IonQ}
\author{Ezekiel Meulbroek}\affiliation{\IonQ}
\author{Andrea Mucchietto}\affiliation{\IonQ}
\author{Daniel Riedel}\affiliation{\IonQ}
\author{Jonah Sachs}\affiliation{\IonQ}
\author{Jeremy Sage}\affiliation{\IonQ}
\author{Mikhail Shalaev}\affiliation{\Duke}
\author{Harriet Shi}\affiliation{\Duke}
\author{Denis Sukachev}\affiliation{\IonQ}
\author{Yichao Yu}\affiliation{\Duke}

\author{Christopher Monroe}\affiliation{\IonQ}\affiliation{\Duke}
\author{Nicholas Mondrik}\affiliation{\IonQ}
\author{Mihir Bhaskar}\affiliation{\IonQ}
\author{Bart Machielse}\affiliation{\IonQ}
\author{Matteo Pompili}\affiliation{\IonQ}
\author{Carsten Robens}\email{Carsten.Robens@IonQ.co}\affiliation{\IonQ}
\author{David Levonian}\affiliation{\IonQ}

\date{\today}

\begin{abstract}
Distributed quantum computing requires rapid, high-fidelity entanglement between qubits in separate processor modules. Photonic links between single-atom qubits provide switchable, long-range connections, but photon loss limits their entanglement rate. Here we entangle a trapped \Ba{} ion with a silicon-vacancy (\SiV{}) center in a diamond nanophotonic cavity. A photon emitted by the ion is converted in wavelength and encoding, reflected from the \SiV{}--cavity system and detected, so that a single detected photon heralds entanglement. Including all conversion losses, we generate ion--\SiV{} Bell pairs at an average rate of 1.03(1)\,kHz with a fidelity of 87.9(7)\%. This is four times the rate of the fastest photonic link between two trapped ions and corresponds to one Bell pair per millisecond, matching the cycle time of planned trapped-ion processors.

\end{abstract}

\maketitle

Modular quantum computing offers a path to large-scale quantum systems by interconnecting smaller quantum processors~\cite{monroe2014modular,main2025,Wu2024,Nickerson2014}, analogous to the role of networking in modern datacenters. To implement fault-tolerant operations between modules, a modular quantum computer must be capable of consistently generating large numbers of high-fidelity entangled qubit pairs between modules. For known algorithms and plausible near-term qubit fidelities and quantum error correction schemes, estimates of the entanglement rate needed to support continuous logical operations between modules range from 10 to 100\,kHz~\cite{sinclair2025modular,xue2026modular,Marqversen2026}, at an infidelity about ten times higher than that of local two-qubit operations~\cite{ramette2024}. The fastest photonic link between trapped ions, which hold some of the highest reported local gate fidelities~\cite{hughes2025}, produces 250 Bell pairs per second~\cite{Oreilly2024}.

Photonically mediated entanglement offers a straightforward path to multiplexing and can be implemented using largely commercially available parts, but is also necessarily probabilistic~\cite{Barrett2005}. The rate of photonic entanglement is a product of the attempt rate, the probability of a successful herald, and the number of multiplexed channels. For most atomic qubit systems, the attempt rate has a fixed upper limit set by the radiative lifetimes involved in state preparation~\cite{stephenson2020,Oreilly2024}. Therefore, most photonic entanglement architectures focus on increasing the number of channels~\cite{li_parallelized_2025} or the probability of successfully extracting a photon per channel~\cite{safari2026atomphoton}.

\begin{figure*}[tb]
  \centering
  \includegraphics[width=\textwidth]{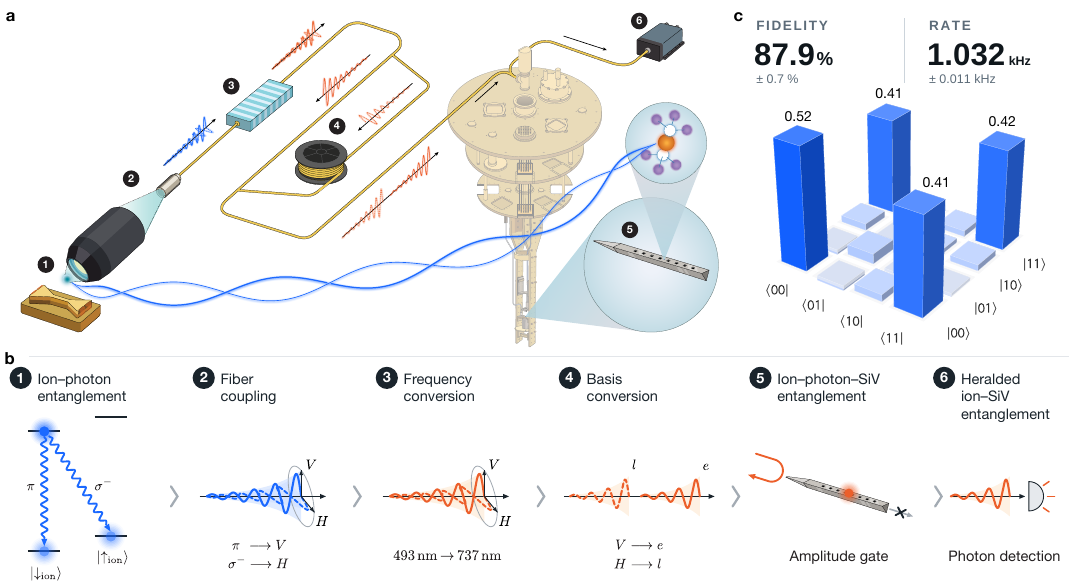}
  \caption{
      \textbf{Entangling a trapped-ion qubit and a solid-state quantum memory.}
      (a, b)~\circnum{1}~A trapped \Ba{} ion emits a $493.5$\,nm photon whose polarization is entangled with the ion spin state; \circnum{2}~fiber coupling through a high-numerical-aperture objective maps $\sigma^-~(\pi)$ polarization onto $H~(V)$; \circnum{3}~difference-frequency generation in a periodically poled lithium niobate (PPLN) crystal, pumped at $1492.3$\,nm, converts the photon to $737.45$\,nm and preserves its polarization;
      \circnum{4}~an unbalanced interferometer converts polarization encoding into time-bin encoding, $V~(H)$ into early (late);
      \circnum{5}~spin-dependent reflection from the \SiV{} imprints the memory spin state onto the time-bin qubit;
      \circnum{6}~the returning photon is routed to the detectors, and its detection heralds a Bell state between the ion and the quantum memory (blue wavy lines).
      (c)~Real part of the reconstructed ion--\SiV{} density matrix, $\Fid = 87.9(7)\%$ with respect to $\ket{\Phi^+} = \ket{00} + \ket{11}$ at an average rate of $1.032(11)$\,kHz. All imaginary elements are below 0.05 in magnitude (see SM, Sec.~\ref{sec:SI:tomography}).
  }
  \label{fig:main}
\end{figure*}

Increasing the number of channels can be costly, because communication and computation then compete for the same resources, such as qubits, cooling and initialization~\cite{Marqversen2026}. Photon extraction efficiencies can be improved with larger numerical aperture (NA) optics for free-space emission~\cite{safari2026atomphoton}, or by Purcell-enhancing qubit emission via cavities~\cite{thomas2022atomphoton,Krutyanskiy2024ioncavity}. The trade-off between NA and field of view makes it difficult to combine high-efficiency free-space collection with highly multiplexed schemes. Multiple imaging zones or cavities with Purcell enhancement offer a potential path around this. However, dielectric surfaces near the qubits degrade gate fidelities and coherence in both trapped-ion and neutral-atom systems~\cite{Ocola2024Rydberg,Teller2021}. Optics therefore need to be placed farther from computational qubits, which requires either shuttling qubits to a communication zone~\cite{Li2024,Hartung2024,Sunami2025} or long cavities, whose narrow linewidths slow the extraction of photons~\cite{krutyanskiy2023}.

Better collection reduces photon loss but cannot remove it, so the heralding scheme sets how strongly the remaining loss limits the rate. In the two-photon schemes used for the fastest ion--ion links~\cite{stephenson2020,Oreilly2024}, photons from both nodes must be detected, so the success probability scales with the square of the photon collection efficiency of a single node. Single-photon heralding between two emitters scales linearly, but it trades rate against fidelity through the excitation probability and requires interferometric phase stability and indistinguishable photons from the two nodes~\cite{Slodicka2013, humphreys2018deterministic, Knollmann2026}. A quantum memory that reflects the photon emitted by the other node avoids these costs: a single detected photon heralds entanglement, the success probability scales linearly with the photon collection efficiency, the emitter can be excited with near-unit probability, and no two-photon interference is needed. Moreover, a reflective quantum memory can be entangled with any photon whose spectral width is smaller than the bandwidth of its spin-dependent reflection.

In this paper, we introduce a hybrid architecture that combines the free-space photon collection from an atomic qubit with the near-perfect efficiency from a solid-state memory in a cavity. We collect photons from ions via a conventional imaging system outside the vacuum chamber, without an optical cavity, reflect them from a qubit-cavity quantum memory system with high photon interaction efficiency, and detect them to produce a herald signal. Such heterogeneous links have been considered difficult because two different platforms must operate together through interfaces such as quantum frequency conversion. \SiV{} memories interact efficiently with photons but have been limited by the rate of their photon sources~\cite{knaut2024}, whereas trapped ions emit photons at a high rate but collect them with low probability~\cite{Oreilly2024}. Entanglement between a trapped ion and a solid-state quantum memory was recently reported with a rare-earth ensemble memory~\cite{wang2026}; here we use a single-spin memory and a single-photon herald, which together allow kilohertz rates.

The quantum memory is implemented as a silicon-vacancy center in a diamond nanophotonic cavity. \SiV{} centers in such cavities interact efficiently with photons~\cite{bhaskar2020}, provide an electron--nuclear spin register in which a nuclear-controlled gate reaches 99.9\% fidelity~\cite{stas2022science}, and can be fabricated in large numbers on a single chip~\cite{Riedel2026}. The memory is paired with an ion--photon interface in \Ba{}~\cite{Hannegan2022}, which has been used for the fastest ion--ion entanglement to date~\cite{Oreilly2024} and is compatible with leading ion quantum computing architectures~\cite{tripier2026, Mehdi2025,hughes2025}. Combining the two systems in a hybrid photonic entanglement scheme, we entangle the ion and the \SiV{} at a rate of 1.03\,kHz, i.e.\ one entangled pair per millisecond on average, comparable to the computational cycle time of planned ion quantum computers~\cite{WISE2023}.

\begin{figure*}[tb]
  \centering
  \includegraphics[width=\textwidth]{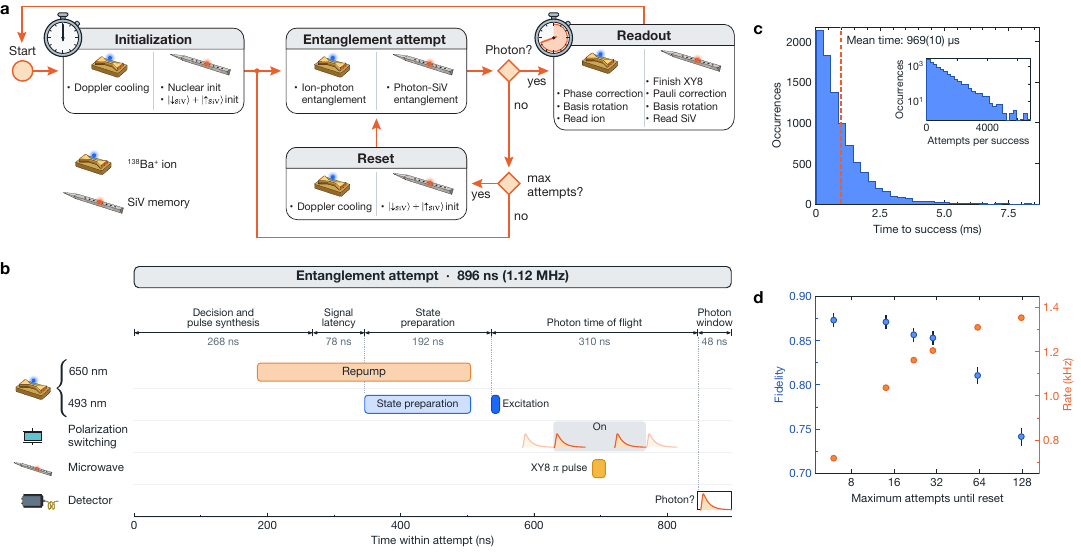}
  \caption{
      \textbf{Establishing heralded ion-memory entanglement.}
      (a)~Control-flow diagram of the sequence used to deliver each entangled state. Each box represents a step, with sub-steps separated by which sub-system they act on (trapped-ion or quantum memory). The steps are repeated as indicated until a photon is detected, heralding success. Upon success the \textit{Readout} sequence compensates for which of the two detectors clicked and for its detection time using local gates on the two qubits (see main text and SM, Sec.~\ref{sec:SI:experimental-sequence}). It also applies the required measurement basis rotation. We measure the time needed to generate the state starting from the beginning of the \textit{Initialization} step until the beginning of the \textit{Readout} step, as indicated by the two stopwatch icons.
      (b)~Pulse sequence for the \textit{Entanglement attempt} step. A microwave $\pi$-pulse flips the state of the \SiV{} in between the two time bins. We choose the phase of the pulses between consecutive attempts to implement an XY8 decoupling sequence on the quantum memory and preserve its coherence before it interacts with a photon (see SM, Sec.~\ref{sec:SI:experimental-sequence}).
      (c)~Histogram of the time to success for the data presented in Fig.~\ref{fig:main}. The dashed orange line shows the mean of the distribution, corresponding to a rate of $1.032(11)$\,kHz. Inset, histogram of the number of attempts for each success from the same dataset.
      (d)~Rate-fidelity tradeoff as a function of the maximum attempts until reset. A higher number of attempts reduces the relative overhead of initialization and reset, increasing rate, but leads to more errors on the \SiV{}, decreasing fidelity (see main text and SM, Sec.~\ref{sec:SI:siv_reset}). Here fidelity is estimated as $\Fid = (1 + \braket{XX} - \braket{YY} + \braket{ZZ})/4$.
      Error bars are one standard deviation; where not visible, they are smaller than the markers.
  }
  \label{fig:sequence}
\end{figure*}

A schematic of the experiment is illustrated in Figure~\ref{fig:main}(a,b). A single trapped \Ba{} ion~\cite{carter2024} is Doppler cooled and optically pumped into $\ket{\uparrow_{\mathrm{ion}}} = \ket{S_{1/2},m_J=+1/2}$.
Each ion--photon entanglement generation attempt starts with a picosecond $\sigma^-$-polarized pulse at $493.5$\,nm that excites the ion with near-unity probability to $\ket{P_{1/2},m_J=-1/2}$, from which the ion spontaneously decays either to $\ket{\downarrow_{\mathrm{ion}}} = \ket{S_{1/2},m_J=-1/2}$ by emitting a $\pi$-polarized photon or back to $\ket{\uparrow_{\mathrm{ion}}}$ by emitting a $\sigma^-$-polarized photon, thereby entangling the ion's electron spin with the photon polarization (decay to the $D_{3/2}$ manifold at 650\,nm constitutes a loss channel and is repumped in the next attempt).

The entanglement is maximal only if the two decay channels contribute equally
to the photon in the fiber. We collect the photon perpendicular to the quantization axis with an $\mathrm{NA}=0.75$ objective (see Supplemental Material (SM), Sec.~\ref{sec:SI:ion}). Coupling into a single-mode optical fiber maps \(\sigma^-\) and \(\pi\) emission onto orthogonal linear polarizations \(H\) and \(V\), respectively~\cite{stephenson2020}. The smaller branching probability of the \(\pi\) transition is nearly compensated by its stronger transverse emission and higher fiber-coupling efficiency (see SM, Sec.~\ref{sec:SI:microscope}). The resulting state is close to maximally entangled, $\ket{\downarrow_{\mathrm{ion}}V_{493}}+\ket{\uparrow_{\mathrm{ion}}H_{493}}$ (normalization omitted throughout).

This polarization-encoded $493.5$\,nm photon is not directly compatible with the \SiV{} memory, which requires a time-bin-encoded photon at $737.45$\,nm~\cite{bhaskar2020,knaut2024}.
We therefore first convert the photon wavelength from $493.5$\,nm to $737.45$\,nm using polarization-insensitive quantum frequency conversion (QFC) in a Sagnac geometry (see SM, Sec.~\ref{sec:SI:qfc}), $\ket{H_{493}} \xrightarrow{\mathrm{QFC}} \ket{H_{737}}$, $\ket{V_{493}} \xrightarrow{\mathrm{QFC}} \ket{V_{737}}$, and then convert the encoding with a basis converter (BC) consisting of an unbalanced Mach--Zehnder interferometer and a fast Pockels cell (see SM, Sec.~\ref{sec:SI:tb}), $\ket{V_{737}}\xrightarrow{\mathrm{BC}}\ket{e}$, $\ket{H_{737}}\xrightarrow{\mathrm{BC}}\ket{l}$, where $\ket{e}$ and $\ket{l}$ denote the early and late time bins that traverse the short and long arms of the interferometer, respectively.
The ion--photon state entering the memory is thus $\ket{\downarrow_{\mathrm{ion}}}\ket{e}+\ket{\uparrow_{\mathrm{ion}}}\ket{l}$.

The ion--photon entanglement is transferred to the \SiV{} via a heralded spin--photon gate based on cavity-enhanced, spin-dependent reflection~\cite{nguyen2019prl,bhaskar2020,knaut2024}.
Owing to the strong coupling of the \SiV{} to the nanophotonic cavity at the operating frequency, an incident photon is reflected only if the \SiV{} electron spin is in $\ket{\uparrow_{\mathrm{SiV}}}$, whereas for $\ket{\downarrow_{\mathrm{SiV}}}$ it is not reflected and lost. High-fidelity operation therefore requires the spectrum of the incident photon to lie within the bandwidth over which the cavity exhibits spin-dependent reflection.
Before the photon arrives, the \SiV{} is prepared in an equal superposition of $\ket{\downarrow_{\mathrm{SiV}}}$ and $\ket{\uparrow_{\mathrm{SiV}}}$, and a microwave $\pi$ pulse resonant with the \SiV{} electron transition is applied between the two photon time bins.
The early bin is thus reflected only if the spin was initially in $\ket{\uparrow_{\mathrm{SiV}}}$ (and is flipped to $\ket{\downarrow_{\mathrm{SiV}}}$ by the $\pi$ pulse), and the late bin only if the spin was initially in $\ket{\downarrow_{\mathrm{SiV}}}$ (and is flipped to $\ket{\uparrow_{\mathrm{SiV}}}$ before the late bin arrives), so that, conditioned on reflection, the joint state becomes
\begin{equation}\label{eq:ionphotonsiv}
  \ket{\downarrow_{\mathrm{ion}}}\ket{e}\ket{\downarrow_{\mathrm{SiV}}}+\ket{\uparrow_{\mathrm{ion}}}\ket{l}\ket{\uparrow_{\mathrm{SiV}}}.
\end{equation}
To herald the gate, the time-bin information must be erased by measuring the reflected photon in a superposition basis of $\ket{e}$ and $\ket{l}$.
For this purpose the reflected photon is sent back through the BC, where Faraday rotators separate the return path from the input and the Pockels cell now routes the early bin through the long arm and the late bin through the short arm.
Both bins therefore acquire the same total delay and exit the BC simultaneously in a common spatial mode, with the time-bin qubit mapped back onto polarization, $\ket{e}\to\ket{H}$, $\ket{l}\to\ket{V}$.
Since every photon traverses both arms exactly once, the interferometer phase cancels and no active stabilization of the BC is required.
A projective measurement in the diagonal polarization basis, $\ket{D/A}=\ket{H}\pm\ket{V}$, then erases the which-path information, and a detection event heralds the successful gate and projects the ion and the \SiV{} onto one of the two Bell states
\begin{equation}
  \ket{\downarrow_\mathrm{ion}\downarrow_\mathrm{SiV}} \pm \ket{\uparrow_\mathrm{ion}\uparrow_\mathrm{SiV}},
  \label{eq:heralded}
\end{equation}
with the sign determined by which superconducting nanowire single-photon
detector (SNSPD) fires.
Throughout, $\ket{0}$ ($\ket{1}$) denotes the dark (bright) readout state of
each qubit: for the ion $\ket{0}\equiv\ket{\uparrow_\mathrm{ion}}$ (shelved) and
$\ket{1}\equiv\ket{\downarrow_\mathrm{ion}}$; for the \SiV{}
$\ket{0}\equiv\ket{\downarrow_\mathrm{SiV}}$ and $\ket{1}\equiv\ket{\uparrow_\mathrm{SiV}}$.
We write two-qubit states as $\ket{\mathrm{ion},\mathrm{SiV}}$.
Spin flips that we apply to the \SiV{} before readout map the states of
Eq.~\eqref{eq:heralded} onto the target Bell state
$\ket{\Phi^+} = \ket{00}+\ket{11}$, as described below.

To characterize the entanglement between the \Ba{} ion and the \SiV{} memory, we perform tomography of the combined ion--\SiV{} state, which is summarized in Fig.~\ref{fig:main}(c).
With a $1.1$\,MHz attempt rate we obtain an entanglement rate of $\Rate = 1.032(11)$\,kHz and $\Fid =87.9(7)\,\%$.

The entanglement rate $\Rate$ is a product of the attempt rate $f_{\mathrm{rep}}$ and the individual subsystem efficiencies: $\etaion$, $\etasig$, $\etamem$, and $\etaprot$:
\begin{equation}\label{eq:rate}
  \Rate = f_{\mathrm{rep}}\;\etaion\,\etasig\,\etamem\,\etaprot .
\end{equation}
Multiplying the independently calibrated subsystem efficiencies (for definitions and values of $\eta_i$ see Table~\ref{tab:RateBudget}) predicts $\Rate = 1.21(20)$\,kHz.
The measured rate of $1.032(11)$\,kHz lies $14\,\%$ below this prediction, within its uncertainty.
We suspect that part of this difference arises from birefringence in the basis converter, which spatially separates the two polarization modes and thereby reduces the average fiber-coupling efficiency.

To identify what limits the fidelity, we combine independently calibrated error contributions of each subsystem (Table~\ref{tab:Fid_summary}; SM, Table~\ref{tab:Fid_error_budget}).
The ion contributes most ($\Delta\mathcal{F} = 4.3\,\%$), mainly from optical-pumping light that leaks through the electro-optic modulators after a herald ($\approx 1.6\,\%$) and from residual errors that we attribute to ion temperature and aberrations of the collection optics ($\approx 1.5\,\%$).
The memory contributes $3.9\,\%$, about $3\,\%$ of which arises from the finite spin-dependent reflection contrast of the cavity-coupled \SiV{} and from nuclear-spin flips during the sequence, while microwave pulse errors and residual dephasing under XY8 decoupling each remain below $0.4\,\%$. Frequency conversion contributes $2.9\,\%$, almost entirely through conversion noise that produces false heralds.
Photons that reach the memory but are lost before detection dephase the \SiV{} and contribute $1.9\,\%$; because this error accumulates over the attempts in a batch, it sets the rate--fidelity trade-off in Fig.~\ref{fig:sequence}(d).
Together, these contributions predict $\Fid = 87.4\,\%$ with $\langle XX\rangle = 0.81$ and $\langle ZZ\rangle = 0.87$, in agreement with the tomographically reconstructed state ($\Fid = 87.9(7)\,\%$). None of the dominant errors is fundamental: narrower spectral filtering of the conversion noise, higher EOM extinction, higher cavity--\SiV{} contrast and a higher return efficiency from the memory to the detectors would each directly reduce the infidelity.

\begin{table}[tb]
  \caption{\textbf{Summary of the ion--\SiV{} fidelity budget.}
  Fidelity reduction $\Delta\mathcal{F}$ from each subsystem, i.e.\ the infidelity if that
  subsystem were the only source of error, obtained by combining the entries of
  Table~\ref{tab:Fid_error_budget}. The predicted fidelity combines all error sources per correlator, so it is not exactly the product of the subsystem terms.}
  \label{tab:Fid_summary}
  \centering
  \begin{tabular}{lc}
    \toprule
    Subsystem & $\Delta\mathcal{F}$ (\%) \\
    \midrule
    Ion                                      & 4.3 \\
    QFC                                      & 2.9 \\
    Memory                                   & 3.9 \\
    Detection (undetected heralding photons) & 1.9 \\
    \midrule
    Predicted fidelity $\mathcal{F}$         & 87.4 \\
    \bottomrule
  \end{tabular}
\end{table}

\begin{figure*}[t]
  \centering
  \includegraphics[width=0.75\textwidth]{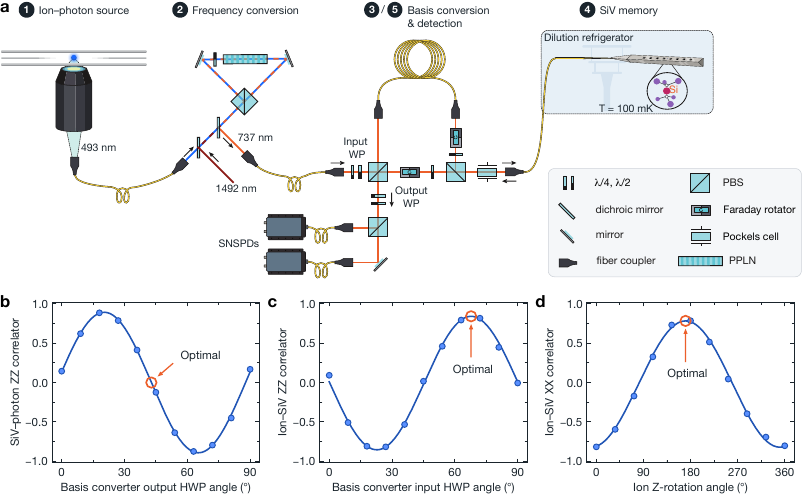}
  \caption{
    \textbf{Schematic of the apparatus and calibration procedure.}
    (a)~\circnum{1}~A microscope objective with high numerical aperture (NA~$=0.75$) collects $493.5$\,nm photons from a single \Ba{} ion in a four-rod Paul trap and couples them into a single-mode fiber;
    \circnum{2}~difference-frequency generation in a PPLN crystal, pumped at $1492.3$\,nm in a Sagnac geometry, converts the photons to $737.45$\,nm and preserves their polarization;
    \circnum{3}~an unbalanced interferometer (the basis converter) re-encodes the photonic qubit in early and late time bins, and a Pockels cell rotates the polarization of the late time bin so both bins reach the memory with the same polarization;
    \circnum{4}~a quantum memory, which consists of a nanophotonic cavity coupled to a \SiV{} center in a dilution refrigerator at $T\approx100$\,mK, reflects the photon and imprints the spin state onto it;
    \circnum{5}~on the way back, the Pockels cell rotates the polarization of the early time bin so each bin takes the other arm, and Faraday rotators route the recombined photon toward the detection path, where waveplates (WP, $\lambda/2$ and $\lambda/4$), a polarizing beam splitter (PBS), and superconducting nanowire single-photon detectors (SNSPDs) measure its polarization.
    (b)~$\braket{ZZ}$ correlator between the \SiV{} and the photon (i.e., which detector clicked). To generate the target ion--\SiV{} Bell state, the photon detection must herald entanglement without revealing the \SiV{} state. We therefore sweep the basis converter output half-wave plate (HWP) and select the angle where $\braket{ZZ}=0$.
    (c)~$\braket{ZZ}$ correlator between the ion and the \SiV{}. We sweep the basis converter input HWP and select the angle of maximum positive parity.
    (d)~$\braket{XX}$ correlator between the ion and the \SiV{}. We measure both qubits in the $X$ basis, and we sweep the phase of a $Z$ rotation applied to the ion before readout. We select the phase that maximizes $\braket{XX}$, which gives the largest overlap with $\ket{\Phi^+} = \ket{00} + \ket{11}$.
    Error bars are one standard deviation; where not visible, they are smaller than the markers.
    }
  \label{fig:setup_and_calibration}
\end{figure*}

We implement the experimental sequence with a commercial modular control system. Its FPGA-based sequencer executes the pulse sequences and can branch in real time. Figure~\ref{fig:sequence}(a, b) shows a control-flow diagram of the sequence (see SM, Sec.~\ref{sec:SI:experimental-sequence} for additional details).

The delivery of each entangled state starts with an initialization step. It consists of Doppler cooling of the ion, initialization of the \SiV{} nuclear spin, and preparation of the \SiV{} electron spin in a balanced superposition state.
We then start a batch of up to $15$ entanglement attempts. Each attempt consists of ion repumping, pulsed excitation of the ion to emit a photon, polarization switching with the Pockels cell, a microwave $\pi$ pulse that flips the \SiV{} spin between the two time bins, and detection and time-tagging of the photon.
The microwave $\pi$ pulses alternate between the $X$ and $Y$ axes according to the XY8 sequence, which preserves the \SiV{} coherence better than pulses about a single axis.
If a photon is detected in the $48$\,ns detection window, which spans the photon wave packet (SM, Sec.~\ref{sec:SI:experimental-sequence}), the sequence exits the batch and starts the readout step. If no attempt in the batch succeeds, we apply a reset step, which consists of Doppler cooling of the ion and preparation of the \SiV{} electron spin in a balanced superposition state. We then start a new batch of entanglement attempts.

During the readout step, we apply two real-time corrections (SM, Sec.~\ref{sec:SI:experimental-sequence} and Fig.~\ref{fig:SI:photonshape}): a phase correction on the ion compensates for the variable photon emission time, and a correction on the \SiV{} compensates for which detector clicked and for the \SiV{} phase at the photon interaction.
We merge the latter with the remaining pulses of the XY8 sequence, whose number depends on how many attempts preceded the herald.
Completing the sequence, with an extra $\pi$ pulse when needed, returns the \SiV{} spin to its initial frame, which interchanges $\ket{\uparrow_\mathrm{SiV}}$ and $\ket{\downarrow_\mathrm{SiV}}$ in Eq.~\eqref{eq:heralded}.
The delivered state is therefore $\ket{\Phi^+} = \ket{\uparrow_\mathrm{ion}\downarrow_\mathrm{SiV}} + \ket{\downarrow_\mathrm{ion}\uparrow_\mathrm{SiV}} = \ket{00}+\ket{11}$. These corrections could also be applied after the measurement, by relabeling the outcomes according to which detector clicked and when. Applying them in real time makes every herald deliver the same state $\ket{\Phi^+}$, which is required when the entangled pair is used as a resource, as in distributed quantum computing.

Finally, we measure the two qubits in the selected bases: for every heralded
event, the control system applies the basis rotations for the current
measurement setting, together with the real-time corrections described above. For full state tomography, we use the $9$ measurement settings $\{X, Y, Z\}_\text{ion}\otimes\{X, Y, Z\}_\text{\SiV{}}$. From these, we calculate the two-qubit correlators $\braket{ij} \equiv \braket{\sigma_i \otimes \sigma_j}_\rho = \text{Tr}(\rho~\sigma_i \otimes \sigma_j)$ (see SM, Sec.~\ref{sec:SI:tomography}).
We then reconstruct the density matrix of the generated entangled state using maximum likelihood estimation, and bootstrapping for error bars on the fidelity.
Occasionally, the ion is lost from the trap or the \SiV{} changes charge state. Both failures cause an abrupt and complete loss of signal. In these cases we stop the acquisition, restore the system, and discard the full interrupted dataset. This decision depends only on the presence of the ion and of the \SiV{} readout signal, never on the measured correlations.

We measure the time needed to generate the entangled state from the beginning of the initialization step to the beginning of the readout step using the FPGA on-board clock ($4$\,ns resolution), obtaining a mean time to success of $969(10)$\,µs, and a mean number of attempts of $731(8)$ (see Fig.~\ref{fig:sequence}(c)). This interval excludes the readout step ($664$\,µs per event), which characterizes the state and is not optimized for speed; in a distributed quantum computer, gates that consume the pair would take its place. Including the readout step, the wall-clock rate of generated and measured pairs is $612(4)$\,Hz.
By increasing the number of attempts per batch, we can reduce the relative overhead that the initialization and reset steps have on the time to success, therefore increasing the overall rate (see Fig.~\ref{fig:sequence}(d)). This comes at the expense of a reduced entangled state fidelity: the leading source of error is the increased chance per batch that a photon from the ion reaches the quantum memory but gets lost on the way back and is not detected, therefore dephasing the \SiV{} which will not be in the correct state for the next attempt (see SM, Sec.~\ref{sec:SI:siv_reset}).
This effect can be mitigated with increased detector efficiencies and reduced return loss from the nanophotonic cavity and the basis converter, and by collecting and heralding on photons transmitted through the cavity, in addition to reflected photons.
Several pre-initialized memories addressed in turn would also allow each memory to be reset after a few attempts without pausing the ion.

Figure~\ref{fig:setup_and_calibration} (a) shows a schematic of the experimental apparatus, as well as elements of its calibration to deliver the target ion-\SiV{} entangled state (Fig.~\ref{fig:setup_and_calibration}(b--d)).

\textbf{Discussion.}\label{sec:discussion}
We have established an entangled link between a trapped-ion qubit and an efficient quantum memory with a fidelity of $87.9(7)\,\%$ at a rate comparable to local operations between ions, a step toward the 10--100\,kHz rates required for distributed quantum computation. On the ion side, the link needs no cavity and no optics inside the vacuum chamber: an objective outside the vacuum window collects the photons and couples them into a single-mode fiber.

The dilution refrigerator used here is not a fundamental requirement either: strained \SiV{} centers have been operated at temperatures up to $1.5$\,K~\cite{stas2022science}, which compact closed-cycle cryostats can reach. Furthermore, the memory provides the operations needed for a network layer~\cite{dahlberg2019linklayer} that distributes entanglement between processors: the $^{29}$Si nuclear spin of the \SiV{} can store entanglement while new links are generated~\cite{stas2022science}, \SiV{} nodes have performed Bell-state measurements on multiple photons~\cite{bhaskar2020}, and two \SiV{} nodes have been entangled over deployed fiber~\cite{knaut2024}. All stages of the link (frequency conversion, basis conversion and the memory) are compatible with integrated photonics~\cite{Riedel2026,assumpcao2024}, which would reduce loss and improve stability.
We expect that a higher ion-to-fiber coupling, a more efficient signal chain, and replacing the amplitude-based spin--photon gate of the \SiV{} by a phase gate, which reflects the photon in both spin states and therefore works with strongly overcoupled cavities~\cite{duan2004}, will raise the entanglement rate to $>5$\,kHz.

These abilities can be combined to produce ion-ion entanglement between quantum processors for extended error-corrected circuits, providing a scaling route for quantum computers~\cite{raskop2026}. In addition, because SiVs interact efficiently with a range of photons, they can serve as a photonic front-end that receives photons for ion processors. For example, they can receive sparse photons from a weak optical sensing target and teleport the photon state into an ion processor for algorithms such as quantum principal component analysis (PCA)~\cite{Mokeev2026,stas2026}. Similarly, SiVs can receive weak photonic signals over fiber in quantum communication or blind quantum computing~\cite{baranes2025} and pass the resulting states to ion processors for error correction. We expect this heterogeneous system to benefit applications that combine low-loss interaction with photons and quantum processing.

\begin{acknowledgments}
\textit{Acknowledgments.}
We thank Ashish Kalakuntla, Brendan Pankovich, Isabella Goetting, Jesse Amato-Grill, Johannes Borregaard, John Hannegan, Mayer Feldman and Sergey Dushenko for helpful discussions.

\textit{Author contributions.}
CR and DL conceived the experiment.
AA, AM, BM, ChaD, ChrD, DR, JoD, JoS, PG, RF and SD designed, fabricated and characterized the nanophotonic cavity.
GT, HS, IC, MiS, PB, SS and YY built and operated the ion-trap node.
JaD built the quantum frequency converter.
LH built the basis-conversion interferometer.
AC, JB, MaS, MC and MH built and operated the memory setup.
AD, MP and NG built and operated the control hardware and software infrastructure.
DS, EM and JH built and operated cryogenic and vacuum systems.
AC, GT, JaD, MaS, MC, MP, NG, PB and SS took and analyzed the data and prepared the figures.
CR, DL and MP wrote the manuscript with input from all authors.
AC, CR, DL, GT, JaD, LH, MC, MP, NG, PB and SS wrote the Supplemental Material with input from all authors.
CM, JC, JeS, MB and NM contributed to the design of the experiment and the interpretation of the results.
CR, DL and MP supervised the project.

\textit{Competing interests.}
Some of the authors are employees of IonQ, Inc. (as described in the author list), which has filed patent
applications related to this work. The remaining authors declare no competing
interests.

\textit{Data and code availability.}
The data that support the findings of this study, with the source data and analysis code for all figures, are available at \url{https://github.com/ionq-publications/khz-heterogeneous-entanglement}.

\textit{AI usage disclosure.}
During the preparation of this work, we used AI tools from Anthropic and OpenAI to improve readability and language, and to generate code for data visualization. After using these tools, we reviewed and edited the content and code and we take full responsibility for the content of the publication.

\end{acknowledgments}

\bibliography{references}

\begin{thebibliography}{13}%
\makeatletter
\providecommand \@ifxundefined [1]{%
 \@ifx{#1\undefined}
}%
\providecommand \@ifnum [1]{%
 \ifnum #1\expandafter \@firstoftwo
 \else \expandafter \@secondoftwo
 \fi
}%
\providecommand \@ifx [1]{%
 \ifx #1\expandafter \@firstoftwo
 \else \expandafter \@secondoftwo
 \fi
}%
\providecommand \natexlab [1]{#1}%
\providecommand \enquote  [1]{``#1''}%
\providecommand \bibnamefont  [1]{#1}%
\providecommand \bibfnamefont [1]{#1}%
\providecommand \citenamefont [1]{#1}%
\providecommand \href@noop [0]{\@secondoftwo}%
\providecommand \href [0]{\begingroup \@sanitize@url \@href}%
\providecommand \@href[1]{\@@startlink{#1}\@@href}%
\providecommand \@@href[1]{\endgroup#1\@@endlink}%
\providecommand \@sanitize@url [0]{\catcode `\\12\catcode `\$12\catcode `\&12\catcode `\#12\catcode `\^12\catcode `\_12\catcode `\%12\relax}%
\providecommand \@@startlink[1]{}%
\providecommand \@@endlink[0]{}%
\providecommand \url  [0]{\begingroup\@sanitize@url \@url }%
\providecommand \@url [1]{\endgroup\@href {#1}{\urlprefix }}%
\providecommand \urlprefix  [0]{URL }%
\providecommand \Eprint [0]{\href }%
\providecommand \doibase [0]{https://doi.org/}%
\providecommand \selectlanguage [0]{\@gobble}%
\providecommand \bibinfo  [0]{\@secondoftwo}%
\providecommand \bibfield  [0]{\@secondoftwo}%
\providecommand \translation [1]{[#1]}%
\providecommand \BibitemOpen [0]{}%
\providecommand \bibitemStop [0]{}%
\providecommand \bibitemNoStop [0]{.\EOS\space}%
\providecommand \EOS [0]{\spacefactor3000\relax}%
\providecommand \BibitemShut  [1]{\csname bibitem#1\endcsname}%
\let\auto@bib@innerbib\@empty
\bibitem [{\citenamefont {Carter}\ \emph {et~al.}(2024)\citenamefont {Carter}, \citenamefont {O’Reilly}, \citenamefont {Toh}, \citenamefont {Saha}, \citenamefont {Shalaev}, \citenamefont {Goetting},\ and\ \citenamefont {Monroe}}]{carter2024}%
  \BibitemOpen
  \bibfield  {author} {\bibinfo {author} {\bibfnamefont {A.~L.}\ \bibnamefont {Carter}}, \bibinfo {author} {\bibfnamefont {J.}~\bibnamefont {O’Reilly}}, \bibinfo {author} {\bibfnamefont {G.}~\bibnamefont {Toh}}, \bibinfo {author} {\bibfnamefont {S.}~\bibnamefont {Saha}}, \bibinfo {author} {\bibfnamefont {M.}~\bibnamefont {Shalaev}}, \bibinfo {author} {\bibfnamefont {I.}~\bibnamefont {Goetting}},\ and\ \bibinfo {author} {\bibfnamefont {C.}~\bibnamefont {Monroe}},\ }\bibfield  {title} {\bibinfo {title} {Ion trap with in-vacuum high numerical aperture imaging for a dual-species modular quantum computer},\ }\href {https://doi.org/10.1063/5.0180732} {\bibfield  {journal} {\bibinfo  {journal} {Review of Scientific Instruments}\ }\textbf {\bibinfo {volume} {95}},\ \bibinfo {pages} {033201} (\bibinfo {year} {2024})}\BibitemShut {NoStop}%
\bibitem [{\citenamefont {Arnold}\ \emph {et~al.}(2019)\citenamefont {Arnold}, \citenamefont {Chanu}, \citenamefont {Kaewuam}, \citenamefont {Tan}, \citenamefont {Yeo}, \citenamefont {Zhang}, \citenamefont {Safronova},\ and\ \citenamefont {Barrett}}]{arnold2019}%
  \BibitemOpen
  \bibfield  {author} {\bibinfo {author} {\bibfnamefont {K.~J.}\ \bibnamefont {Arnold}}, \bibinfo {author} {\bibfnamefont {S.~R.}\ \bibnamefont {Chanu}}, \bibinfo {author} {\bibfnamefont {R.}~\bibnamefont {Kaewuam}}, \bibinfo {author} {\bibfnamefont {T.~R.}\ \bibnamefont {Tan}}, \bibinfo {author} {\bibfnamefont {L.}~\bibnamefont {Yeo}}, \bibinfo {author} {\bibfnamefont {Z.}~\bibnamefont {Zhang}}, \bibinfo {author} {\bibfnamefont {M.~S.}\ \bibnamefont {Safronova}},\ and\ \bibinfo {author} {\bibfnamefont {M.~D.}\ \bibnamefont {Barrett}},\ }\bibfield  {title} {\bibinfo {title} {Measurements of the branching ratios for {$6P_{1/2}$} decays in $^{138}\mathrm{Ba}^+$},\ }\href {https://doi.org/10.1103/PhysRevA.100.032503} {\bibfield  {journal} {\bibinfo  {journal} {Phys. Rev. A}\ }\textbf {\bibinfo {volume} {100}},\ \bibinfo {pages} {032503} (\bibinfo {year} {2019})}\BibitemShut {NoStop}%
\bibitem [{\citenamefont {{SCHOTT AG}}()}]{schott_datasheet}%
  \BibitemOpen
  \bibfield  {author} {\bibinfo {author} {\bibnamefont {{SCHOTT AG}}},\ }\href@noop {} {\bibinfo {title} {Optical glass -- datasheet collection}},\ \bibinfo {howpublished} {\url{https://media.schott.com/api/public/content/820eba3413cc4e788433a3751f8edba9?v=97b3ea2b}},\ \bibinfo {note} {{SF57} stress-optical coefficient}\BibitemShut {NoStop}%
\bibitem [{\citenamefont {Nadlinger}(2022)}]{nadlinger2022thesis}%
  \BibitemOpen
  \bibfield  {author} {\bibinfo {author} {\bibfnamefont {D.~P.}\ \bibnamefont {Nadlinger}},\ }\emph {\bibinfo {title} {Device-independent key distribution between trapped-ion quantum network nodes}},\ \href@noop {} {Ph.D. thesis},\ \bibinfo  {school} {University of Oxford} (\bibinfo {year} {2022})\BibitemShut {NoStop}%
\bibitem [{\citenamefont {{Thorlabs, Inc.}}()}]{thorlabs_S405XP}%
  \BibitemOpen
  \bibfield  {author} {\bibinfo {author} {\bibnamefont {{Thorlabs, Inc.}}},\ }\href@noop {} {\bibinfo {title} {{S405-XP} single-mode optical fiber, specification sheet}},\ \bibinfo {howpublished} {\url{https://media.thorlabs.com/globalassets/items/s/s4/s40/s405-xp/ttn020911-s01.pdf?v=0116024102}}\BibitemShut {NoStop}%
\bibitem [{\citenamefont {Br{\"u}ning}\ \emph {et~al.}(2015)\citenamefont {Br{\"u}ning}, \citenamefont {Zhang}, \citenamefont {McLaren}, \citenamefont {Duparr{\'e}},\ and\ \citenamefont {Forbes}}]{bruning2015}%
  \BibitemOpen
  \bibfield  {author} {\bibinfo {author} {\bibfnamefont {R.}~\bibnamefont {Br{\"u}ning}}, \bibinfo {author} {\bibfnamefont {Y.}~\bibnamefont {Zhang}}, \bibinfo {author} {\bibfnamefont {M.}~\bibnamefont {McLaren}}, \bibinfo {author} {\bibfnamefont {M.}~\bibnamefont {Duparr{\'e}}},\ and\ \bibinfo {author} {\bibfnamefont {A.}~\bibnamefont {Forbes}},\ }\bibfield  {title} {\bibinfo {title} {Overlap relation between free-space {Laguerre} {Gaussian} modes and step-index fiber modes},\ }\href {https://doi.org/10.1364/JOSAA.32.001678} {\bibfield  {journal} {\bibinfo  {journal} {J. Opt. Soc. Am. A}\ }\textbf {\bibinfo {volume} {32}},\ \bibinfo {pages} {1678} (\bibinfo {year} {2015})}\BibitemShut {NoStop}%
\bibitem [{\citenamefont {Bock}\ \emph {et~al.}(2018)\citenamefont {Bock}, \citenamefont {Eich}, \citenamefont {Kucera}, \citenamefont {Kreis}, \citenamefont {Lenhard}, \citenamefont {Becher},\ and\ \citenamefont {Eschner}}]{bock2018}%
  \BibitemOpen
  \bibfield  {author} {\bibinfo {author} {\bibfnamefont {M.}~\bibnamefont {Bock}}, \bibinfo {author} {\bibfnamefont {P.}~\bibnamefont {Eich}}, \bibinfo {author} {\bibfnamefont {S.}~\bibnamefont {Kucera}}, \bibinfo {author} {\bibfnamefont {M.}~\bibnamefont {Kreis}}, \bibinfo {author} {\bibfnamefont {A.}~\bibnamefont {Lenhard}}, \bibinfo {author} {\bibfnamefont {C.}~\bibnamefont {Becher}},\ and\ \bibinfo {author} {\bibfnamefont {J.}~\bibnamefont {Eschner}},\ }\bibfield  {title} {\bibinfo {title} {High-fidelity entanglement between a trapped ion and a telecom photon via quantum frequency conversion},\ }\href {https://doi.org/10.1038/s41467-018-04341-2} {\bibfield  {journal} {\bibinfo  {journal} {Nat. Commun.}\ }\textbf {\bibinfo {volume} {9}},\ \bibinfo {pages} {1998} (\bibinfo {year} {2018})}\BibitemShut {NoStop}%
\bibitem [{\citenamefont {B{\"u}ki}\ \emph {et~al.}(2026)\citenamefont {B{\"u}ki}, \citenamefont {Malik}, \citenamefont {Fertig}, \citenamefont {Frank}, \citenamefont {Scholz}, \citenamefont {Block}, \citenamefont {Chiarella}, \citenamefont {Zhou}, \citenamefont {Distante}, \citenamefont {Farrera}, \citenamefont {Rempe},\ and\ \citenamefont {Weinfurter}}]{buki2026}%
  \BibitemOpen
  \bibfield  {author} {\bibinfo {author} {\bibfnamefont {M.}~\bibnamefont {B{\"u}ki}}, \bibinfo {author} {\bibfnamefont {P.}~\bibnamefont {Malik}}, \bibinfo {author} {\bibfnamefont {F.}~\bibnamefont {Fertig}}, \bibinfo {author} {\bibfnamefont {T.}~\bibnamefont {Frank}}, \bibinfo {author} {\bibfnamefont {M.}~\bibnamefont {Scholz}}, \bibinfo {author} {\bibfnamefont {T.}~\bibnamefont {Block}}, \bibinfo {author} {\bibfnamefont {G.}~\bibnamefont {Chiarella}}, \bibinfo {author} {\bibfnamefont {Y.}~\bibnamefont {Zhou}}, \bibinfo {author} {\bibfnamefont {E.}~\bibnamefont {Distante}}, \bibinfo {author} {\bibfnamefont {P.}~\bibnamefont {Farrera}}, \bibinfo {author} {\bibfnamefont {G.}~\bibnamefont {Rempe}},\ and\ \bibinfo {author} {\bibfnamefont {H.}~\bibnamefont {Weinfurter}},\ }\bibfield  {title} {\bibinfo {title} {Metropolitan entanglement distribution between an atom and a near-visible photon},\ }\href {https://doi.org/10.1103/94hz-xtht} {\bibfield  {journal} {\bibinfo  {journal} {Phys. Rev. Lett.}\ }\textbf
  {\bibinfo {volume} {137}},\ \bibinfo {pages} {090803} (\bibinfo {year} {2026})}\BibitemShut {NoStop}%
\bibitem [{\citenamefont {Stas}\ \emph {et~al.}(2022)\citenamefont {Stas}, \citenamefont {Huan}, \citenamefont {Machielse}, \citenamefont {Knall}, \citenamefont {Suleymanzade}, \citenamefont {Pingault}, \citenamefont {Sutula}, \citenamefont {Ding}, \citenamefont {Knaut}, \citenamefont {Assumpcao}, \citenamefont {Wei}, \citenamefont {Bhaskar}, \citenamefont {Riedinger}, \citenamefont {Sukachev}, \citenamefont {Park}, \citenamefont {Lončar}, \citenamefont {Levonian},\ and\ \citenamefont {Lukin}}]{stas2022science}%
  \BibitemOpen
  \bibfield  {author} {\bibinfo {author} {\bibfnamefont {P.-J.}\ \bibnamefont {Stas}}, \bibinfo {author} {\bibfnamefont {Y.~Q.}\ \bibnamefont {Huan}}, \bibinfo {author} {\bibfnamefont {B.}~\bibnamefont {Machielse}}, \bibinfo {author} {\bibfnamefont {E.~N.}\ \bibnamefont {Knall}}, \bibinfo {author} {\bibfnamefont {A.}~\bibnamefont {Suleymanzade}}, \bibinfo {author} {\bibfnamefont {B.}~\bibnamefont {Pingault}}, \bibinfo {author} {\bibfnamefont {M.}~\bibnamefont {Sutula}}, \bibinfo {author} {\bibfnamefont {S.~W.}\ \bibnamefont {Ding}}, \bibinfo {author} {\bibfnamefont {C.~M.}\ \bibnamefont {Knaut}}, \bibinfo {author} {\bibfnamefont {D.~R.}\ \bibnamefont {Assumpcao}}, \bibinfo {author} {\bibfnamefont {Y.-C.}\ \bibnamefont {Wei}}, \bibinfo {author} {\bibfnamefont {M.~K.}\ \bibnamefont {Bhaskar}}, \bibinfo {author} {\bibfnamefont {R.}~\bibnamefont {Riedinger}}, \bibinfo {author} {\bibfnamefont {D.~D.}\ \bibnamefont {Sukachev}}, \bibinfo {author} {\bibfnamefont {H.}~\bibnamefont {Park}}, \bibinfo {author}
  {\bibfnamefont {M.}~\bibnamefont {Lončar}}, \bibinfo {author} {\bibfnamefont {D.~S.}\ \bibnamefont {Levonian}},\ and\ \bibinfo {author} {\bibfnamefont {M.~D.}\ \bibnamefont {Lukin}},\ }\bibfield  {title} {\bibinfo {title} {Robust multi-qubit quantum network node with integrated error detection},\ }\href {https://doi.org/10.1126/science.add9771} {\bibfield  {journal} {\bibinfo  {journal} {Science}\ }\textbf {\bibinfo {volume} {378}},\ \bibinfo {pages} {557} (\bibinfo {year} {2022})}\BibitemShut {NoStop}%
\bibitem [{\citenamefont {Nguyen}\ \emph {et~al.}(2019{\natexlab{a}})\citenamefont {Nguyen}, \citenamefont {Sukachev}, \citenamefont {Bhaskar}, \citenamefont {Machielse}, \citenamefont {Levonian}, \citenamefont {Knall}, \citenamefont {Stroganov}, \citenamefont {Chia}, \citenamefont {Burek}, \citenamefont {Riedinger}, \citenamefont {Park}, \citenamefont {Lon{\v c}ar},\ and\ \citenamefont {Lukin}}]{nguyen2019prl}%
  \BibitemOpen
  \bibfield  {author} {\bibinfo {author} {\bibfnamefont {C.~T.}\ \bibnamefont {Nguyen}}, \bibinfo {author} {\bibfnamefont {D.~D.}\ \bibnamefont {Sukachev}}, \bibinfo {author} {\bibfnamefont {M.~K.}\ \bibnamefont {Bhaskar}}, \bibinfo {author} {\bibfnamefont {B.}~\bibnamefont {Machielse}}, \bibinfo {author} {\bibfnamefont {D.~S.}\ \bibnamefont {Levonian}}, \bibinfo {author} {\bibfnamefont {E.~N.}\ \bibnamefont {Knall}}, \bibinfo {author} {\bibfnamefont {P.}~\bibnamefont {Stroganov}}, \bibinfo {author} {\bibfnamefont {C.}~\bibnamefont {Chia}}, \bibinfo {author} {\bibfnamefont {M.~J.}\ \bibnamefont {Burek}}, \bibinfo {author} {\bibfnamefont {R.}~\bibnamefont {Riedinger}}, \bibinfo {author} {\bibfnamefont {H.}~\bibnamefont {Park}}, \bibinfo {author} {\bibfnamefont {M.}~\bibnamefont {Lon{\v c}ar}},\ and\ \bibinfo {author} {\bibfnamefont {M.~D.}\ \bibnamefont {Lukin}},\ }\bibfield  {title} {\bibinfo {title} {Quantum network nodes based on diamond qubits with an efficient nanophotonic interface},\ }\href
  {https://doi.org/10.1103/PhysRevLett.123.183602} {\bibfield  {journal} {\bibinfo  {journal} {Phys. Rev. Lett.}\ }\textbf {\bibinfo {volume} {123}},\ \bibinfo {pages} {183602} (\bibinfo {year} {2019}{\natexlab{a}})}\BibitemShut {NoStop}%
\bibitem [{\citenamefont {Nguyen}\ \emph {et~al.}(2019{\natexlab{b}})\citenamefont {Nguyen}, \citenamefont {Sukachev}, \citenamefont {Bhaskar}, \citenamefont {Machielse}, \citenamefont {Levonian}, \citenamefont {Knall}, \citenamefont {Stroganov}, \citenamefont {Chia}, \citenamefont {Burek}, \citenamefont {Riedinger}, \citenamefont {Park}, \citenamefont {Lon{\v c}ar},\ and\ \citenamefont {Lukin}}]{nguyen2019prb}%
  \BibitemOpen
  \bibfield  {author} {\bibinfo {author} {\bibfnamefont {C.~T.}\ \bibnamefont {Nguyen}}, \bibinfo {author} {\bibfnamefont {D.~D.}\ \bibnamefont {Sukachev}}, \bibinfo {author} {\bibfnamefont {M.~K.}\ \bibnamefont {Bhaskar}}, \bibinfo {author} {\bibfnamefont {B.}~\bibnamefont {Machielse}}, \bibinfo {author} {\bibfnamefont {D.~S.}\ \bibnamefont {Levonian}}, \bibinfo {author} {\bibfnamefont {E.~N.}\ \bibnamefont {Knall}}, \bibinfo {author} {\bibfnamefont {P.}~\bibnamefont {Stroganov}}, \bibinfo {author} {\bibfnamefont {C.}~\bibnamefont {Chia}}, \bibinfo {author} {\bibfnamefont {M.~J.}\ \bibnamefont {Burek}}, \bibinfo {author} {\bibfnamefont {R.}~\bibnamefont {Riedinger}}, \bibinfo {author} {\bibfnamefont {H.}~\bibnamefont {Park}}, \bibinfo {author} {\bibfnamefont {M.}~\bibnamefont {Lon{\v c}ar}},\ and\ \bibinfo {author} {\bibfnamefont {M.~D.}\ \bibnamefont {Lukin}},\ }\bibfield  {title} {\bibinfo {title} {An integrated nanophotonic quantum register based on silicon-vacancy spins in diamond},\ }\href@noop {}
  {\bibfield  {journal} {\bibinfo  {journal} {Phys. Rev. B}\ }\textbf {\bibinfo {volume} {100}},\ \bibinfo {pages} {165428} (\bibinfo {year} {2019}{\natexlab{b}})}\BibitemShut {NoStop}%
\bibitem [{\citenamefont {James}\ \emph {et~al.}(2001)\citenamefont {James}, \citenamefont {Kwiat}, \citenamefont {Munro},\ and\ \citenamefont {White}}]{James2001_PRA64_052312_MeasurementOfQubits}%
  \BibitemOpen
  \bibfield  {author} {\bibinfo {author} {\bibfnamefont {D.~F.~V.}\ \bibnamefont {James}}, \bibinfo {author} {\bibfnamefont {P.~G.}\ \bibnamefont {Kwiat}}, \bibinfo {author} {\bibfnamefont {W.~J.}\ \bibnamefont {Munro}},\ and\ \bibinfo {author} {\bibfnamefont {A.~G.}\ \bibnamefont {White}},\ }\bibfield  {title} {\bibinfo {title} {Measurement of qubits},\ }\href {https://doi.org/10.1103/PhysRevA.64.052312} {\bibfield  {journal} {\bibinfo  {journal} {Phys. Rev. A}\ }\textbf {\bibinfo {volume} {64}},\ \bibinfo {pages} {052312} (\bibinfo {year} {2001})}\BibitemShut {NoStop}%
\bibitem [{\citenamefont {Diamond}\ and\ \citenamefont {Boyd}(2016)}]{Diamond2016_JMLR17_83_CVXPY}%
  \BibitemOpen
  \bibfield  {author} {\bibinfo {author} {\bibfnamefont {S.}~\bibnamefont {Diamond}}\ and\ \bibinfo {author} {\bibfnamefont {S.}~\bibnamefont {Boyd}},\ }\bibfield  {title} {\bibinfo {title} {{CVXPY}: A {Python}-embedded modeling language for convex optimization},\ }\href {https://jmlr.org/papers/v17/15-408.html} {\bibfield  {journal} {\bibinfo  {journal} {J. Mach. Learn. Res.}\ }\textbf {\bibinfo {volume} {17}},\ \bibinfo {pages} {1} (\bibinfo {year} {2016})}\BibitemShut {NoStop}%
\end{thebibliography}%


\begin{thebibliography}{43}%
\makeatletter
\providecommand \@ifxundefined [1]{%
 \@ifx{#1\undefined}
}%
\providecommand \@ifnum [1]{%
 \ifnum #1\expandafter \@firstoftwo
 \else \expandafter \@secondoftwo
 \fi
}%
\providecommand \@ifx [1]{%
 \ifx #1\expandafter \@firstoftwo
 \else \expandafter \@secondoftwo
 \fi
}%
\providecommand \natexlab [1]{#1}%
\providecommand \enquote  [1]{``#1''}%
\providecommand \bibnamefont  [1]{#1}%
\providecommand \bibfnamefont [1]{#1}%
\providecommand \citenamefont [1]{#1}%
\providecommand \href@noop [0]{\@secondoftwo}%
\providecommand \href [0]{\begingroup \@sanitize@url \@href}%
\providecommand \@href[1]{\@@startlink{#1}\@@href}%
\providecommand \@@href[1]{\endgroup#1\@@endlink}%
\providecommand \@sanitize@url [0]{\catcode `\\12\catcode `\$12\catcode `\&12\catcode `\#12\catcode `\^12\catcode `\_12\catcode `\%12\relax}%
\providecommand \@@startlink[1]{}%
\providecommand \@@endlink[0]{}%
\providecommand \url  [0]{\begingroup\@sanitize@url \@url }%
\providecommand \@url [1]{\endgroup\@href {#1}{\urlprefix }}%
\providecommand \urlprefix  [0]{URL }%
\providecommand \Eprint [0]{\href }%
\providecommand \doibase [0]{https://doi.org/}%
\providecommand \selectlanguage [0]{\@gobble}%
\providecommand \bibinfo  [0]{\@secondoftwo}%
\providecommand \bibfield  [0]{\@secondoftwo}%
\providecommand \translation [1]{[#1]}%
\providecommand \BibitemOpen [0]{}%
\providecommand \bibitemStop [0]{}%
\providecommand \bibitemNoStop [0]{.\EOS\space}%
\providecommand \EOS [0]{\spacefactor3000\relax}%
\providecommand \BibitemShut  [1]{\csname bibitem#1\endcsname}%
\let\auto@bib@innerbib\@empty
\bibitem [{\citenamefont {Monroe}\ \emph {et~al.}(2014)\citenamefont {Monroe}, \citenamefont {Raussendorf}, \citenamefont {Ruthven}, \citenamefont {Brown}, \citenamefont {Maunz}, \citenamefont {Duan},\ and\ \citenamefont {Kim}}]{monroe2014modular}%
  \BibitemOpen
  \bibfield  {author} {\bibinfo {author} {\bibfnamefont {C.}~\bibnamefont {Monroe}}, \bibinfo {author} {\bibfnamefont {R.}~\bibnamefont {Raussendorf}}, \bibinfo {author} {\bibfnamefont {A.}~\bibnamefont {Ruthven}}, \bibinfo {author} {\bibfnamefont {K.~R.}\ \bibnamefont {Brown}}, \bibinfo {author} {\bibfnamefont {P.}~\bibnamefont {Maunz}}, \bibinfo {author} {\bibfnamefont {L.-M.}\ \bibnamefont {Duan}},\ and\ \bibinfo {author} {\bibfnamefont {J.}~\bibnamefont {Kim}},\ }\bibfield  {title} {\bibinfo {title} {Large-scale modular quantum-computer architecture with atomic memory and photonic interconnects},\ }\href {https://doi.org/10.1103/PhysRevA.89.022317} {\bibfield  {journal} {\bibinfo  {journal} {Phys. Rev. A}\ }\textbf {\bibinfo {volume} {89}},\ \bibinfo {pages} {022317} (\bibinfo {year} {2014})}\BibitemShut {NoStop}%
\bibitem [{\citenamefont {Main}\ \emph {et~al.}(2025)\citenamefont {Main}, \citenamefont {Drmota}, \citenamefont {Nadlinger}, \citenamefont {Ainley}, \citenamefont {Agrawal}, \citenamefont {Nichol}, \citenamefont {Srinivas}, \citenamefont {Araneda},\ and\ \citenamefont {Lucas}}]{main2025}%
  \BibitemOpen
  \bibfield  {author} {\bibinfo {author} {\bibfnamefont {D.}~\bibnamefont {Main}}, \bibinfo {author} {\bibfnamefont {P.}~\bibnamefont {Drmota}}, \bibinfo {author} {\bibfnamefont {D.~P.}\ \bibnamefont {Nadlinger}}, \bibinfo {author} {\bibfnamefont {E.~M.}\ \bibnamefont {Ainley}}, \bibinfo {author} {\bibfnamefont {A.}~\bibnamefont {Agrawal}}, \bibinfo {author} {\bibfnamefont {B.~C.}\ \bibnamefont {Nichol}}, \bibinfo {author} {\bibfnamefont {R.}~\bibnamefont {Srinivas}}, \bibinfo {author} {\bibfnamefont {G.}~\bibnamefont {Araneda}},\ and\ \bibinfo {author} {\bibfnamefont {D.~M.}\ \bibnamefont {Lucas}},\ }\bibfield  {title} {\bibinfo {title} {Distributed quantum computing across an optical network link},\ }\href {https://doi.org/10.1038/s41586-024-08404-x} {\bibfield  {journal} {\bibinfo  {journal} {Nature}\ }\textbf {\bibinfo {volume} {638}},\ \bibinfo {pages} {383} (\bibinfo {year} {2025})}\BibitemShut {NoStop}%
\bibitem [{\citenamefont {Wu}\ \emph {et~al.}(2024)\citenamefont {Wu}, \citenamefont {Yan}, \citenamefont {Andersson}, \citenamefont {Anferov}, \citenamefont {Chou}, \citenamefont {Conner}, \citenamefont {Grebel}, \citenamefont {Joshi}, \citenamefont {Li}, \citenamefont {Miller}, \citenamefont {Povey}, \citenamefont {Qiao},\ and\ \citenamefont {Cleland}}]{Wu2024}%
  \BibitemOpen
  \bibfield  {author} {\bibinfo {author} {\bibfnamefont {X.}~\bibnamefont {Wu}}, \bibinfo {author} {\bibfnamefont {H.}~\bibnamefont {Yan}}, \bibinfo {author} {\bibfnamefont {G.}~\bibnamefont {Andersson}}, \bibinfo {author} {\bibfnamefont {A.}~\bibnamefont {Anferov}}, \bibinfo {author} {\bibfnamefont {M.-H.}\ \bibnamefont {Chou}}, \bibinfo {author} {\bibfnamefont {C.~R.}\ \bibnamefont {Conner}}, \bibinfo {author} {\bibfnamefont {J.}~\bibnamefont {Grebel}}, \bibinfo {author} {\bibfnamefont {Y.~J.}\ \bibnamefont {Joshi}}, \bibinfo {author} {\bibfnamefont {S.}~\bibnamefont {Li}}, \bibinfo {author} {\bibfnamefont {J.~M.}\ \bibnamefont {Miller}}, \bibinfo {author} {\bibfnamefont {R.~G.}\ \bibnamefont {Povey}}, \bibinfo {author} {\bibfnamefont {H.}~\bibnamefont {Qiao}},\ and\ \bibinfo {author} {\bibfnamefont {A.~N.}\ \bibnamefont {Cleland}},\ }\bibfield  {title} {\bibinfo {title} {Modular quantum processor with an all-to-all reconfigurable router},\ }\href {https://doi.org/10.1103/PhysRevX.14.041030} {\bibfield
  {journal} {\bibinfo  {journal} {Phys. Rev. X}\ }\textbf {\bibinfo {volume} {14}},\ \bibinfo {pages} {041030} (\bibinfo {year} {2024})}\BibitemShut {NoStop}%
\bibitem [{\citenamefont {Nickerson}\ \emph {et~al.}(2014)\citenamefont {Nickerson}, \citenamefont {Fitzsimons},\ and\ \citenamefont {Benjamin}}]{Nickerson2014}%
  \BibitemOpen
  \bibfield  {author} {\bibinfo {author} {\bibfnamefont {N.~H.}\ \bibnamefont {Nickerson}}, \bibinfo {author} {\bibfnamefont {J.~F.}\ \bibnamefont {Fitzsimons}},\ and\ \bibinfo {author} {\bibfnamefont {S.~C.}\ \bibnamefont {Benjamin}},\ }\bibfield  {title} {\bibinfo {title} {Freely scalable quantum technologies using cells of 5-to-50 qubits with very lossy and noisy photonic links},\ }\href {https://doi.org/10.1103/PhysRevX.4.041041} {\bibfield  {journal} {\bibinfo  {journal} {Phys. Rev. X}\ }\textbf {\bibinfo {volume} {4}},\ \bibinfo {pages} {041041} (\bibinfo {year} {2014})}\BibitemShut {NoStop}%
\bibitem [{\citenamefont {Sinclair}\ \emph {et~al.}(2025)\citenamefont {Sinclair}, \citenamefont {Ramette}, \citenamefont {Grinkemeyer}, \citenamefont {Bluvstein}, \citenamefont {Lukin},\ and\ \citenamefont {Vuleti{\ifmmode \acute{c}\else \'{c}\fi{}}}}]{sinclair2025modular}%
  \BibitemOpen
  \bibfield  {author} {\bibinfo {author} {\bibfnamefont {J.}~\bibnamefont {Sinclair}}, \bibinfo {author} {\bibfnamefont {J.}~\bibnamefont {Ramette}}, \bibinfo {author} {\bibfnamefont {B.}~\bibnamefont {Grinkemeyer}}, \bibinfo {author} {\bibfnamefont {D.}~\bibnamefont {Bluvstein}}, \bibinfo {author} {\bibfnamefont {M.~D.}\ \bibnamefont {Lukin}},\ and\ \bibinfo {author} {\bibfnamefont {V.}~\bibnamefont {Vuleti{\ifmmode \acute{c}\else \'{c}\fi{}}}},\ }\bibfield  {title} {\bibinfo {title} {Fault-tolerant optical interconnects for neutral-atom arrays},\ }\href {https://doi.org/10.1103/PhysRevResearch.7.013313} {\bibfield  {journal} {\bibinfo  {journal} {Phys. Rev. Res.}\ }\textbf {\bibinfo {volume} {7}},\ \bibinfo {pages} {013313} (\bibinfo {year} {2025})}\BibitemShut {NoStop}%
\bibitem [{\citenamefont {Xue}\ and\ \citenamefont {Covey}(2026)}]{xue2026modular}%
  \BibitemOpen
  \bibfield  {author} {\bibinfo {author} {\bibfnamefont {T.}~\bibnamefont {Xue}}\ and\ \bibinfo {author} {\bibfnamefont {J.~P.}\ \bibnamefont {Covey}},\ }\href {https://arxiv.org/abs/2605.03951} {\bibinfo {title} {Factoring $2048$ bit {RSA} integers with a half-million-qubit modular atomic processor}} (\bibinfo {year} {2026}),\ \Eprint {https://arxiv.org/abs/2605.03951} {arXiv:2605.03951 [quant-ph]} \BibitemShut {NoStop}%
\bibitem [{\citenamefont {Marqversen}\ \emph {et~al.}(2026)\citenamefont {Marqversen}, \citenamefont {Baranes}, \citenamefont {Sirotin},\ and\ \citenamefont {Borregaard}}]{Marqversen2026}%
  \BibitemOpen
  \bibfield  {author} {\bibinfo {author} {\bibfnamefont {F.~K.}\ \bibnamefont {Marqversen}}, \bibinfo {author} {\bibfnamefont {G.}~\bibnamefont {Baranes}}, \bibinfo {author} {\bibfnamefont {M.}~\bibnamefont {Sirotin}},\ and\ \bibinfo {author} {\bibfnamefont {J.}~\bibnamefont {Borregaard}},\ }\bibfield  {title} {\bibinfo {title} {Fault-tolerant interfaces for modular quantum computing on diverse qubit platforms},\ }\href {https://doi.org/10.1103/ds45-fm9n} {\bibfield  {journal} {\bibinfo  {journal} {Phys. Rev. Res.}\ }\textbf {\bibinfo {volume} {8}},\ \bibinfo {pages} {023097} (\bibinfo {year} {2026})}\BibitemShut {NoStop}%
\bibitem [{\citenamefont {Ramette}\ \emph {et~al.}(2024)\citenamefont {Ramette}, \citenamefont {Sinclair}, \citenamefont {Breuckmann},\ and\ \citenamefont {Vuletić}}]{ramette2024}%
  \BibitemOpen
  \bibfield  {author} {\bibinfo {author} {\bibfnamefont {J.}~\bibnamefont {Ramette}}, \bibinfo {author} {\bibfnamefont {J.}~\bibnamefont {Sinclair}}, \bibinfo {author} {\bibfnamefont {N.~P.}\ \bibnamefont {Breuckmann}},\ and\ \bibinfo {author} {\bibfnamefont {V.}~\bibnamefont {Vuletić}},\ }\bibfield  {title} {\bibinfo {title} {Fault-tolerant connection of error-corrected qubits with noisy links},\ }\href {https://doi.org/10.1038/s41534-024-00855-4} {\bibfield  {journal} {\bibinfo  {journal} {npj Quantum Information}\ }\textbf {\bibinfo {volume} {10}},\ \bibinfo {pages} {58} (\bibinfo {year} {2024})}\BibitemShut {NoStop}%
\bibitem [{\citenamefont {Hughes}\ \emph {et~al.}(2025)\citenamefont {Hughes}, \citenamefont {Srinivas}, \citenamefont {Löschnauer}, \citenamefont {Knaack}, \citenamefont {Matt}, \citenamefont {Ballance}, \citenamefont {Malinowski}, \citenamefont {Harty},\ and\ \citenamefont {Sutherland}}]{hughes2025}%
  \BibitemOpen
  \bibfield  {author} {\bibinfo {author} {\bibfnamefont {A.~C.}\ \bibnamefont {Hughes}}, \bibinfo {author} {\bibfnamefont {R.}~\bibnamefont {Srinivas}}, \bibinfo {author} {\bibfnamefont {C.~M.}\ \bibnamefont {Löschnauer}}, \bibinfo {author} {\bibfnamefont {H.~M.}\ \bibnamefont {Knaack}}, \bibinfo {author} {\bibfnamefont {R.}~\bibnamefont {Matt}}, \bibinfo {author} {\bibfnamefont {C.~J.}\ \bibnamefont {Ballance}}, \bibinfo {author} {\bibfnamefont {M.}~\bibnamefont {Malinowski}}, \bibinfo {author} {\bibfnamefont {T.~P.}\ \bibnamefont {Harty}},\ and\ \bibinfo {author} {\bibfnamefont {R.~T.}\ \bibnamefont {Sutherland}},\ }\href {https://arxiv.org/abs/2510.17286} {\bibinfo {title} {Trapped-ion two-qubit gates with $>99.99\%$ fidelity without ground-state cooling}} (\bibinfo {year} {2025}),\ \Eprint {https://arxiv.org/abs/2510.17286} {arXiv:2510.17286 [quant-ph]} \BibitemShut {NoStop}%
\bibitem [{\citenamefont {O'Reilly}\ \emph {et~al.}(2024)\citenamefont {O'Reilly}, \citenamefont {Toh}, \citenamefont {Goetting}, \citenamefont {Saha}, \citenamefont {Shalaev}, \citenamefont {Carter}, \citenamefont {Risinger}, \citenamefont {Kalakuntla}, \citenamefont {Li}, \citenamefont {Verma},\ and\ \citenamefont {Monroe}}]{Oreilly2024}%
  \BibitemOpen
  \bibfield  {author} {\bibinfo {author} {\bibfnamefont {J.}~\bibnamefont {O'Reilly}}, \bibinfo {author} {\bibfnamefont {G.}~\bibnamefont {Toh}}, \bibinfo {author} {\bibfnamefont {I.}~\bibnamefont {Goetting}}, \bibinfo {author} {\bibfnamefont {S.}~\bibnamefont {Saha}}, \bibinfo {author} {\bibfnamefont {M.}~\bibnamefont {Shalaev}}, \bibinfo {author} {\bibfnamefont {A.~L.}\ \bibnamefont {Carter}}, \bibinfo {author} {\bibfnamefont {A.}~\bibnamefont {Risinger}}, \bibinfo {author} {\bibfnamefont {A.}~\bibnamefont {Kalakuntla}}, \bibinfo {author} {\bibfnamefont {T.}~\bibnamefont {Li}}, \bibinfo {author} {\bibfnamefont {A.}~\bibnamefont {Verma}},\ and\ \bibinfo {author} {\bibfnamefont {C.}~\bibnamefont {Monroe}},\ }\bibfield  {title} {\bibinfo {title} {Fast photon-mediated entanglement of continuously cooled trapped ions for quantum networking},\ }\href {https://doi.org/10.1103/PhysRevLett.133.090802} {\bibfield  {journal} {\bibinfo  {journal} {Phys. Rev. Lett.}\ }\textbf {\bibinfo {volume} {133}},\ \bibinfo {pages}
  {090802} (\bibinfo {year} {2024})}\BibitemShut {NoStop}%
\bibitem [{\citenamefont {Barrett}\ and\ \citenamefont {Kok}(2005)}]{Barrett2005}%
  \BibitemOpen
  \bibfield  {author} {\bibinfo {author} {\bibfnamefont {S.~D.}\ \bibnamefont {Barrett}}\ and\ \bibinfo {author} {\bibfnamefont {P.}~\bibnamefont {Kok}},\ }\bibfield  {title} {\bibinfo {title} {Efficient high-fidelity quantum computation using matter qubits and linear optics},\ }\href {https://doi.org/10.1103/PhysRevA.71.060310} {\bibfield  {journal} {\bibinfo  {journal} {Phys. Rev. A}\ }\textbf {\bibinfo {volume} {71}},\ \bibinfo {pages} {060310(R)} (\bibinfo {year} {2005})}\BibitemShut {NoStop}%
\bibitem [{\citenamefont {Stephenson}\ \emph {et~al.}(2020)\citenamefont {Stephenson}, \citenamefont {Nadlinger}, \citenamefont {Nichol}, \citenamefont {An}, \citenamefont {Drmota}, \citenamefont {Ballance}, \citenamefont {Thirumalai}, \citenamefont {Goodwin}, \citenamefont {Lucas},\ and\ \citenamefont {Ballance}}]{stephenson2020}%
  \BibitemOpen
  \bibfield  {author} {\bibinfo {author} {\bibfnamefont {L.~J.}\ \bibnamefont {Stephenson}}, \bibinfo {author} {\bibfnamefont {D.~P.}\ \bibnamefont {Nadlinger}}, \bibinfo {author} {\bibfnamefont {B.~C.}\ \bibnamefont {Nichol}}, \bibinfo {author} {\bibfnamefont {S.}~\bibnamefont {An}}, \bibinfo {author} {\bibfnamefont {P.}~\bibnamefont {Drmota}}, \bibinfo {author} {\bibfnamefont {T.~G.}\ \bibnamefont {Ballance}}, \bibinfo {author} {\bibfnamefont {K.}~\bibnamefont {Thirumalai}}, \bibinfo {author} {\bibfnamefont {J.~F.}\ \bibnamefont {Goodwin}}, \bibinfo {author} {\bibfnamefont {D.~M.}\ \bibnamefont {Lucas}},\ and\ \bibinfo {author} {\bibfnamefont {C.~J.}\ \bibnamefont {Ballance}},\ }\bibfield  {title} {\bibinfo {title} {High-rate, high-fidelity entanglement of qubits across an elementary quantum network},\ }\href {https://doi.org/10.1103/PhysRevLett.124.110501} {\bibfield  {journal} {\bibinfo  {journal} {Phys. Rev. Lett.}\ }\textbf {\bibinfo {volume} {124}},\ \bibinfo {pages} {110501} (\bibinfo {year}
  {2020})}\BibitemShut {NoStop}%
\bibitem [{\citenamefont {Li}\ \emph {et~al.}(2025)\citenamefont {Li}, \citenamefont {Hu}, \citenamefont {Jia}, \citenamefont {Huie}, \citenamefont {Sun}, \citenamefont {{Aakash}}, \citenamefont {Dong}, \citenamefont {Hiri-O-Tuppa},\ and\ \citenamefont {Covey}}]{li_parallelized_2025}%
  \BibitemOpen
  \bibfield  {author} {\bibinfo {author} {\bibfnamefont {L.}~\bibnamefont {Li}}, \bibinfo {author} {\bibfnamefont {X.}~\bibnamefont {Hu}}, \bibinfo {author} {\bibfnamefont {Z.}~\bibnamefont {Jia}}, \bibinfo {author} {\bibfnamefont {W.}~\bibnamefont {Huie}}, \bibinfo {author} {\bibfnamefont {W.~K.~C.}\ \bibnamefont {Sun}}, \bibinfo {author} {\bibnamefont {{Aakash}}}, \bibinfo {author} {\bibfnamefont {Y.}~\bibnamefont {Dong}}, \bibinfo {author} {\bibfnamefont {N.}~\bibnamefont {Hiri-O-Tuppa}},\ and\ \bibinfo {author} {\bibfnamefont {J.~P.}\ \bibnamefont {Covey}},\ }\bibfield  {title} {\bibinfo {title} {Parallelized telecom quantum networking with an ytterbium-171 atom array},\ }\href {https://doi.org/10.1038/s41567-025-03022-4} {\bibfield  {journal} {\bibinfo  {journal} {Nature Physics}\ }\textbf {\bibinfo {volume} {21}},\ \bibinfo {pages} {1826} (\bibinfo {year} {2025})}\BibitemShut {NoStop}%
\bibitem [{\citenamefont {Safari}\ \emph {et~al.}(2026)\citenamefont {Safari}, \citenamefont {Oh}, \citenamefont {Huft}, \citenamefont {Chase}, \citenamefont {Zhang},\ and\ \citenamefont {Saffman}}]{safari2026atomphoton}%
  \BibitemOpen
  \bibfield  {author} {\bibinfo {author} {\bibfnamefont {A.}~\bibnamefont {Safari}}, \bibinfo {author} {\bibfnamefont {E.}~\bibnamefont {Oh}}, \bibinfo {author} {\bibfnamefont {P.}~\bibnamefont {Huft}}, \bibinfo {author} {\bibfnamefont {G.}~\bibnamefont {Chase}}, \bibinfo {author} {\bibfnamefont {J.}~\bibnamefont {Zhang}},\ and\ \bibinfo {author} {\bibfnamefont {M.}~\bibnamefont {Saffman}},\ }\href {https://arxiv.org/abs/2601.13420} {\bibinfo {title} {Efficient and compact quantum network node based on a parabolic mirror on an optical chip}} (\bibinfo {year} {2026}),\ \Eprint {https://arxiv.org/abs/2601.13420} {arXiv:2601.13420 [quant-ph]} \BibitemShut {NoStop}%
\bibitem [{\citenamefont {Thomas}\ \emph {et~al.}(2022)\citenamefont {Thomas}, \citenamefont {Ruscio}, \citenamefont {Morin},\ and\ \citenamefont {Rempe}}]{thomas2022atomphoton}%
  \BibitemOpen
  \bibfield  {author} {\bibinfo {author} {\bibfnamefont {P.}~\bibnamefont {Thomas}}, \bibinfo {author} {\bibfnamefont {L.}~\bibnamefont {Ruscio}}, \bibinfo {author} {\bibfnamefont {O.}~\bibnamefont {Morin}},\ and\ \bibinfo {author} {\bibfnamefont {G.}~\bibnamefont {Rempe}},\ }\bibfield  {title} {\bibinfo {title} {Efficient generation of entangled multiphoton graph states from a single atom},\ }\href {https://doi.org/10.1038/s41586-022-04987-5} {\bibfield  {journal} {\bibinfo  {journal} {Nature}\ }\textbf {\bibinfo {volume} {608}},\ \bibinfo {pages} {677} (\bibinfo {year} {2022})}\BibitemShut {NoStop}%
\bibitem [{\citenamefont {Krutyanskiy}\ \emph {et~al.}(2024)\citenamefont {Krutyanskiy}, \citenamefont {Canteri}, \citenamefont {Meraner}, \citenamefont {Krcmarsky},\ and\ \citenamefont {Lanyon}}]{Krutyanskiy2024ioncavity}%
  \BibitemOpen
  \bibfield  {author} {\bibinfo {author} {\bibfnamefont {V.}~\bibnamefont {Krutyanskiy}}, \bibinfo {author} {\bibfnamefont {M.}~\bibnamefont {Canteri}}, \bibinfo {author} {\bibfnamefont {M.}~\bibnamefont {Meraner}}, \bibinfo {author} {\bibfnamefont {V.}~\bibnamefont {Krcmarsky}},\ and\ \bibinfo {author} {\bibfnamefont {B.}~\bibnamefont {Lanyon}},\ }\bibfield  {title} {\bibinfo {title} {Multimode ion-photon entanglement over 101 kilometers},\ }\href {https://doi.org/10.1103/PRXQuantum.5.020308} {\bibfield  {journal} {\bibinfo  {journal} {PRX Quantum}\ }\textbf {\bibinfo {volume} {5}},\ \bibinfo {pages} {020308} (\bibinfo {year} {2024})}\BibitemShut {NoStop}%
\bibitem [{\citenamefont {Ocola}\ \emph {et~al.}(2024)\citenamefont {Ocola}, \citenamefont {Dimitrova}, \citenamefont {Grinkemeyer}, \citenamefont {Guardado-Sanchez}, \citenamefont {Dordevic}, \citenamefont {Samutpraphoot}, \citenamefont {Vuletic},\ and\ \citenamefont {Lukin}}]{Ocola2024Rydberg}%
  \BibitemOpen
  \bibfield  {author} {\bibinfo {author} {\bibfnamefont {P.~L.}\ \bibnamefont {Ocola}}, \bibinfo {author} {\bibfnamefont {I.}~\bibnamefont {Dimitrova}}, \bibinfo {author} {\bibfnamefont {B.}~\bibnamefont {Grinkemeyer}}, \bibinfo {author} {\bibfnamefont {E.}~\bibnamefont {Guardado-Sanchez}}, \bibinfo {author} {\bibfnamefont {T.}~\bibnamefont {Dordevic}}, \bibinfo {author} {\bibfnamefont {P.}~\bibnamefont {Samutpraphoot}}, \bibinfo {author} {\bibfnamefont {V.}~\bibnamefont {Vuletic}},\ and\ \bibinfo {author} {\bibfnamefont {M.~D.}\ \bibnamefont {Lukin}},\ }\bibfield  {title} {\bibinfo {title} {Control and entanglement of individual rydberg atoms near a nanoscale device},\ }\href {https://doi.org/10.1103/PhysRevLett.132.113601} {\bibfield  {journal} {\bibinfo  {journal} {Phys. Rev. Lett.}\ }\textbf {\bibinfo {volume} {132}},\ \bibinfo {pages} {113601} (\bibinfo {year} {2024})}\BibitemShut {NoStop}%
\bibitem [{\citenamefont {Teller}\ \emph {et~al.}(2021)\citenamefont {Teller}, \citenamefont {Fioretto}, \citenamefont {Holz}, \citenamefont {Schindler}, \citenamefont {Messerer}, \citenamefont {Sch{\"u}ppert}, \citenamefont {Zou}, \citenamefont {Blatt}, \citenamefont {Chiaverini}, \citenamefont {Sage},\ and\ \citenamefont {Northup}}]{Teller2021}%
  \BibitemOpen
  \bibfield  {author} {\bibinfo {author} {\bibfnamefont {M.}~\bibnamefont {Teller}}, \bibinfo {author} {\bibfnamefont {D.~A.}\ \bibnamefont {Fioretto}}, \bibinfo {author} {\bibfnamefont {P.~C.}\ \bibnamefont {Holz}}, \bibinfo {author} {\bibfnamefont {P.}~\bibnamefont {Schindler}}, \bibinfo {author} {\bibfnamefont {V.}~\bibnamefont {Messerer}}, \bibinfo {author} {\bibfnamefont {K.}~\bibnamefont {Sch{\"u}ppert}}, \bibinfo {author} {\bibfnamefont {Y.}~\bibnamefont {Zou}}, \bibinfo {author} {\bibfnamefont {R.}~\bibnamefont {Blatt}}, \bibinfo {author} {\bibfnamefont {J.}~\bibnamefont {Chiaverini}}, \bibinfo {author} {\bibfnamefont {J.}~\bibnamefont {Sage}},\ and\ \bibinfo {author} {\bibfnamefont {T.~E.}\ \bibnamefont {Northup}},\ }\bibfield  {title} {\bibinfo {title} {Heating of a trapped ion induced by dielectric materials},\ }\href {https://doi.org/10.1103/PhysRevLett.126.230505} {\bibfield  {journal} {\bibinfo  {journal} {Phys. Rev. Lett.}\ }\textbf {\bibinfo {volume} {126}},\ \bibinfo {pages} {230505}
  (\bibinfo {year} {2021})}\BibitemShut {NoStop}%
\bibitem [{\citenamefont {Li}\ and\ \citenamefont {Thompson}(2024)}]{Li2024}%
  \BibitemOpen
  \bibfield  {author} {\bibinfo {author} {\bibfnamefont {Y.}~\bibnamefont {Li}}\ and\ \bibinfo {author} {\bibfnamefont {J.~D.}\ \bibnamefont {Thompson}},\ }\bibfield  {title} {\bibinfo {title} {High-rate and high-fidelity modular interconnects between neutral atom quantum processors},\ }\href {https://doi.org/10.1103/PRXQuantum.5.020363} {\bibfield  {journal} {\bibinfo  {journal} {PRX Quantum}\ }\textbf {\bibinfo {volume} {5}},\ \bibinfo {pages} {020363} (\bibinfo {year} {2024})}\BibitemShut {NoStop}%
\bibitem [{\citenamefont {Hartung}\ \emph {et~al.}(2024)\citenamefont {Hartung}, \citenamefont {Seubert}, \citenamefont {Welte}, \citenamefont {Distante},\ and\ \citenamefont {Rempe}}]{Hartung2024}%
  \BibitemOpen
  \bibfield  {author} {\bibinfo {author} {\bibfnamefont {L.}~\bibnamefont {Hartung}}, \bibinfo {author} {\bibfnamefont {M.}~\bibnamefont {Seubert}}, \bibinfo {author} {\bibfnamefont {S.}~\bibnamefont {Welte}}, \bibinfo {author} {\bibfnamefont {E.}~\bibnamefont {Distante}},\ and\ \bibinfo {author} {\bibfnamefont {G.}~\bibnamefont {Rempe}},\ }\bibfield  {title} {\bibinfo {title} {A quantum-network register assembled with optical tweezers in an optical cavity},\ }\href {https://doi.org/10.1126/science.ado6471} {\bibfield  {journal} {\bibinfo  {journal} {Science}\ }\textbf {\bibinfo {volume} {385}},\ \bibinfo {pages} {179} (\bibinfo {year} {2024})}\BibitemShut {NoStop}%
\bibitem [{\citenamefont {Sunami}\ \emph {et~al.}(2025)\citenamefont {Sunami}, \citenamefont {Tamiya}, \citenamefont {Inoue}, \citenamefont {Yamasaki},\ and\ \citenamefont {Goban}}]{Sunami2025}%
  \BibitemOpen
  \bibfield  {author} {\bibinfo {author} {\bibfnamefont {S.}~\bibnamefont {Sunami}}, \bibinfo {author} {\bibfnamefont {S.}~\bibnamefont {Tamiya}}, \bibinfo {author} {\bibfnamefont {R.}~\bibnamefont {Inoue}}, \bibinfo {author} {\bibfnamefont {H.}~\bibnamefont {Yamasaki}},\ and\ \bibinfo {author} {\bibfnamefont {A.}~\bibnamefont {Goban}},\ }\bibfield  {title} {\bibinfo {title} {Scalable networking of neutral-atom qubits: Nanofiber-based approach for multiprocessor fault-tolerant quantum computers},\ }\href {https://doi.org/10.1103/PRXQuantum.6.010101} {\bibfield  {journal} {\bibinfo  {journal} {PRX Quantum}\ }\textbf {\bibinfo {volume} {6}},\ \bibinfo {pages} {010101} (\bibinfo {year} {2025})}\BibitemShut {NoStop}%
\bibitem [{\citenamefont {Krutyanskiy}\ \emph {et~al.}(2023)\citenamefont {Krutyanskiy}, \citenamefont {Galli}, \citenamefont {Krcmarsky}, \citenamefont {Baier}, \citenamefont {Fioretto}, \citenamefont {Pu}, \citenamefont {Mazloom}, \citenamefont {Sekatski}, \citenamefont {Canteri}, \citenamefont {Teller}, \citenamefont {Schupp}, \citenamefont {Bate}, \citenamefont {Meraner}, \citenamefont {Sangouard}, \citenamefont {Northup},\ and\ \citenamefont {Lanyon}}]{krutyanskiy2023}%
  \BibitemOpen
  \bibfield  {author} {\bibinfo {author} {\bibfnamefont {V.}~\bibnamefont {Krutyanskiy}}, \bibinfo {author} {\bibfnamefont {M.}~\bibnamefont {Galli}}, \bibinfo {author} {\bibfnamefont {V.}~\bibnamefont {Krcmarsky}}, \bibinfo {author} {\bibfnamefont {S.}~\bibnamefont {Baier}}, \bibinfo {author} {\bibfnamefont {D.~A.}\ \bibnamefont {Fioretto}}, \bibinfo {author} {\bibfnamefont {Y.}~\bibnamefont {Pu}}, \bibinfo {author} {\bibfnamefont {A.}~\bibnamefont {Mazloom}}, \bibinfo {author} {\bibfnamefont {P.}~\bibnamefont {Sekatski}}, \bibinfo {author} {\bibfnamefont {M.}~\bibnamefont {Canteri}}, \bibinfo {author} {\bibfnamefont {M.}~\bibnamefont {Teller}}, \bibinfo {author} {\bibfnamefont {J.}~\bibnamefont {Schupp}}, \bibinfo {author} {\bibfnamefont {J.}~\bibnamefont {Bate}}, \bibinfo {author} {\bibfnamefont {M.}~\bibnamefont {Meraner}}, \bibinfo {author} {\bibfnamefont {N.}~\bibnamefont {Sangouard}}, \bibinfo {author} {\bibfnamefont {T.~E.}\ \bibnamefont {Northup}},\ and\ \bibinfo {author} {\bibfnamefont {B.~P.}\
  \bibnamefont {Lanyon}},\ }\bibfield  {title} {\bibinfo {title} {Entanglement of trapped-ion qubits separated by 230 metres},\ }\href {https://doi.org/10.1103/PhysRevLett.130.050803} {\bibfield  {journal} {\bibinfo  {journal} {Phys. Rev. Lett.}\ }\textbf {\bibinfo {volume} {130}},\ \bibinfo {pages} {050803} (\bibinfo {year} {2023})}\BibitemShut {NoStop}%
\bibitem [{\citenamefont {Slodi{\v c}ka}\ \emph {et~al.}(2013)\citenamefont {Slodi{\v c}ka}, \citenamefont {H{\'e}tet}, \citenamefont {R{\"o}ck}, \citenamefont {Schindler}, \citenamefont {Hennrich},\ and\ \citenamefont {Blatt}}]{Slodicka2013}%
  \BibitemOpen
  \bibfield  {author} {\bibinfo {author} {\bibfnamefont {L.}~\bibnamefont {Slodi{\v c}ka}}, \bibinfo {author} {\bibfnamefont {G.}~\bibnamefont {H{\'e}tet}}, \bibinfo {author} {\bibfnamefont {N.}~\bibnamefont {R{\"o}ck}}, \bibinfo {author} {\bibfnamefont {P.}~\bibnamefont {Schindler}}, \bibinfo {author} {\bibfnamefont {M.}~\bibnamefont {Hennrich}},\ and\ \bibinfo {author} {\bibfnamefont {R.}~\bibnamefont {Blatt}},\ }\bibfield  {title} {\bibinfo {title} {Atom-atom entanglement by single-photon detection},\ }\href@noop {} {\bibfield  {journal} {\bibinfo  {journal} {Phys. Rev. Lett.}\ }\textbf {\bibinfo {volume} {110}},\ \bibinfo {pages} {083603} (\bibinfo {year} {2013})}\BibitemShut {NoStop}%
\bibitem [{\citenamefont {Humphreys}\ \emph {et~al.}(2018)\citenamefont {Humphreys}, \citenamefont {Kalb}, \citenamefont {Morits}, \citenamefont {Schouten}, \citenamefont {Vermeulen}, \citenamefont {Twitchen}, \citenamefont {Markham},\ and\ \citenamefont {Hanson}}]{humphreys2018deterministic}%
  \BibitemOpen
  \bibfield  {author} {\bibinfo {author} {\bibfnamefont {P.~C.}\ \bibnamefont {Humphreys}}, \bibinfo {author} {\bibfnamefont {N.}~\bibnamefont {Kalb}}, \bibinfo {author} {\bibfnamefont {J.~P.~J.}\ \bibnamefont {Morits}}, \bibinfo {author} {\bibfnamefont {R.~N.}\ \bibnamefont {Schouten}}, \bibinfo {author} {\bibfnamefont {R.~F.~L.}\ \bibnamefont {Vermeulen}}, \bibinfo {author} {\bibfnamefont {D.~J.}\ \bibnamefont {Twitchen}}, \bibinfo {author} {\bibfnamefont {M.}~\bibnamefont {Markham}},\ and\ \bibinfo {author} {\bibfnamefont {R.}~\bibnamefont {Hanson}},\ }\bibfield  {title} {\bibinfo {title} {Deterministic delivery of remote entanglement on a quantum network},\ }\href {https://doi.org/10.1038/s41586-018-0200-5} {\bibfield  {journal} {\bibinfo  {journal} {Nature}\ }\textbf {\bibinfo {volume} {558}},\ \bibinfo {pages} {268} (\bibinfo {year} {2018})}\BibitemShut {NoStop}%
\bibitem [{\citenamefont {Knollmann}\ \emph {et~al.}(2026)\citenamefont {Knollmann}, \citenamefont {Nadlinger}, \citenamefont {Blue}, \citenamefont {Corsetti}, \citenamefont {Bishop}, \citenamefont {Martinez}, \citenamefont {Notaros}, \citenamefont {Bruzewicz}, \citenamefont {McConnell},\ and\ \citenamefont {Chuang}}]{Knollmann2026}%
  \BibitemOpen
  \bibfield  {author} {\bibinfo {author} {\bibfnamefont {F.~W.}\ \bibnamefont {Knollmann}}, \bibinfo {author} {\bibfnamefont {D.~P.}\ \bibnamefont {Nadlinger}}, \bibinfo {author} {\bibfnamefont {J.}~\bibnamefont {Blue}}, \bibinfo {author} {\bibfnamefont {S.~M.}\ \bibnamefont {Corsetti}}, \bibinfo {author} {\bibfnamefont {S.~J.}\ \bibnamefont {Bishop}}, \bibinfo {author} {\bibfnamefont {A.~R.}\ \bibnamefont {Martinez}}, \bibinfo {author} {\bibfnamefont {J.}~\bibnamefont {Notaros}}, \bibinfo {author} {\bibfnamefont {C.~D.}\ \bibnamefont {Bruzewicz}}, \bibinfo {author} {\bibfnamefont {R.}~\bibnamefont {McConnell}},\ and\ \bibinfo {author} {\bibfnamefont {I.~L.}\ \bibnamefont {Chuang}},\ }\href@noop {} {\bibinfo {title} {Remote entanglement need not be the bottleneck for modular trapped-ion quantum computing}} (\bibinfo {year} {2026}),\ \Eprint {https://arxiv.org/abs/2607.18387} {arXiv:2607.18387 [quant-ph]} \BibitemShut {NoStop}%
\bibitem [{\citenamefont {Knaut}\ \emph {et~al.}(2024)\citenamefont {Knaut}, \citenamefont {Suleymanzade}, \citenamefont {Wei}, \citenamefont {Assumpcao}, \citenamefont {Stas}, \citenamefont {Huan}, \citenamefont {Machielse}, \citenamefont {Knall}, \citenamefont {Sutula}, \citenamefont {Baranes}, \citenamefont {Sinclair}, \citenamefont {De-Eknamkul}, \citenamefont {Levonian}, \citenamefont {Bhaskar}, \citenamefont {Park}, \citenamefont {Lon{\v c}ar},\ and\ \citenamefont {Lukin}}]{knaut2024}%
  \BibitemOpen
  \bibfield  {author} {\bibinfo {author} {\bibfnamefont {C.~M.}\ \bibnamefont {Knaut}}, \bibinfo {author} {\bibfnamefont {A.}~\bibnamefont {Suleymanzade}}, \bibinfo {author} {\bibfnamefont {Y.-C.}\ \bibnamefont {Wei}}, \bibinfo {author} {\bibfnamefont {D.~R.}\ \bibnamefont {Assumpcao}}, \bibinfo {author} {\bibfnamefont {P.-J.}\ \bibnamefont {Stas}}, \bibinfo {author} {\bibfnamefont {Y.~Q.}\ \bibnamefont {Huan}}, \bibinfo {author} {\bibfnamefont {B.}~\bibnamefont {Machielse}}, \bibinfo {author} {\bibfnamefont {E.~N.}\ \bibnamefont {Knall}}, \bibinfo {author} {\bibfnamefont {M.}~\bibnamefont {Sutula}}, \bibinfo {author} {\bibfnamefont {G.}~\bibnamefont {Baranes}}, \bibinfo {author} {\bibfnamefont {N.}~\bibnamefont {Sinclair}}, \bibinfo {author} {\bibfnamefont {C.}~\bibnamefont {De-Eknamkul}}, \bibinfo {author} {\bibfnamefont {D.~S.}\ \bibnamefont {Levonian}}, \bibinfo {author} {\bibfnamefont {M.~K.}\ \bibnamefont {Bhaskar}}, \bibinfo {author} {\bibfnamefont {H.}~\bibnamefont {Park}}, \bibinfo {author}
  {\bibfnamefont {M.}~\bibnamefont {Lon{\v c}ar}},\ and\ \bibinfo {author} {\bibfnamefont {M.~D.}\ \bibnamefont {Lukin}},\ }\bibfield  {title} {\bibinfo {title} {Entanglement of nanophotonic quantum memory nodes in a telecom network},\ }\href {https://doi.org/10.1038/s41586-024-07252-z} {\bibfield  {journal} {\bibinfo  {journal} {Nature}\ }\textbf {\bibinfo {volume} {629}},\ \bibinfo {pages} {573} (\bibinfo {year} {2024})}\BibitemShut {NoStop}%
\bibitem [{\citenamefont {Wang}\ \emph {et~al.}(2026)\citenamefont {Wang}, \citenamefont {Wang}, \citenamefont {Zhu}, \citenamefont {Yao}, \citenamefont {Liang}, \citenamefont {Li}, \citenamefont {Liu}, \citenamefont {He}, \citenamefont {Han}, \citenamefont {Cui}, \citenamefont {Zhou}, \citenamefont {Huang}, \citenamefont {Li},\ and\ \citenamefont {Guo}}]{wang2026}%
  \BibitemOpen
  \bibfield  {author} {\bibinfo {author} {\bibfnamefont {C.-X.}\ \bibnamefont {Wang}}, \bibinfo {author} {\bibfnamefont {Y.-Y.}\ \bibnamefont {Wang}}, \bibinfo {author} {\bibfnamefont {T.-X.}\ \bibnamefont {Zhu}}, \bibinfo {author} {\bibfnamefont {Q.-Q.}\ \bibnamefont {Yao}}, \bibinfo {author} {\bibfnamefont {P.-J.}\ \bibnamefont {Liang}}, \bibinfo {author} {\bibfnamefont {Y.-C.}\ \bibnamefont {Li}}, \bibinfo {author} {\bibfnamefont {Z.-P.}\ \bibnamefont {Liu}}, \bibinfo {author} {\bibfnamefont {R.}~\bibnamefont {He}}, \bibinfo {author} {\bibfnamefont {Y.-J.}\ \bibnamefont {Han}}, \bibinfo {author} {\bibfnamefont {J.-M.}\ \bibnamefont {Cui}}, \bibinfo {author} {\bibfnamefont {Z.-Q.}\ \bibnamefont {Zhou}}, \bibinfo {author} {\bibfnamefont {Y.-F.}\ \bibnamefont {Huang}}, \bibinfo {author} {\bibfnamefont {C.-F.}\ \bibnamefont {Li}},\ and\ \bibinfo {author} {\bibfnamefont {G.-C.}\ \bibnamefont {Guo}},\ }\href {https://arxiv.org/abs/2603.05836} {\bibinfo {title} {Heterogeneous entanglement between a trapped ion
  and a solid-state quantum memory}} (\bibinfo {year} {2026}),\ \Eprint {https://arxiv.org/abs/2603.05836} {arXiv:2603.05836 [quant-ph]} \BibitemShut {NoStop}%
\bibitem [{\citenamefont {Bhaskar}\ \emph {et~al.}(2020)\citenamefont {Bhaskar}, \citenamefont {Riedinger}, \citenamefont {Machielse}, \citenamefont {Levonian}, \citenamefont {Nguyen}, \citenamefont {Knall}, \citenamefont {Park}, \citenamefont {Englund}, \citenamefont {Lon{\v c}ar}, \citenamefont {Sukachev},\ and\ \citenamefont {Lukin}}]{bhaskar2020}%
  \BibitemOpen
  \bibfield  {author} {\bibinfo {author} {\bibfnamefont {M.~K.}\ \bibnamefont {Bhaskar}}, \bibinfo {author} {\bibfnamefont {R.}~\bibnamefont {Riedinger}}, \bibinfo {author} {\bibfnamefont {B.}~\bibnamefont {Machielse}}, \bibinfo {author} {\bibfnamefont {D.~S.}\ \bibnamefont {Levonian}}, \bibinfo {author} {\bibfnamefont {C.~T.}\ \bibnamefont {Nguyen}}, \bibinfo {author} {\bibfnamefont {E.~N.}\ \bibnamefont {Knall}}, \bibinfo {author} {\bibfnamefont {H.}~\bibnamefont {Park}}, \bibinfo {author} {\bibfnamefont {D.}~\bibnamefont {Englund}}, \bibinfo {author} {\bibfnamefont {M.}~\bibnamefont {Lon{\v c}ar}}, \bibinfo {author} {\bibfnamefont {D.~D.}\ \bibnamefont {Sukachev}},\ and\ \bibinfo {author} {\bibfnamefont {M.~D.}\ \bibnamefont {Lukin}},\ }\bibfield  {title} {\bibinfo {title} {Experimental demonstration of memory-enhanced quantum communication},\ }\href {https://doi.org/10.1038/s41586-020-2103-5} {\bibfield  {journal} {\bibinfo  {journal} {Nature}\ }\textbf {\bibinfo {volume} {580}},\ \bibinfo {pages} {60}
  (\bibinfo {year} {2020})}\BibitemShut {NoStop}%
\bibitem [{\citenamefont {Stas}\ \emph {et~al.}(2022)\citenamefont {Stas}, \citenamefont {Huan}, \citenamefont {Machielse}, \citenamefont {Knall}, \citenamefont {Suleymanzade}, \citenamefont {Pingault}, \citenamefont {Sutula}, \citenamefont {Ding}, \citenamefont {Knaut}, \citenamefont {Assumpcao}, \citenamefont {Wei}, \citenamefont {Bhaskar}, \citenamefont {Riedinger}, \citenamefont {Sukachev}, \citenamefont {Park}, \citenamefont {Lončar}, \citenamefont {Levonian},\ and\ \citenamefont {Lukin}}]{stas2022science}%
  \BibitemOpen
  \bibfield  {author} {\bibinfo {author} {\bibfnamefont {P.-J.}\ \bibnamefont {Stas}}, \bibinfo {author} {\bibfnamefont {Y.~Q.}\ \bibnamefont {Huan}}, \bibinfo {author} {\bibfnamefont {B.}~\bibnamefont {Machielse}}, \bibinfo {author} {\bibfnamefont {E.~N.}\ \bibnamefont {Knall}}, \bibinfo {author} {\bibfnamefont {A.}~\bibnamefont {Suleymanzade}}, \bibinfo {author} {\bibfnamefont {B.}~\bibnamefont {Pingault}}, \bibinfo {author} {\bibfnamefont {M.}~\bibnamefont {Sutula}}, \bibinfo {author} {\bibfnamefont {S.~W.}\ \bibnamefont {Ding}}, \bibinfo {author} {\bibfnamefont {C.~M.}\ \bibnamefont {Knaut}}, \bibinfo {author} {\bibfnamefont {D.~R.}\ \bibnamefont {Assumpcao}}, \bibinfo {author} {\bibfnamefont {Y.-C.}\ \bibnamefont {Wei}}, \bibinfo {author} {\bibfnamefont {M.~K.}\ \bibnamefont {Bhaskar}}, \bibinfo {author} {\bibfnamefont {R.}~\bibnamefont {Riedinger}}, \bibinfo {author} {\bibfnamefont {D.~D.}\ \bibnamefont {Sukachev}}, \bibinfo {author} {\bibfnamefont {H.}~\bibnamefont {Park}}, \bibinfo {author}
  {\bibfnamefont {M.}~\bibnamefont {Lončar}}, \bibinfo {author} {\bibfnamefont {D.~S.}\ \bibnamefont {Levonian}},\ and\ \bibinfo {author} {\bibfnamefont {M.~D.}\ \bibnamefont {Lukin}},\ }\bibfield  {title} {\bibinfo {title} {Robust multi-qubit quantum network node with integrated error detection},\ }\href {https://doi.org/10.1126/science.add9771} {\bibfield  {journal} {\bibinfo  {journal} {Science}\ }\textbf {\bibinfo {volume} {378}},\ \bibinfo {pages} {557} (\bibinfo {year} {2022})}\BibitemShut {NoStop}%
\bibitem [{\citenamefont {Riedel}\ \emph {et~al.}(2026)\citenamefont {Riedel}, \citenamefont {Graziosi}, \citenamefont {Wang}, \citenamefont {De-Eknamkul}, \citenamefont {Abulnaga}, \citenamefont {Dietz}, \citenamefont {Mucchietto}, \citenamefont {Haas}, \citenamefont {Sutula}, \citenamefont {Barral}, \citenamefont {Pompili}, \citenamefont {Raha}, \citenamefont {Robens}, \citenamefont {Ha}, \citenamefont {Sukachev}, \citenamefont {Levonian}, \citenamefont {Bhaskar}, \citenamefont {Markham},\ and\ \citenamefont {Machielse}}]{Riedel2026}%
  \BibitemOpen
  \bibfield  {author} {\bibinfo {author} {\bibfnamefont {D.}~\bibnamefont {Riedel}}, \bibinfo {author} {\bibfnamefont {T.}~\bibnamefont {Graziosi}}, \bibinfo {author} {\bibfnamefont {Z.}~\bibnamefont {Wang}}, \bibinfo {author} {\bibfnamefont {C.}~\bibnamefont {De-Eknamkul}}, \bibinfo {author} {\bibfnamefont {A.}~\bibnamefont {Abulnaga}}, \bibinfo {author} {\bibfnamefont {J.}~\bibnamefont {Dietz}}, \bibinfo {author} {\bibfnamefont {A.}~\bibnamefont {Mucchietto}}, \bibinfo {author} {\bibfnamefont {M.}~\bibnamefont {Haas}}, \bibinfo {author} {\bibfnamefont {M.}~\bibnamefont {Sutula}}, \bibinfo {author} {\bibfnamefont {P.}~\bibnamefont {Barral}}, \bibinfo {author} {\bibfnamefont {M.}~\bibnamefont {Pompili}}, \bibinfo {author} {\bibfnamefont {M.}~\bibnamefont {Raha}}, \bibinfo {author} {\bibfnamefont {C.}~\bibnamefont {Robens}}, \bibinfo {author} {\bibfnamefont {J.}~\bibnamefont {Ha}}, \bibinfo {author} {\bibfnamefont {D.}~\bibnamefont {Sukachev}}, \bibinfo {author} {\bibfnamefont {D.}~\bibnamefont {Levonian}},
  \bibinfo {author} {\bibfnamefont {M.}~\bibnamefont {Bhaskar}}, \bibinfo {author} {\bibfnamefont {M.}~\bibnamefont {Markham}},\ and\ \bibinfo {author} {\bibfnamefont {B.}~\bibnamefont {Machielse}},\ }\bibfield  {title} {\bibinfo {title} {Scalable photonic quantum interconnect platform},\ }\href {https://doi.org/10.1103/nfrg-zsts} {\bibfield  {journal} {\bibinfo  {journal} {Phys. Rev. X}\ }\textbf {\bibinfo {volume} {16}},\ \bibinfo {pages} {011063} (\bibinfo {year} {2026})}\BibitemShut {NoStop}%
\bibitem [{\citenamefont {Hannegan}\ \emph {et~al.}(2022)\citenamefont {Hannegan}, \citenamefont {Siverns},\ and\ \citenamefont {Quraishi}}]{Hannegan2022}%
  \BibitemOpen
  \bibfield  {author} {\bibinfo {author} {\bibfnamefont {J.}~\bibnamefont {Hannegan}}, \bibinfo {author} {\bibfnamefont {J.~D.}\ \bibnamefont {Siverns}},\ and\ \bibinfo {author} {\bibfnamefont {Q.}~\bibnamefont {Quraishi}},\ }\bibfield  {title} {\bibinfo {title} {Entanglement between a trapped-ion qubit and a 780-nm photon via quantum frequency conversion},\ }\href {https://doi.org/10.1103/PhysRevA.106.042441} {\bibfield  {journal} {\bibinfo  {journal} {Phys. Rev. A}\ }\textbf {\bibinfo {volume} {106}},\ \bibinfo {pages} {042441} (\bibinfo {year} {2022})}\BibitemShut {NoStop}%
\bibitem [{\citenamefont {Tripier}\ \emph {et~al.}(2026)\citenamefont {Tripier}, \citenamefont {Chung}, \citenamefont {Young}, \citenamefont {Alam}, \citenamefont {Bjork}, \citenamefont {Brodutch}, \citenamefont {Buessen}, \citenamefont {Coble}, \citenamefont {Dellaert}, \citenamefont {Maslov}, \citenamefont {Roetteler}, \citenamefont {Tham}, \citenamefont {Webster}, \citenamefont {Ye}, \citenamefont {Gamble}, \citenamefont {Maksymov}, \citenamefont {Marceaux},\ and\ \citenamefont {Delfosse}}]{tripier2026}%
  \BibitemOpen
  \bibfield  {author} {\bibinfo {author} {\bibfnamefont {F.}~\bibnamefont {Tripier}}, \bibinfo {author} {\bibfnamefont {W.~C.}\ \bibnamefont {Chung}}, \bibinfo {author} {\bibfnamefont {J.}~\bibnamefont {Young}}, \bibinfo {author} {\bibfnamefont {S.}~\bibnamefont {Alam}}, \bibinfo {author} {\bibfnamefont {B.}~\bibnamefont {Bjork}}, \bibinfo {author} {\bibfnamefont {A.}~\bibnamefont {Brodutch}}, \bibinfo {author} {\bibfnamefont {F.~L.}\ \bibnamefont {Buessen}}, \bibinfo {author} {\bibfnamefont {N.~J.}\ \bibnamefont {Coble}}, \bibinfo {author} {\bibfnamefont {T.}~\bibnamefont {Dellaert}}, \bibinfo {author} {\bibfnamefont {D.}~\bibnamefont {Maslov}}, \bibinfo {author} {\bibfnamefont {M.}~\bibnamefont {Roetteler}}, \bibinfo {author} {\bibfnamefont {E.}~\bibnamefont {Tham}}, \bibinfo {author} {\bibfnamefont {M.}~\bibnamefont {Webster}}, \bibinfo {author} {\bibfnamefont {M.}~\bibnamefont {Ye}}, \bibinfo {author} {\bibfnamefont {J.}~\bibnamefont {Gamble}}, \bibinfo {author} {\bibfnamefont {A.}~\bibnamefont
  {Maksymov}}, \bibinfo {author} {\bibfnamefont {J.~P.}\ \bibnamefont {Marceaux}},\ and\ \bibinfo {author} {\bibfnamefont {N.}~\bibnamefont {Delfosse}},\ }\href {https://arxiv.org/abs/2604.19481} {\bibinfo {title} {Fault-tolerant quantum computing with trapped ions: The walking cat architecture}} (\bibinfo {year} {2026}),\ \Eprint {https://arxiv.org/abs/2604.19481} {arXiv:2604.19481 [quant-ph]} \BibitemShut {NoStop}%
\bibitem [{\citenamefont {Mehdi}\ \emph {et~al.}(2025)\citenamefont {Mehdi}, \citenamefont {Vaidya}, \citenamefont {Savill-Brown}, \citenamefont {Grosser}, \citenamefont {Ratcliffe}, \citenamefont {Liu}, \citenamefont {Haine}, \citenamefont {Hope},\ and\ \citenamefont {Viteri}}]{Mehdi2025}%
  \BibitemOpen
  \bibfield  {author} {\bibinfo {author} {\bibfnamefont {Z.}~\bibnamefont {Mehdi}}, \bibinfo {author} {\bibfnamefont {V.~D.}\ \bibnamefont {Vaidya}}, \bibinfo {author} {\bibfnamefont {I.}~\bibnamefont {Savill-Brown}}, \bibinfo {author} {\bibfnamefont {P.}~\bibnamefont {Grosser}}, \bibinfo {author} {\bibfnamefont {A.~K.}\ \bibnamefont {Ratcliffe}}, \bibinfo {author} {\bibfnamefont {H.}~\bibnamefont {Liu}}, \bibinfo {author} {\bibfnamefont {S.~A.}\ \bibnamefont {Haine}}, \bibinfo {author} {\bibfnamefont {J.~J.}\ \bibnamefont {Hope}},\ and\ \bibinfo {author} {\bibfnamefont {C.~R.}\ \bibnamefont {Viteri}},\ }\bibfield  {title} {\bibinfo {title} {Fast mixed-species quantum logic gates for trapped-ion quantum networks},\ }\href {https://doi.org/10.1103/gyls-g198} {\bibfield  {journal} {\bibinfo  {journal} {Phys. Rev. A}\ }\textbf {\bibinfo {volume} {112}},\ \bibinfo {pages} {L050601} (\bibinfo {year} {2025})}\BibitemShut {NoStop}%
\bibitem [{\citenamefont {Malinowski}\ \emph {et~al.}(2023)\citenamefont {Malinowski}, \citenamefont {Allcock},\ and\ \citenamefont {Ballance}}]{WISE2023}%
  \BibitemOpen
  \bibfield  {author} {\bibinfo {author} {\bibfnamefont {M.}~\bibnamefont {Malinowski}}, \bibinfo {author} {\bibfnamefont {D.}~\bibnamefont {Allcock}},\ and\ \bibinfo {author} {\bibfnamefont {C.}~\bibnamefont {Ballance}},\ }\bibfield  {title} {\bibinfo {title} {How to wire a $1000$-qubit trapped-ion quantum computer},\ }\href {https://doi.org/10.1103/PRXQuantum.4.040313} {\bibfield  {journal} {\bibinfo  {journal} {PRX Quantum}\ }\textbf {\bibinfo {volume} {4}},\ \bibinfo {pages} {040313} (\bibinfo {year} {2023})}\BibitemShut {NoStop}%
\bibitem [{\citenamefont {Carter}\ \emph {et~al.}(2024)\citenamefont {Carter}, \citenamefont {O’Reilly}, \citenamefont {Toh}, \citenamefont {Saha}, \citenamefont {Shalaev}, \citenamefont {Goetting},\ and\ \citenamefont {Monroe}}]{carter2024}%
  \BibitemOpen
  \bibfield  {author} {\bibinfo {author} {\bibfnamefont {A.~L.}\ \bibnamefont {Carter}}, \bibinfo {author} {\bibfnamefont {J.}~\bibnamefont {O’Reilly}}, \bibinfo {author} {\bibfnamefont {G.}~\bibnamefont {Toh}}, \bibinfo {author} {\bibfnamefont {S.}~\bibnamefont {Saha}}, \bibinfo {author} {\bibfnamefont {M.}~\bibnamefont {Shalaev}}, \bibinfo {author} {\bibfnamefont {I.}~\bibnamefont {Goetting}},\ and\ \bibinfo {author} {\bibfnamefont {C.}~\bibnamefont {Monroe}},\ }\bibfield  {title} {\bibinfo {title} {Ion trap with in-vacuum high numerical aperture imaging for a dual-species modular quantum computer},\ }\href {https://doi.org/10.1063/5.0180732} {\bibfield  {journal} {\bibinfo  {journal} {Review of Scientific Instruments}\ }\textbf {\bibinfo {volume} {95}},\ \bibinfo {pages} {033201} (\bibinfo {year} {2024})}\BibitemShut {NoStop}%
\bibitem [{\citenamefont {Nguyen}\ \emph {et~al.}(2019)\citenamefont {Nguyen}, \citenamefont {Sukachev}, \citenamefont {Bhaskar}, \citenamefont {Machielse}, \citenamefont {Levonian}, \citenamefont {Knall}, \citenamefont {Stroganov}, \citenamefont {Chia}, \citenamefont {Burek}, \citenamefont {Riedinger}, \citenamefont {Park}, \citenamefont {Lon{\v c}ar},\ and\ \citenamefont {Lukin}}]{nguyen2019prl}%
  \BibitemOpen
  \bibfield  {author} {\bibinfo {author} {\bibfnamefont {C.~T.}\ \bibnamefont {Nguyen}}, \bibinfo {author} {\bibfnamefont {D.~D.}\ \bibnamefont {Sukachev}}, \bibinfo {author} {\bibfnamefont {M.~K.}\ \bibnamefont {Bhaskar}}, \bibinfo {author} {\bibfnamefont {B.}~\bibnamefont {Machielse}}, \bibinfo {author} {\bibfnamefont {D.~S.}\ \bibnamefont {Levonian}}, \bibinfo {author} {\bibfnamefont {E.~N.}\ \bibnamefont {Knall}}, \bibinfo {author} {\bibfnamefont {P.}~\bibnamefont {Stroganov}}, \bibinfo {author} {\bibfnamefont {C.}~\bibnamefont {Chia}}, \bibinfo {author} {\bibfnamefont {M.~J.}\ \bibnamefont {Burek}}, \bibinfo {author} {\bibfnamefont {R.}~\bibnamefont {Riedinger}}, \bibinfo {author} {\bibfnamefont {H.}~\bibnamefont {Park}}, \bibinfo {author} {\bibfnamefont {M.}~\bibnamefont {Lon{\v c}ar}},\ and\ \bibinfo {author} {\bibfnamefont {M.~D.}\ \bibnamefont {Lukin}},\ }\bibfield  {title} {\bibinfo {title} {Quantum network nodes based on diamond qubits with an efficient nanophotonic interface},\ }\href
  {https://doi.org/10.1103/PhysRevLett.123.183602} {\bibfield  {journal} {\bibinfo  {journal} {Phys. Rev. Lett.}\ }\textbf {\bibinfo {volume} {123}},\ \bibinfo {pages} {183602} (\bibinfo {year} {2019})}\BibitemShut {NoStop}%
\bibitem [{\citenamefont {Dahlberg}\ \emph {et~al.}(2019)\citenamefont {Dahlberg}, \citenamefont {Skrzypczyk}, \citenamefont {Coopmans}, \citenamefont {Wubben}, \citenamefont {Rozpedek}, \citenamefont {Pompili}, \citenamefont {Stolk}, \citenamefont {Pawelczak}, \citenamefont {Knegjens}, \citenamefont {de~Oliveira~Filho}, \citenamefont {Hanson},\ and\ \citenamefont {Wehner}}]{dahlberg2019linklayer}%
  \BibitemOpen
  \bibfield  {author} {\bibinfo {author} {\bibfnamefont {A.}~\bibnamefont {Dahlberg}}, \bibinfo {author} {\bibfnamefont {M.}~\bibnamefont {Skrzypczyk}}, \bibinfo {author} {\bibfnamefont {T.}~\bibnamefont {Coopmans}}, \bibinfo {author} {\bibfnamefont {L.}~\bibnamefont {Wubben}}, \bibinfo {author} {\bibfnamefont {F.}~\bibnamefont {Rozpedek}}, \bibinfo {author} {\bibfnamefont {M.}~\bibnamefont {Pompili}}, \bibinfo {author} {\bibfnamefont {A.}~\bibnamefont {Stolk}}, \bibinfo {author} {\bibfnamefont {P.}~\bibnamefont {Pawelczak}}, \bibinfo {author} {\bibfnamefont {R.}~\bibnamefont {Knegjens}}, \bibinfo {author} {\bibfnamefont {J.}~\bibnamefont {de~Oliveira~Filho}}, \bibinfo {author} {\bibfnamefont {R.}~\bibnamefont {Hanson}},\ and\ \bibinfo {author} {\bibfnamefont {S.}~\bibnamefont {Wehner}},\ }\bibfield  {title} {\bibinfo {title} {A link layer protocol for quantum networks},\ }in\ \href {https://doi.org/10.1145/3341302.3342070} {\emph {\bibinfo {booktitle} {Proceedings of the ACM Special Interest Group on Data
  Communication (SIGCOMM '19)}}}\ (\bibinfo  {publisher} {ACM},\ \bibinfo {year} {2019})\ pp.\ \bibinfo {pages} {159--173}\BibitemShut {NoStop}%
\bibitem [{\citenamefont {Assumpcao}\ \emph {et~al.}(2024)\citenamefont {Assumpcao}, \citenamefont {Renaud}, \citenamefont {Baradari}, \citenamefont {Zeng}, \citenamefont {De-Eknamkul}, \citenamefont {Xin}, \citenamefont {Shams-Ansari}, \citenamefont {Barton}, \citenamefont {Machielse},\ and\ \citenamefont {Loncar}}]{assumpcao2024}%
  \BibitemOpen
  \bibfield  {author} {\bibinfo {author} {\bibfnamefont {D.}~\bibnamefont {Assumpcao}}, \bibinfo {author} {\bibfnamefont {D.}~\bibnamefont {Renaud}}, \bibinfo {author} {\bibfnamefont {A.}~\bibnamefont {Baradari}}, \bibinfo {author} {\bibfnamefont {B.}~\bibnamefont {Zeng}}, \bibinfo {author} {\bibfnamefont {C.}~\bibnamefont {De-Eknamkul}}, \bibinfo {author} {\bibfnamefont {C.~J.}\ \bibnamefont {Xin}}, \bibinfo {author} {\bibfnamefont {A.}~\bibnamefont {Shams-Ansari}}, \bibinfo {author} {\bibfnamefont {D.}~\bibnamefont {Barton}}, \bibinfo {author} {\bibfnamefont {B.}~\bibnamefont {Machielse}},\ and\ \bibinfo {author} {\bibfnamefont {M.}~\bibnamefont {Loncar}},\ }\bibfield  {title} {\bibinfo {title} {A thin film lithium niobate near-infrared platform for multiplexing quantum nodes},\ }\href {https://doi.org/10.1038/s41467-024-54541-2} {\bibfield  {journal} {\bibinfo  {journal} {Nature Communications}\ }\textbf {\bibinfo {volume} {15}},\ \bibinfo {pages} {10459} (\bibinfo {year} {2024})}\BibitemShut {NoStop}%
\bibitem [{\citenamefont {Duan}\ and\ \citenamefont {Kimble}(2004)}]{duan2004}%
  \BibitemOpen
  \bibfield  {author} {\bibinfo {author} {\bibfnamefont {L.-M.}\ \bibnamefont {Duan}}\ and\ \bibinfo {author} {\bibfnamefont {H.~J.}\ \bibnamefont {Kimble}},\ }\bibfield  {title} {\bibinfo {title} {Scalable photonic quantum computation through cavity-assisted interactions},\ }\href {https://doi.org/10.1103/PhysRevLett.92.127902} {\bibfield  {journal} {\bibinfo  {journal} {Phys. Rev. Lett.}\ }\textbf {\bibinfo {volume} {92}},\ \bibinfo {pages} {127902} (\bibinfo {year} {2004})}\BibitemShut {NoStop}%
\bibitem [{\citenamefont {Raskop}\ \emph {et~al.}(2026)\citenamefont {Raskop}, \citenamefont {Kandel}, \citenamefont {Arwas}, \citenamefont {Amichy}, \citenamefont {Jarach}, \citenamefont {Kanonich}, \citenamefont {Militaru}, \citenamefont {Fink},\ and\ \citenamefont {Dayan}}]{raskop2026}%
  \BibitemOpen
  \bibfield  {author} {\bibinfo {author} {\bibfnamefont {J.}~\bibnamefont {Raskop}}, \bibinfo {author} {\bibfnamefont {N.}~\bibnamefont {Kandel}}, \bibinfo {author} {\bibfnamefont {G.}~\bibnamefont {Arwas}}, \bibinfo {author} {\bibfnamefont {Y.}~\bibnamefont {Amichy}}, \bibinfo {author} {\bibfnamefont {Y.}~\bibnamefont {Jarach}}, \bibinfo {author} {\bibfnamefont {T.}~\bibnamefont {Kanonich}}, \bibinfo {author} {\bibfnamefont {A.}~\bibnamefont {Militaru}}, \bibinfo {author} {\bibfnamefont {J.}~\bibnamefont {Fink}},\ and\ \bibinfo {author} {\bibfnamefont {B.}~\bibnamefont {Dayan}},\ }\href {https://arxiv.org/abs/2609.21961} {\bibinfo {title} {Single-atom-based asynchronous photonic interconnect for scalable modular quantum computing}} (\bibinfo {year} {2026}),\ \Eprint {https://arxiv.org/abs/2609.21961} {arXiv:2609.21961 [quant-ph]} \BibitemShut {NoStop}%
\bibitem [{\citenamefont {Mokeev}\ \emph {et~al.}(2026)\citenamefont {Mokeev}, \citenamefont {Saif}, \citenamefont {Lukin},\ and\ \citenamefont {Borregaard}}]{Mokeev2026}%
  \BibitemOpen
  \bibfield  {author} {\bibinfo {author} {\bibfnamefont {A.}~\bibnamefont {Mokeev}}, \bibinfo {author} {\bibfnamefont {B.}~\bibnamefont {Saif}}, \bibinfo {author} {\bibfnamefont {M.~D.}\ \bibnamefont {Lukin}},\ and\ \bibinfo {author} {\bibfnamefont {J.}~\bibnamefont {Borregaard}},\ }\bibfield  {title} {\bibinfo {title} {Enhancing optical imaging via quantum computation},\ }\href {https://doi.org/10.1103/s94k-929p} {\bibfield  {journal} {\bibinfo  {journal} {PRX Quantum}\ }\textbf {\bibinfo {volume} {7}},\ \bibinfo {pages} {010318} (\bibinfo {year} {2026})}\BibitemShut {NoStop}%
\bibitem [{\citenamefont {Stas}\ \emph {et~al.}(2026)\citenamefont {Stas}, \citenamefont {Wei}, \citenamefont {Sirotin}, \citenamefont {Huan}, \citenamefont {Yazlar}, \citenamefont {Abdo~Arias}, \citenamefont {Knyazev}, \citenamefont {Baranes}, \citenamefont {Machielse}, \citenamefont {Grandi}, \citenamefont {Riedel}, \citenamefont {Borregaard}, \citenamefont {Park}, \citenamefont {Lončar}, \citenamefont {Suleymanzade},\ and\ \citenamefont {Lukin}}]{stas2026}%
  \BibitemOpen
  \bibfield  {author} {\bibinfo {author} {\bibfnamefont {P.-J.}\ \bibnamefont {Stas}}, \bibinfo {author} {\bibfnamefont {Y.-C.}\ \bibnamefont {Wei}}, \bibinfo {author} {\bibfnamefont {M.}~\bibnamefont {Sirotin}}, \bibinfo {author} {\bibfnamefont {Y.~Q.}\ \bibnamefont {Huan}}, \bibinfo {author} {\bibfnamefont {U.}~\bibnamefont {Yazlar}}, \bibinfo {author} {\bibfnamefont {F.}~\bibnamefont {Abdo~Arias}}, \bibinfo {author} {\bibfnamefont {E.}~\bibnamefont {Knyazev}}, \bibinfo {author} {\bibfnamefont {G.}~\bibnamefont {Baranes}}, \bibinfo {author} {\bibfnamefont {B.}~\bibnamefont {Machielse}}, \bibinfo {author} {\bibfnamefont {S.}~\bibnamefont {Grandi}}, \bibinfo {author} {\bibfnamefont {D.}~\bibnamefont {Riedel}}, \bibinfo {author} {\bibfnamefont {J.}~\bibnamefont {Borregaard}}, \bibinfo {author} {\bibfnamefont {H.}~\bibnamefont {Park}}, \bibinfo {author} {\bibfnamefont {M.}~\bibnamefont {Lončar}}, \bibinfo {author} {\bibfnamefont {A.}~\bibnamefont {Suleymanzade}},\ and\ \bibinfo {author} {\bibfnamefont {M.~D.}\
  \bibnamefont {Lukin}},\ }\bibfield  {title} {\bibinfo {title} {Entanglement-assisted non-local optical interferometry in a quantum network},\ }\href {https://doi.org/10.1038/s41586-026-10171-w} {\bibfield  {journal} {\bibinfo  {journal} {Nature}\ }\textbf {\bibinfo {volume} {651}},\ \bibinfo {pages} {326} (\bibinfo {year} {2026})}\BibitemShut {NoStop}%
\bibitem [{\citenamefont {Baranes}\ \emph {et~al.}(2025)\citenamefont {Baranes}, \citenamefont {Wang}, \citenamefont {Machado}, \citenamefont {Suleymanzade}, \citenamefont {Stas}, \citenamefont {Wei}, \citenamefont {Yelin}, \citenamefont {Borregaard},\ and\ \citenamefont {Lukin}}]{baranes2025}%
  \BibitemOpen
  \bibfield  {author} {\bibinfo {author} {\bibfnamefont {G.}~\bibnamefont {Baranes}}, \bibinfo {author} {\bibfnamefont {I.~W.}\ \bibnamefont {Wang}}, \bibinfo {author} {\bibfnamefont {F.}~\bibnamefont {Machado}}, \bibinfo {author} {\bibfnamefont {A.}~\bibnamefont {Suleymanzade}}, \bibinfo {author} {\bibfnamefont {P.-J.}\ \bibnamefont {Stas}}, \bibinfo {author} {\bibfnamefont {Y.-C.}\ \bibnamefont {Wei}}, \bibinfo {author} {\bibfnamefont {S.~F.}\ \bibnamefont {Yelin}}, \bibinfo {author} {\bibfnamefont {J.}~\bibnamefont {Borregaard}},\ and\ \bibinfo {author} {\bibfnamefont {M.~D.}\ \bibnamefont {Lukin}},\ }\href {https://arxiv.org/abs/2505.21621} {\bibinfo {title} {Designing fault-tolerant blind quantum computation}} (\bibinfo {year} {2025}),\ \Eprint {https://arxiv.org/abs/2505.21621} {arXiv:2505.21621 [quant-ph]} \BibitemShut {NoStop}%
\end{thebibliography}%

\clearpage
\begin{supplement}

\onecolumngrid
\begin{center}
  {\large\bfseries Supplemental Material for\\[2pt]
  ``Quantum Networking at the Speed of Quantum Computation: \\Kilohertz Entanglement in a Heterogeneous Quantum System''\par}
  \vspace{10pt}
\end{center}
\twocolumngrid

\section{Ion trap, qubit control and photon collection}\label{sec:SI:ion}
\subsection{Ion trap and qubit control}\label{sec:methods:ion}

The \Ba{} ion is trapped in a four-rod Paul trap with a design similar to that of Ref.~\cite{carter2024}. 
The ion trap is wider along one dimension to maximize the numerical aperture for light collection ($\mathrm{NA}=0.75$, Sec.~\ref{sec:SI:microscope}).
The ion secular frequencies along the three principal axes are $(0.57,\,0.72,\,1.07)$\,MHz.
We use two rings of permanent magnets and shim coils to apply a magnetic field of 3.68\,G at the ion.
This gives a ground-state Zeeman qubit splitting of $\omega_\mathrm{ion} = 2\pi \times 10.30\,\mathrm{MHz}$.

At the beginning of each entanglement sequence, we Doppler cool the ion for $20\,\mu$s.
We then execute repeated entanglement attempts, with a $3.1\,\mu$s recooling period after every batch of one lead-in and up to $8N-1$ attempts, as shown in Fig.~\ref{fig:sequence}(a).
Before each excitation pulse, we prepare the ion in $\ket{\uparrow_{\mathrm{ion}}}=\ket{S_{1/2},m_J=+1/2}$ by optical pumping with the $493.5$\,nm $\sigma^+$ beam for 160\,ns.
To repump the ion out of the $D_{3/2}$ manifold, we turn on the 650\,nm beam 160\,ns before the $493.5$\,nm optical-pumping light and keep it on for 320\,ns.
With this pumping duration, the measured probability of preparing the ion in $\ket{\uparrow_{\mathrm{ion}}}$ is $0.82(2)$ (Table~\ref{tab:RateBudget}).
Incomplete pumping leaves residual population dark to the excitation pulse, so, when false heralds are rare, it primarily reduces the success probability per attempt rather than the heralded-state fidelity. We therefore optimize the pumping duration to maximize the overall entanglement-generation rate, accepting imperfect preparation in exchange for more attempts per second.

We detect the ion state by shelving $\ket{\uparrow_{\mathrm{ion}}}$ in $\ket{D}=\ket{D_{5/2},m_J=+3/2}$ with a narrow-linewidth ($<1\,$kHz) 1762\,nm laser.
The $\ket{\downarrow_{\mathrm{ion}}}$ state fluoresces under $493.5$\,nm and 650\,nm illumination, while the shelved population remains dark.
We assign the dark state ($\ket{\uparrow_\mathrm{ion}}$) to $\ket{0}$ and the
bright state ($\ket{\downarrow_\mathrm{ion}}$) to $\ket{1}$.
We apply the fluorescence beams for $500\,\mu$s and detect the scattered photons with a Hamamatsu ORCA-Quest~2 camera.
A histogram of the detected photoelectron counts is shown in Fig.~\ref{fig:SI:readout}.
The measured detection fidelity is $0.9935(6)$ for the shelved state and $0.9993(2)$ for the bright state, and most of the error is caused by incomplete 1762\,nm transfer to $\ket{D}$.
For some diagnostic measurements, we reduce this error with a second shelving pulse.
After the first 1762\,nm pulse transfers $\ket{\uparrow_{\mathrm{ion}}}$ to $\ket{D}$, a second 1762\,nm pulse shelves the residual $\ket{\uparrow_{\mathrm{ion}}}$ population to $\ket{D_{5/2},m_J=-3/2}$.
This double shelving squares the shelving failure probability and makes the dark-state readout error negligible.

We also use the 1762\,nm laser for coherent operations between the qubit states.
We first convert the ground-state qubit to an optical qubit by shelving $\ket{\uparrow_{\mathrm{ion}}}$ in $\ket{D}$.
A $\pi/2$ pulse with variable phase $\phi_{1762}$ then couples $\ket{\downarrow_{\mathrm{ion}}}$ and $\ket{D}$ for single-qubit operations.
After photon emission, the ion accrues a phase $\phi = \omega_\text{ion}\,\Delta t$, where $\Delta t$ is the time since emission. Because $\Delta t$ changes from shot to shot, the phase also changes.
For each herald, the controller reads the photon time tag and adds a compensating phase to $\phi_{1762}$ in real time (Fig.~\ref{fig:SI:photonshape}).

We switch most laser beams with double-pass AOMs.
To reduce the turn-on latency, we pass the $493.5$\,nm $\sigma^+$ optical-pumping light through two Mach--Zehnder (MZ) EOMs in series.
The MZ EOM paths have turn-on latencies below 100\,ns and sub-nanosecond rise times.
We condition the $493.5$\,nm MZ EOM pulses on whether a photon is detected during the current attempt (Fig.~\ref{fig:si:detailed_sequence}).
The AOM paths have latencies greater than 500\,ns, so we cannot turn them off in time after a photon detection.
Thus, after a herald, the suppression of the $493.5$\,nm optical-pumping light depends only on the combined extinction of the two MZ EOMs ($\sim$47\,dB).
Leakage light incident on the ion after a herald can optically pump the ion and thereby degrade the heralded ion--photon state.
From ion--photon measurements at $493.5$\,nm with and without pre-triggering the AOMs, we find that this leakage flips the measured correlation with probability $0.009$ in $XX$ and $0.015$ in $ZZ$ (see below and Table~\ref{tab:Fid_error_budget}).

\subsection{Pulsed excitation and ion--photon entanglement}\label{sec:methods:excite}

A mode-locked Ti:sapphire laser (Spectra-Physics Tsunami) at 987\,nm generates 10\,ps pulses at a 79.3\,MHz repetition rate.
An externally triggered pulse-picker selects pulses from this train.
We adjust the pulse-picker timing to transmit the selected pulse and suppress the neighboring pulses.
Since the attempt rate is not commensurate with the repetition rate of the pulsed laser, the selected pulse arrives at a random time within a 12.6\,ns interval.
This results in a photon emission shape where the exponential decay of the excited state is convolved with a uniform distribution of 12.6\,ns width (Fig.~\ref{fig:SI:photonshape}).
We frequency-double the selected pulses in a periodically poled lithium niobate (PPLN) waveguide (NTT Innovative Devices Corporation) to generate $493.5$\,nm excitation light.
An AOM gates this light before delivery to the ion.
We time the falling edge of the AOM gate to suppress subsequent pulses.
We calibrate the pulse energy by scanning the AOM RF amplitude and set it to maximize the excitation probability.
Conditional on preparation in $\ket{\uparrow_{\mathrm{ion}}}$, the excitation probability per selected pulse is $0.97(2)$ (Fig.~\ref{fig:SI:excitation}).

After excitation, the ion decays to $S_{1/2}$ with a probability of $\approx 0.73$~\cite{arnold2019}, summed over the two useful Zeeman decay channels.
Decay to $\ket{\downarrow_{\mathrm{ion}}}=\ket{S_{1/2},m_J=-1/2}$ emits a $\pi$-polarized photon, and decay back to $\ket{\uparrow_{\mathrm{ion}}}$ emits a $\sigma^-$-polarized photon.
The Clebsch--Gordan coefficients give total emission probabilities of $2/3$ for $\sigma^-$ and $1/3$ for $\pi$.
Perpendicular to the quantization axis, the $\pi$ emission is twice as strong as the $\sigma^-$ emission, so the two components have equal amplitudes along the collection axis.
Over the full $\mathrm{NA}=0.75$ aperture, the collected light contains fewer $\pi$ than $\sigma^-$ photons (43:57), but the lower fiber-coupling efficiency for $\sigma^-$ light compensates this difference (Sec.~\ref{sec:SI:microscope}).
The single-mode fiber maps $\sigma^-$ and $\pi$ onto $H_{493}$ and $V_{493}$, which gives the ion--photon state $\ket{\downarrow_{\mathrm{ion}}V_{493}}+\ket{\uparrow_{\mathrm{ion}}H_{493}}$.

We characterize the ion--photon state via joint measurements of the ion's internal state and the photon's polarization. To this end, the photons are directed to a polarization analyzer at 493\,nm, consisting of a quarter-wave plate, a half-wave plate, and a polarizing beam splitter (PBS), with a single-photon counting module (SPCM) at each PBS output port.
We rotate the photon-analysis waveplates to measure the polarization correlations.
We also scan the phase of the 1762\,nm ion-analysis pulse to probe the coherence between the two components of the entangled state.
Figure~\ref{fig:SI:ionphoton493} shows the resulting correlations and coherence measurements.
From the measured correlations $\langle ZZ\rangle = 0.949(6)$ and $\langle XX\rangle = 0.939(6)$, we obtain a lower bound on the ion--photon Bell-state fidelity of $F \ge (\langle ZZ\rangle + \langle XX\rangle)/2 = 0.944(4)$, without correction for state preparation and measurement (SPAM) errors.
To test the rest of the QFC and BC apparatus, we shift the converted light frequency and use the cavity as a bare reflector. For photons converted to $737.45$\,nm and reflected off-resonantly from the cavity (Fig.~\ref{fig:SI:ionphoton737}), we measure $\langle ZZ\rangle = 0.923(7)$ and $\langle XX\rangle = 0.935(7)$, giving $F \ge 0.929(5)$ without SPAM correction.

The ion contributions to the combined ion--\SiV{} error budget are listed in Table~\ref{tab:Fid_error_budget}.
Each entry is the probability that an error flips the measured correlation in the given basis, and we assume equal errors for $XX$ and $YY$.
Together, the ion entries reproduce the measured ion--$493.5$\,nm photon correlators $\langle XX\rangle = 0.939(6)$ and $\langle ZZ\rangle = 0.949(6)$.

The largest ion error comes from optical-pumping light that reaches the ion after a herald. The AOMs cannot switch off in time, so after a herald only the two MZ EOMs suppress the pre-triggered 493.5~nm pumping light.
To quantify this error, we compare the above with measurements without AOM pre-triggers (by modifying the experimental sequence, which then runs at roughly half the repetition rate; Fig.~\ref{fig:SI:ionphoton493clean}), which give $\langle ZZ\rangle = 0.984(5)$ and $\langle XX\rangle = 0.956(5)$.
We obtain the leakage error probability in each basis by comparing the correlators measured with and without AOM pre-triggers, which differ only in whether pumping light can reach the ion after a herald.
The $ZZ$ measurement without pre-triggers uses double shelving: after the first 1762\,nm pulse transfers $\ket{\uparrow_{\mathrm{ion}}}$ to $\ket{D}$, a second 1762\,nm pulse shelves the residual $\ket{\uparrow_{\mathrm{ion}}}$ population to $\ket{D_{5/2}, m_J=-3/2}$.
This makes the dark-state readout error negligible, which we account for
in the comparison, giving leakage error probabilities of $0.009$ in $XX$ and $0.015$ in $ZZ$.

Each ion readout error flips the measured correlation.
The shelved state reads bright with probability $0.0065$ and the bright state reads dark with probability $0.0007$ (Fig.~\ref{fig:SI:readout}).
Because the entangled state has equal populations in the two ion states, the ion state detection gives an error probability of $(0.0065+0.0007)/2 \approx 0.004$ in each basis.

Photons detected in the first $\approx 10$\,ns of the detection window herald a lower-fidelity state: post-selecting on arrival times $t > 10$\,ns (73\% of events) increases $\langle XX\rangle$ from $0.956(5)$ to $0.967(6)$, while $\langle ZZ\rangle$ changes only from $0.984(5)$ to $0.987(5)$.
For excitations early in the pulse-picker cycle, the AOM gate is still open when the next laser pulse arrives, and this residual pulse can re-excite the ion.
Re-excitation removes the coherence of the entangled state but changes the $ZZ$ correlation only in the $\ket{\uparrow_{\mathrm{ion}}}$ branch, which gives error probabilities of $0.006$ in $XX$ and $0.002$ in $ZZ$.
Background counts, amplitude imbalance of the collected state, time-tag quantization, magnetic-field dephasing and polarization mixing in the collection optics each contribute less than $0.002$.
We attribute the remaining error to ion temperature during fast-loop attempts and residual aberrations of the collection optics.

\begin{figure*}[tb]
  \centering
    \includegraphics[width=\linewidth]{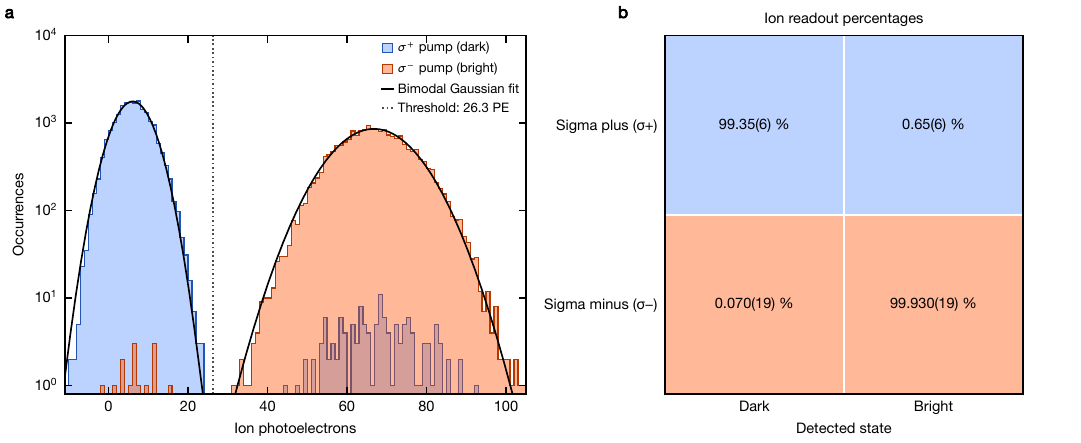}
  \caption{\textbf{Ion state readout.} After optical pumping with either $\sigma^+$ or $\sigma^-$ polarized light, we apply a 1762\,nm pulse to shelve $\ket{\uparrow_{\mathrm{ion}}}$ in $\ket{D}$. 
  (a) Histogram of the number of photoelectrons read out on the camera for $500\,\mu$s of ion fluorescence, for optical pumping with $\sigma^+$ and $\sigma^-$ polarizations of light. On average we detect 67 photoelectrons in the bright state and 6.6 photoelectrons in the dark state.
  (b) Probabilities of the ion state to be detected dark (bright) when we pump with $493.5$\,nm $\sigma^+$ ($\sigma^-$) light.}
  \label{fig:SI:readout}
\end{figure*}

\begin{figure}[tb]
  \centering
  \includegraphics[width=\columnwidth]{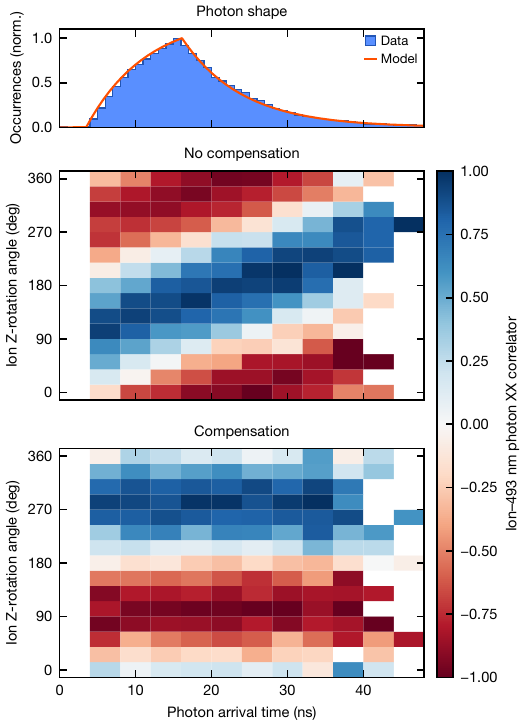}
    \caption{\textbf{Photon temporal shape and accrued ion phase.}
      (top) Histogram of photon arrival times, pooled across both measurements. The shape is the convolution of a uniform excitation window of 12.6\,ns with the exponential decay of the excited state (lifetime 7.9\,ns). The solid line shows this model with a fixed delay of 3.5\,ns. (middle, bottom) Ion--493\,nm photon $XX$ correlator as a function of photon arrival time and ion $Z$-rotation angle before readout, without and with phase compensation, respectively. After photon emission, the ion accrues a phase $\phi = \omega_\text{ion}\,\Delta t$, where $\Delta t$ is the time since emission. Because $\Delta t$ changes from shot to shot, the phase also changes. We bin the data by the photon arrival time recorded by the FPGA time tagger, using 4\,ns bins. Cells with fewer than 10 detected events are left blank. Without compensation, the result is diagonal fringes with a slope set by $\omega_\text{ion}$. We compensate for this phase in real time with a $Z$ rotation of the ion before readout that cancels $\phi$. The resulting horizontal fringes show that the correlation phase is independent of photon arrival time, suppressing dephasing of the ion--photon state.
  }
  \label{fig:SI:photonshape}
\end{figure}

\begin{figure}[tb]
  \centering
  \includegraphics[width=\columnwidth]{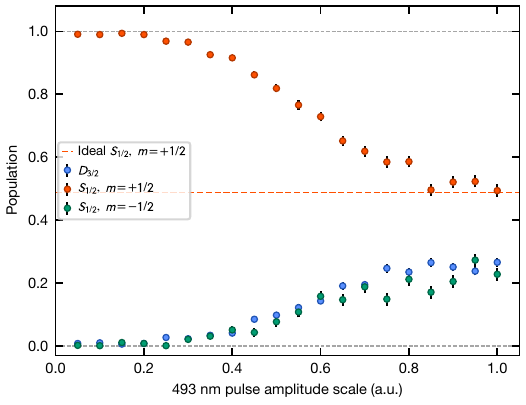}
  \caption{\textbf{Ion excitation probability.} A single $493.5$\,nm pulse excites the ion from $\ket{\uparrow_{\mathrm{ion}}}$ to $\ket{P_{1/2}, m_J=-1/2}$. The excited state can decay to the $D_{3/2}$ manifold with a probability of 0.27, to the $\ket{\downarrow_{\mathrm{ion}}} = \ket{S_{1/2}, m_J=-1/2}$ state with probability 0.24 and to the $\ket{\uparrow_{\mathrm{ion}}} = \ket{S_{1/2}, m_J=+1/2}$ state with probability 0.49. As we scan the power of the excitation pulse using an AOM, we see the ion populations approach the expected distribution; at the highest pulse amplitude the excitation probability is $0.97(2)$.}
  \label{fig:SI:excitation}
\end{figure}

\begin{figure*}[tb]
  \centering
  \begin{minipage}[b]{0.44\textwidth}
    \raggedright
    \textbf{a}\\
    \includegraphics[width=\linewidth]{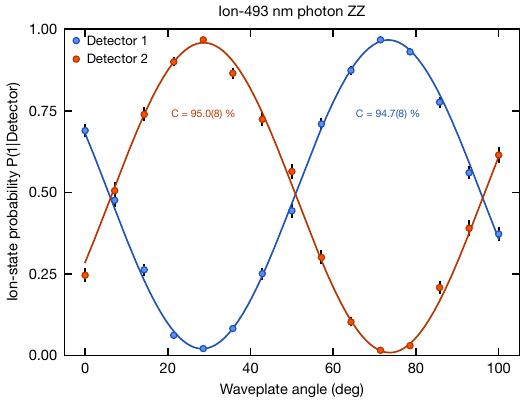}
  \end{minipage}\hfill
  \begin{minipage}[b]{0.44\textwidth}
    \raggedright
    \textbf{b}\\
    \includegraphics[width=\linewidth]{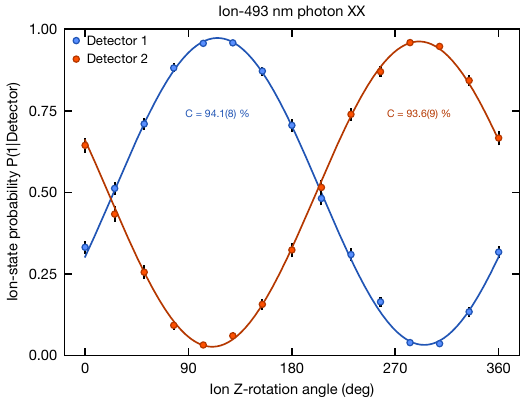}
  \end{minipage}
  \caption{\textbf{Joint ion--$493.5$\,nm photon state measurement.} (a) Conditional populations of the ion state as the measurement basis of the $493.5$\,nm photon is scanned using a half-wave plate. (b) To measure the coherence of the ion--photon state, the half-wave plate is parked to measure the photon in the equator of the Poincar\'e sphere. After this the ion is measured at different points on the Bloch sphere by applying a 1762\,nm $\pi/2$ analysis pulse with varying laser phase. From the measured correlations $\langle ZZ\rangle = 0.949(6)$ and coherences $\langle XX\rangle = 0.939(6)$, we obtain a lower bound on the ion--$493.5$\,nm photon Bell-state fidelity of $F \ge 0.944(4)$. $P(1|\mathrm{Detector})$ is the probability to detect the ion in the bright state, $\ket{1}\equiv\ket{\downarrow_\mathrm{ion}}$.}
  \label{fig:SI:ionphoton493}
\end{figure*}

\begin{figure*}[tb]
  \centering
  \begin{minipage}[b]{0.44\textwidth}
    \raggedright
    \textbf{a}\\
    \includegraphics[width=\linewidth]{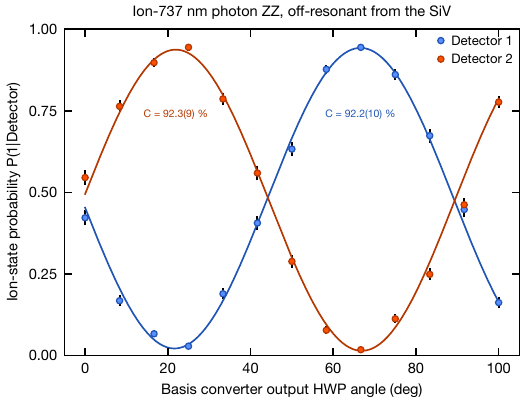}
  \end{minipage}\hfill
  \begin{minipage}[b]{0.44\textwidth}
    \raggedright
    \textbf{b}\\
    \includegraphics[width=\linewidth]{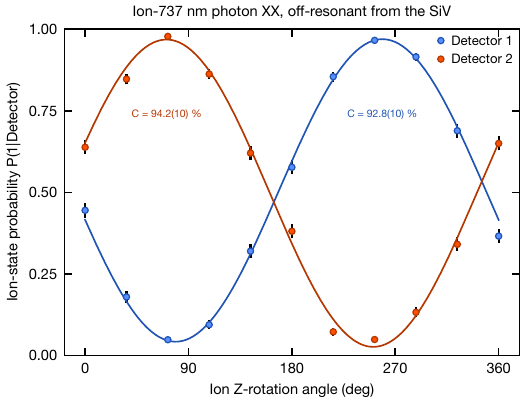}
  \end{minipage}
  \caption{\textbf{Joint ion--$737.45$\,nm photon state measurement.} (a) Correlations between the ion and the frequency-converted photon at $737.45$\,nm. (b) Coherence of the ion--$737.45$\,nm photon state. The photons in both these cases are measured after they travel through the entire signal chain, get reflected from the cavity, pass back through the basis converter and into the SNSPDs. For this measurement the photon is adjusted to be $\sim$5\,GHz off-resonant from the \SiV{} by adjusting the frequency of the $1492.3$\,nm pump light, resulting in the \SiV{}--cavity system being in a mostly reflective state. From these plots, we measure $\langle ZZ\rangle = 0.923(7)$ and $\langle XX\rangle = 0.935(7)$, giving an ion--$737.45$\,nm photon fidelity $F \ge 0.929(5)$.}
  \label{fig:SI:ionphoton737}
\end{figure*}

\begin{figure*}[tb]
  \centering
  \begin{minipage}[b]{0.44\textwidth}
    \raggedright
    \textbf{a}\\
    \includegraphics[width=\linewidth]{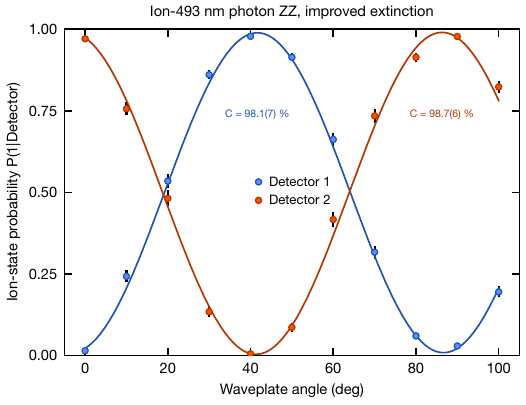}
  \end{minipage}\hfill
  \begin{minipage}[b]{0.44\textwidth}
    \raggedright
    \textbf{b}\\
    \includegraphics[width=\linewidth]{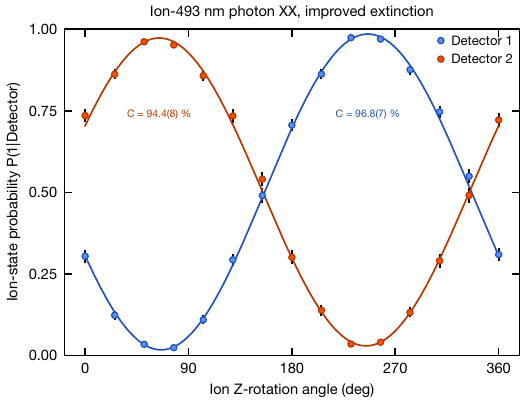}
  \end{minipage}
  \caption{\textbf{Ion--$493.5$\,nm photon state measurement without AOM pre-triggers.} To identify the sources of ion--$493.5$\,nm photon errors, we also measure the ion--photon state with no AOM pre-triggers, resulting in a slower fast-loop attempt rate. (a) Correlations between the ion and $493.5$\,nm photon. (b) Coherence of the ion--$493.5$\,nm photon state. When we remove the pre-triggers, the pumping light for the next attempt no longer disturbs the ion state after entanglement, and the correlation and coherence curves have higher contrast. From these plots, we get $\langle ZZ\rangle = 0.984(5)$, measured with double shelving to suppress readout error, and $\langle XX\rangle = 0.956(5)$, measured with single shelving, giving an ion--$493.5$\,nm photon fidelity $F \ge 0.970(4)$.}
  \label{fig:SI:ionphoton493clean}
\end{figure*}

\section{Microscope design}\label{sec:SI:microscope}

The microscope assembly comprises two objectives built by Special Optics. The first, with 0.75 NA, collects 16.9\% of the light emitted by the ion and collimates it. Special Optics reports a single-pass wavefront error of $0.068\lambda$ rms at $493.5$\,nm, primarily spherical aberration. The objective is compensated for a 4\,mm SF57 glass vacuum window. All the glass up to the fiber is SF57, chosen for its extremely low stress-optic coefficient of $2\cdot 10^{-8}\text{ mm}^2/\text{N}$~\cite{schott_datasheet}. This mitigates ion-photon entanglement infidelities: inhomogeneous stress in the glass can imprint different polarizations on photons passing through different parts of the large-aperture assembly, contaminating the state. The first objective maximizes numerical aperture within the space set by the re-entrant viewport used for optical access. The second must mode-match the focused beam to the fiber mode to maximize coupling efficiency. To find its optimal numerical aperture, we computed the vectorial electric field the microscope projects onto the fiber~\cite{nadlinger2022thesis}. A simpler scalar method, which assumes the objective Fourier-transforms the collimated field, gives the same optimal NA and is relatively independent of apodization factors. Only the vectorial method, however, captures the differential fiber coupling efficiency: 78\% for $\pi$ light and 57\% for $\sigma$ light due to the orbital angular momentum the photon acquires on emission. The higher fiber-coupling efficiency for $\pi$ light compensates the residual imbalance of the collected $\pi$ and $\sigma$ components (43:57 at NA\,=\,0.75), such that the fiber-coupled state is balanced to within 2\% (51:49), corresponding to a fidelity with the maximally entangled state above 99.9\%~\cite{nadlinger2022thesis}. The transmission from the ion to the fiber facet, through the collection objective, the vacuum window and the focusing objective, is $85(3)\,\%$ (Table~\ref{tab:RateBudget}). This is limited mainly by residual reflections at the many anti-reflection-coated surfaces of the multi-element objectives. We achieve 40\% fiber coupling (Table~\ref{tab:RateBudget}). The Zemax model provided by the manufacturer shows that the objective is not diffraction-limited: 55\% of the light falls in the central lobe of the Airy disk, as opposed to the 84\% expected for a diffraction-limited beam. Including these aberrations, our scalar model finds 52\% fiber coupling efficiency for a centered beam, i.e.\ for $\pi$ light with the aligned objective, instead of the 78\% above. This is, to our knowledge, the main reason why we do not obtain higher fiber coupling. We numerically ruled out excessive ion temperature as a significant contribution.

\subsection{Mode of the S405-XP fiber and its free-space expansion}\label{sec:SI:fibermode}
Estimating fiber coupling requires precise knowledge of the fiber core mode shape. Using the manufacturer's parameters~\cite{thorlabs_S405XP} for the core radius $a=1.5\,\mu$m and numerical aperture $\sqrt{n_\text{core}^2-n_\text{cladding}^2} = 0.12$, we solve the equations of Ref.~\cite{bruning2015} to derive the shape of the LP$_{01}(r)$ single mode. This mode is well approximated by a Gaussian $E_0\exp(-r^2/w_0^2)$. The overlap between the two exceeds 99\% but the modes are not strictly interchangeable. Outside the core, the LP$_{01}$ decays as the Bessel function $K_0(\alpha r)\sim \exp(-\alpha r)/\sqrt{\alpha r}$, where $\alpha$ is a numerical constant. It therefore does not propagate as a Gaussian beam and expands significantly more slowly. Figure~\ref{fig:si_microscope} shows how the LP$_{01}$ mode (solid blue) and its best-matched Gaussian (dashed red), both starting with a waist radius $w_0=1.7\,\mu$m, expand at different rates. The manufacturer's data indicates a mode field diameter (MFD $=2w_0$) of 3.3~$\mu$m at 405\,nm and 4.6\,$\mu$m at 630\,nm. A simple linear scaling would give $w_0=1.9\,\mu$m at $493.5$\,nm. Such a Gaussian beam (dotted red) matches the LP$_{01}$ well at long distances. However, at the facet, the mode is best matched by a Gaussian of 1.7\,$\mu$m, not 1.9\,$\mu$m. We experimentally measured the beam profile and found a far-field size smaller than either prediction, which indicates that the mode field radius is likely greater than 1.7\,$\mu$m. We confirmed this on two different fibers. For simplicity, we used the calculated LP$_{01}$ to compute the mode overlap with the ion field and found the optimal NA for the focusing microscope to be $\approx 0.10$.

\begin{figure}[tb]
  \centering
  \includegraphics[width=\columnwidth]{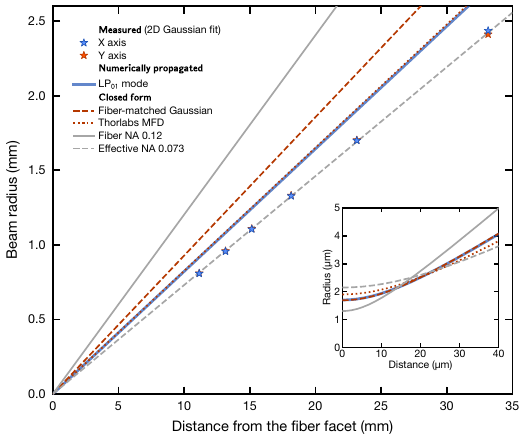}
  \caption{\textbf{Beam radius of the mode emerging from an S405-XP fiber.}
Beam radius versus distance from the fiber facet. The $y$-axis shows the Gaussian beam radius $w$ ($1/e^2$ intensity), not the D4$\sigma$ width ($=2w$ for a Gaussian). Red and blue stars are the principal-axis radii from 2D fits of a $493.5$\,nm beam imaged on a camera. The fiber mode is simulated and then propagated. Its radius is the solid blue line. Fitting this initial mode to a Gaussian, which expands as $w(z) = w_0\sqrt{1+(z/z_R)^2}$ (with $w_0$ the initial beam waist and $z_R$ the Rayleigh range), gives the dashed red curve. The mode field diameter from the Thorlabs datasheet gives the expansion shown in dotted red. Dashed gray is the 0.073~NA beam that best fits the measurements. Solid gray is a beam with NA equal to the fiber NA (0.12). Their initial radius is $\lambda/(\pi\arcsin(\text{NA}))$. Inset: the first $40\,\mu$m from the facet.} 
\label{fig:si_microscope}
\end{figure}

\section{Polarization-insensitive frequency conversion}\label{sec:SI:qfc}\label{sec:methods:qfc}

We convert single photons at $493.5$\,nm from $^{138}$Ba$^+$ to the emission band of silicon-vacancy (SiV) centers in diamond near $737.45$\,nm by difference-frequency generation (DFG) with a continuous-wave pump near $1492.3$\,nm in a temperature-stabilized periodically poled lithium niobate (PPLN) ridge waveguide (NTT Innovative Devices Corporation; 22\,mm long, $11.3 \times 11.6\,\mu$m cross section, poling period $\Lambda = 5.9\,\mu$m). Because DFG in PPLN acts on only one polarization, the waveguide sits in a Sagnac loop, in which the two polarization components are converted in counter-propagating directions and recombined with intrinsic phase stability, preserving the polarization state of the photon~\cite{bock2018,buki2026}. We refer to this scheme as polarization-insensitive quantum frequency conversion (QFC). Signal photons and pump are split by a tri-color polarizing beam splitter (PBS) into the two arms of the loop, and a dual-wavelength half-wave plate (HWP) inside the loop aligns the polarization of both counter-propagating beams with the PPLN crystal axis. Waveplates before the loop set the relative pump power in the clockwise (CW) and anticlockwise (ACW) arms, balancing the H and V conversion efficiencies.

The wavelengths are fixed by four requirements: the input must match the $493.5$\,nm Ba$^+$ transition; the output must reach $737.45$\,nm with 200--300\,GHz of tunability to cover the inhomogeneous distribution of strained \SiV{} transition frequencies; the pump must be the longest of the three wavelengths, so that noise from spontaneous Raman scattering and spontaneous parametric down-conversion (SPDC) of the pump falls at frequencies below the signal band; and the pump linewidth must be well below the $\sim$20\,MHz signal linewidth to preserve the photon's spectral and temporal shape. Reaching near-unity conversion also requires 200--300\,mW of pump power coupled into the fundamental waveguide mode. We use a single-frequency diode laser (Toptica DL pro) amplified to 5\,W by a Raman fiber amplifier (MPB Communications; 2\,nm bandwidth at $1492.3$\,nm). A software lock to a wavemeter (HighFinesse WS6-200) provides tuning and suppresses long-term drift, while a 200\,MHz acousto-optic modulator allows for fast in-sequence switching of converted signal light and pump noise. The setup spans three fiber-linked breadboards---pump, Sagnac loop, and basis conversion---to minimize distance and latency from the ion.

\subsection{Efficiency}\label{sec:SI:qfc:eff}

\begin{figure}[tb]
  \centering
  \includegraphics[width=\columnwidth]{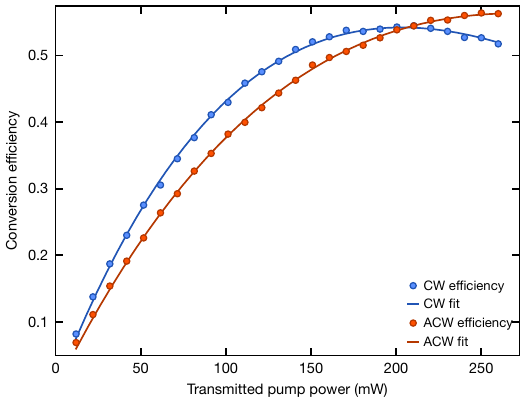}
  \caption{External conversion efficiency of the QFC as a function of transmitted pump power for the clockwise (CW) and anticlockwise (ACW) arms of the Sagnac loop.}
  \label{fig:ppqfc_efficiency}
\end{figure}

The fiber-to-fiber conversion efficiency is 47.5\%, measured from the converter input fiber to the free-space exit of the filtering stage. It is limited by transmission to the chip input facets (92\%), waveguide coupling ($>$90\%), internal transmission at $493.5$\,nm ($>$95\%), internal conversion efficiency ($>$90\%), output fiber coupling ($>$80\%), and spectral filtering (88\%). Light is coupled into and out of the waveguide with $f = 11$\,mm aspheric lenses on three-axis stages (Newport 562-XYZ). Chromatic aberration of the lenses forces a trade-off between pump and signal coupling, so the pump coupling reaches only $\sim$75\%, and residual astigmatism limits the output coupling. Internal conversion reaches 92--94\% when well aligned. The Sagnac board alone reaches 57\% input-to-output efficiency, with a small asymmetry between arms (Fig.~\ref{fig:ppqfc_efficiency}) attributable to the angled chip facets and to chromatic dispersion of the three beams, which prevents simultaneous optimization of both arms. With an efficiency imbalance of 1--2\,\% between the two arms of the Sagnac loop, the polarization-dependent conversion efficiency reduces the fidelity by $\lesssim 10^{-4}$, so the 0.10\,\% entry in Table~\ref{tab:Fid_error_budget} is an upper bound.

\subsection{Noise and spectral filtering}\label{sec:SI:qfc:noise}
Two noise sources limit the signal-to-noise ratio. First, second-harmonic generation (SHG) of the pump produces roughly 1\,$\mu$W near 746\,nm; although this process is poorly phase matched, the high pump intensity makes it significant. Because SHG depends on the poling period and temperature, we survey the 12 waveguides on the chip once the target \SiV{} frequency is chosen and select the one whose DFG phase-matching temperature yields the least SHG. Two angle-tuned 200\,GHz etalon filters (custom, Thorlabs) in series, with transmissions of 96\,\% and 92\,\% (88\,\% combined, Table~\ref{tab:RateBudget}), then remove the residual 746\,nm light. Because their transmission is sensitive to beam collimation, they are placed in the collimated section of the basis converter. Second, a broadband noise pedestal falls directly in the $737.45$\,nm signal band. We attribute it to three potential sources: SHG of the Raman fiber amplifier's noise pedestal, noise accumulated in the long pump delivery fiber, and amplified spontaneous emission near $737.45$\,nm. A 12\,nm bandpass filter on the pump removes $> 95\%$ of these counts. We characterize the effective SNR of the QFC by running the main experiment sequence with the light from the ion blocked. Comparing the herald click rate in this case versus the case in which the ion light is unblocked, we estimate an SNR of 27.

\section{Polarization to time-bin conversion and circulation}\label{sec:SI:tb}\label{sec:methods:tb}
 
The basis conversion is a two-stage process. First, we use an unbalanced Mach--Zehnder interferometer with a polarizing beam splitter (PBS) both at the input and output as a time-delay interferometer (TDI). The input PBS splits the incoming polarization qubit into its $V$ and $H$ components and routes them into the short and long arms of the TDI, respectively ($V\to e$, $H\to l$). This way, we induce a time delay between the two polarization components before recombining them into the same spatial mode at the output PBS. The induced time delay is set by the path length difference between the arms of the interferometer. We induce a delay of 72\,ns using a delay fiber of $\approx 15$\,m length.
The minimum time delay is given by the temporal duration of the photon ($\tau_\mathrm{ph} = 48$\,ns) and the duration of the $\pi$-pulse ($\tau_\pi = 18$\,ns) applied to the electron spin of the \SiV{}.

As the cavity only supports a single linear polarization mode, the polarization of both time bins must be matched to this state. However, after the output PBS of the TDI, both time bins are orthogonally polarized. Therefore, in the second stage of the conversion process, we rotate the polarization of the late time bin (which traversed the long arm) to match that of the early time bin using a Pockels cell. To reduce the effect of spatially inhomogeneous birefringence of the Pockels cell, we reduce the beam size in the Pockels cell by installing a telescope with a magnification of $M=1/2$ in front of it and $M=2$ behind it, each consisting of two spherical singlets with focal lengths of $f=100$\,mm and $f=200$\,mm. After the Pockels cell, the photon is coupled into a single-mode fiber connecting the basis converter to the \SiV{}--cavity system.
 
Overall, this conversion yields the map
\begin{equation}
  \alpha\ket{H} + \beta\ket{V} \;\rightarrow\; \tilde{\alpha}\ket{l} + \tilde{\beta}\ket{e},
  \label{eq:SI:bc_map}
\end{equation}
with $\alpha, \beta, \tilde{\alpha}, \tilde{\beta}$ being complex probability amplitudes ($|\alpha|^2+|\beta|^2 = |\tilde{\alpha}|^2+|\tilde{\beta}|^2 = 1$). As the long and short arms differ in efficiency, mainly because of the fiber coupling in the long arm, the amplitudes of the output qubit are unbalanced relative to the input qubit. However, this rotation is compensated after reflection from the \SiV{}--cavity system, as the photon passes again through the basis converter in the opposite direction and each polarization component now traverses the other arm. On the return path, the conversion is reversed and the time-bin qubit is mapped back onto a polarization qubit. To erase the time-bin information, the Pockels cell is used to set the polarization such that the early (late) time bin is routed through the long (short) arm, so that both polarization components arrive simultaneously at the input PBS. To route the photon at the input PBS to the polarization-resolving detection setup instead of back to the quantum frequency conversion setup, each arm of the TDI includes a Faraday rotator and a set of waveplates, which together function analogously to an optical circulator. A Faraday rotator induces a non-reciprocal $45^\circ$ rotation of linearly polarized light, which in combination with the waveplates gives a propagation-direction-dependent polarization rotation, e.g., in the short arm mapping $V$ onto $V$ for the forward pass vs.\ $V$ onto $H$ for the backward pass.
 
Using the basis converter in double-pass configuration comes with two major advantages. First, it eliminates the need for active phase stabilization, as both time bins accumulate the same total phase along their passage through both the long and short arm, provided the interferometer phase is stable on the timescale of the round trip (which is $\approx 300$\,ns). Second, the arm-dependent (i.e., polarization-dependent) loss is balanced on the second passage and therefore does not affect the fidelity, only the efficiency.
 
The overall single-pass efficiency of the basis converter is $59(2)\%$.
To measure the basis converter efficiency, we inject pulses at the single-photon level, detuned from the cavity resonance, into the basis converter. To calibrate the input photon flux, we use a flip mirror directly at the input of the basis converter. The light reflected off the flip mirror is coupled into an optical fiber and sent onto a superconducting nanowire single-photon detector (SNSPD).
 
Without the flip mirror in the optical path, the pulses pass through the basis converter, are reflected off the cavity and back-propagate through the basis converter before eventually being detected by the SNSPDs of the polarization-resolving detection setup. The ratio of detected counts on these SNSPDs and the detected counts at the input of the basis converter gives us a round-trip efficiency and allows us to calculate the single-pass efficiency.
The single-pass efficiency does not include the detection efficiencies of the SNSPDs, the transmission through the fiber to the cavity (back and forth), or the reflection coefficient of the cavity, which are characterized separately. An overview of the components used in the basis converter and their respective efficiencies is given in Table~\ref{tab:SI:bc_components}.
 
\begin{table*}[tb]
  \caption{\textbf{Components of the basis converter and their transmission at $737.45$\,nm.}
  All values are measured unless marked otherwise.}
  \label{tab:SI:bc_components}
  \centering
  \begin{tabular}{lll}
    \toprule
    Component & Manufacturer, part number & Transmission \\
    \midrule
    AR-coated $\lambda/2$ and $\lambda/4$ waveplates & Resistance optics 
      & $>$99.7\% per surface \\
    Pockels cell & QUBIG, HVOS-NIR\_1.5k100 & $>$98\% \\
    Polarizing beam splitters & Thorlabs, UFPBS10A & $>$99.7\% (in $T$ and $R$) \\
    Fiber coupler (AR coating) & Sch\"after\,+\,Kirchhoff, 60FC-SF-4-M8-10 & $>$99.5\%\footnotemark[1] per surface \\
    Faraday rotator & Ascentta, VFRC-450-737-25-03-19-23 & $>$98\% \\
    Mirrors & FiveNine Optics & $>$99.99\%\footnotemark[1] (reflectivity) \\
    Lenses & Thorlabs, LBF254-100-B / -200-B & $>$99.7\%\footnotemark[1] per surface \\
    \bottomrule
  \end{tabular}
  \footnotetext[1]{Value taken from the manufacturer's data sheet.}
\end{table*}

\section{Diamond nanophotonic SiV memory}
\label{sec:SI:siv}

\subsection{Setup: dilution refrigerator, readout optics, and MW/RF drive electronics}
\label{sec:SI:siv_setup}

\begin{figure*}
\includegraphics[width=0.86\textwidth]{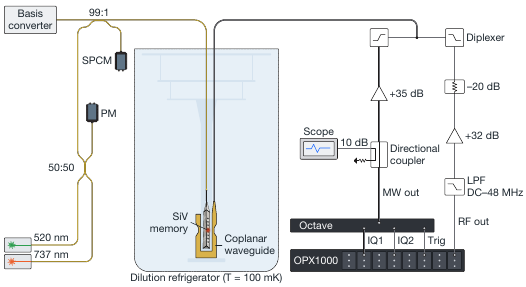}
\caption{\textbf{Diagram of the quantum memory apparatus.} Excitation sources are combined with a 50:50 fiber beamsplitter and coupled into the dilution refrigerator via a 99:1 fiber beamsplitter. The fiber is tapered and coupled to the device at the base of the dilution refrigerator. Reflected photons are detected by SNSPDs after passing through the basis converter. A Quantum Machines OPX1000 and Octave generate the RF and MW pulses that drive electron and nuclear spin transitions on the \SiV{} via a gold coplanar waveguide.}
\label{fig:SI:siv_setup}
\end{figure*}

The quantum memory consists of a silicon-vacancy (\SiV{}) center within a diamond nanophotonic cavity. The \SiV{} used in this work features a nuclear spin-1/2 silicon-29 in addition to an electron spin 1/2. In an applied magnetic field, the \SiV{} effectively comprises a two-qubit system with four spectrally resolved transitions. For the demonstration presented in this work, no quantum information is stored on the nuclear spin: it is initialized into an eigenstate at the beginning of each experimental repetition.

Our setup is similar to that described in Ref.~\cite{stas2022science}. The \SiV{} nanophotonic device is cooled in a dilution refrigerator (Bluefors LD400) to a base temperature of 100\,mK. The dilution refrigerator features fiber, microwave, and RF feedthroughs, a superconducting vector magnet (American Magnetics, Inc., 1\,T, custom model), and a widefield microscope and piezo positioners (attocube XYZ stacks consisting of two ANPx101/LT/HV positioners and an ANPz102/LT/HV) used to couple the tapered fiber to the diamond waveguide. We tune the nanocavity resonance in situ by condensing nitrogen onto the device using a heated gas line.

The optical excitation path (Fig.~\ref{fig:SI:siv_setup}) consists of a $737.45$~nm ECDL (Toptica DL pro), used for resonant readout, and a green diode laser (520\,nm, Thorlabs), used to reset the \SiV{} electron charge state as necessary. The $737.45$\,nm and 520\,nm light sources are combined with a 50:50 fiber beamsplitter (OF-LINK Communications, SMVC-737$\pm$50-22-50-S6-L-15-NE) and coupled into the dilution refrigerator through the 1\% tap port of a 99:1 fiber beamsplitter (KS Photonics FFDC-N-063-1-NO-9901). Inside the dilution refrigerator the fiber is adiabatically tapered for coupling in and out of the nanophotonic device. Reflected photons from the device transmit primarily through the 99\% port of the 99:1 fiber beamsplitter, traverse the basis converter, and are finally collected on SNSPDs (ID Quantique ID281-SYS-16D). The SNSPDs are used both for heralding entanglement generation and for reading out the \SiV{} electron spin state. All fibers used in the basis converter and the dilution refrigerator fiber network are Thorlabs S630-HP single-mode fibers. For diagnostics, the 50:50 fiber beamsplitter is equipped with a power meter (PM, Thorlabs PM101A); the 99:1 fiber beamsplitter is equipped with a single-photon counting module (SPCM, Excelitas SPCM-AQRH-11-FC). Each of these diagnostic ports is terminated with AR-coated FC/APC connectors to avoid unwanted back-reflections.

RF signals for driving the \SiV{} nuclear spin are generated directly by the analog outputs of a Quantum Machines OPX1000 LF-FEM module (see Sec.~\ref{sec:SI:control-system}). Microwave signals for driving the \SiV{} electron spin are generated by mixing RF IQ signals from the OPX1000 with an LO from a Quantum Machines Octave module. Single sideband generation is performed onboard the Octave. The microwave path features a directional coupler (Krytar 1822) and an amplifier (Mini-Circuits ZVE-3W-183+). The RF path features a low-pass filter (Mini-Circuits SLP-50+), an amplifier (Mini-Circuits ZHL-1-2W-S+), and an attenuator (Mini-Circuits BW-S20W5+). The microwave and RF signals are combined with a diplexer (Mini-Circuits ZDSS-3G4G-S+) and sent into the dilution refrigerator through a coaxial feedthrough. Inside the dilution refrigerator, the signals are routed to a custom PCB that is wirebonded to a coplanar waveguide on the diamond chip.

In order to characterize the performance of the basis converter and quantum memory separately from the rest of the experimental subsystems, we generate simulated ion photons using a $737.45$\,nm ECDL (Toptica DL pro) and a Mach--Zehnder EOM (Exail NIR-MX800-LN-20). The pulses are shaped with the EOM and attenuated to roughly match the properties of the light from the ion after frequency conversion. The pulses are injected into the basis converter using a flip mirror that redirects the optical path before the fiber at the output of the QFC. These pulses are also used to characterize the performance of several aspects of the quantum memory.

\subsection{Cavity QED}
\label{sec:SI:siv_cqed}

\begin{figure*}
\includegraphics[width=0.75\textwidth]{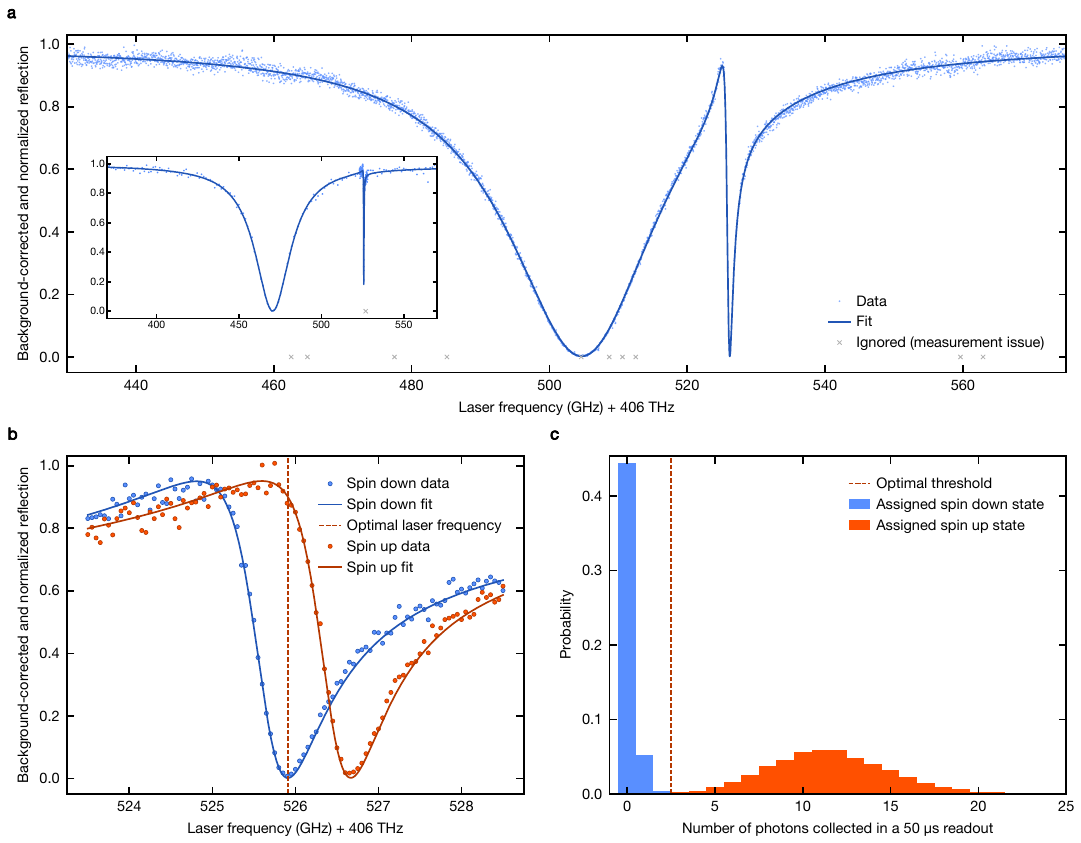}
\caption{\textbf{Optical properties of the quantum memory.}
(a) Cavity QED spectrum of the device and \SiV{} used in the experiment. The broad dip near 406.506\,THz corresponds to the nanophotonic cavity resonance. The narrow feature near 406.525\,THz corresponds to the \SiV{} resonance modulated by the cavity. The spectrum is recorded under zero applied magnetic field. Inset, spectrum of the same device and \SiV{} with the cavity resonance tuned to approximately 406.47\,THz, far detuned from the \SiV{}. At this detuning, the cavity and \SiV{} features are spectrally separated. Thus the fit constrains the \SiV{} linewidth $\gamma$ independently of the coupling rate $g$. Near resonance (main panel), the two modes hybridize and the line shape depends mainly on the cooperativity $C = 4g^2/(\kappa\gamma)$, so $g$ and $\gamma$ are strongly correlated in the fit. Thus we fit the main spectrum with $\gamma$ fixed to the value from the inset. Gray crosses show data points that we exclude from the fit because of a measurement issue.
(b) Spectrum of the \SiV{} center following application of a 6\,kG magnetic field along the \SiV{} axis. Blue (orange) curve shows the spectrum following initialization into spin down (up). The spin contrast is maximized at the frequency indicated by the red line. (c) Histogram of reflected counts under excitation from the readout laser. We achieve a fidelity of 99.96\% for a 50\,$\mu$s readout by thresholding at 2.5 (red line). The measured contrast is 103:1.}
\label{fig:SI:siv_cqed}
\end{figure*}

\begin{table}[tb]
  \caption{Fitted cavity QED parameters.}
  \label{tab:SI:siv_cqed}
  \centering
  \begin{tabular}{lc}
    \toprule
    Parameter & Fitted value \\
    \midrule
    $\kappa_\mathrm{in}/2\pi$ & 14.20(3)\,GHz \\
    $\kappa/2\pi$ & 27.12(3)\,GHz \\
    $\omega_\mathrm{cav}/2\pi$ & 406.505\,597(7)\,THz \\
    $\omega_\mathrm{SiV}/2\pi$ & 406.525\,245(4)\,THz \\
    $\gamma/2\pi$ & 109(4)\,MHz \\
    $g/2\pi$ & 4.361(8)\,GHz \\
    Cooperativity $C = 4g^2/(\kappa\gamma)$ & 25.8(9) \\
    \bottomrule
  \end{tabular}
\end{table}

The reflection coefficient $r$ of a coupled cavity-emitter system as a function of probe frequency $\omega$ is~\cite{nguyen2019prl,nguyen2019prb}
\begin{equation}
r(\omega) = 1 - \frac{\kappa_\mathrm{in}}{i(\omega-\omega_\mathrm{cav}) + \kappa/2 + \dfrac{g^2}{i(\omega-\omega_\mathrm{SiV}) + \gamma/2}},
\label{eq:SI:siv_reflection}
\end{equation}
where $\kappa_\mathrm{in}$ is the coupling rate of the cavity in-coupling port, $\kappa$ is the total cavity decay rate, $g$ is the \SiV{}-cavity coupling, $\omega_\mathrm{cav}$ ($\omega_\mathrm{SiV}$) is the cavity (\SiV{}) resonance frequency, and $\gamma$ is the FWHM linewidth of the bare \SiV{} emitter.

We characterize the device by measuring its reflectivity $|r(\omega)|^2$ as a function of readout laser frequency (Fig.~\ref{fig:SI:siv_cqed}(a)). The extracted fit parameters are shown in Table~\ref{tab:SI:siv_cqed}. The reflectivity near the \SiV{} resonance becomes spin-dependent with the application of an external magnetic field. The \SiV{} electron spin states become spectrally resolved under application of a 6\,kG field along the \SiV{} axis, which splits the line by 790\,MHz (Fig.~\ref{fig:SI:siv_cqed}(b)). The frequencies of the QFC pump laser and the \SiV{} readout laser are tuned to maximize the optical contrast between the two spin states. Under these conditions we can perform fast readouts of the \SiV{} electron spin state. Figure~\ref{fig:SI:siv_cqed}(c) displays a representative histogram of reflected photon counts recorded over 50\,$\mu$s. We assign counts above threshold to the bright, reflective state $\ket{1}\equiv\ket{\uparrow_\mathrm{SiV}}$ and the others to the dark state $\ket{0}\equiv\ket{\downarrow_\mathrm{SiV}}$.
Readout speed is limited in practice by the rate at which the \SiV{} ionizes under illumination, which increases nonlinearly with the optical power. 

\subsection{Microwave driving and nuclear spin initialization}
\label{sec:SI:siv_mw}

\begin{figure*}
\includegraphics[width=\textwidth]{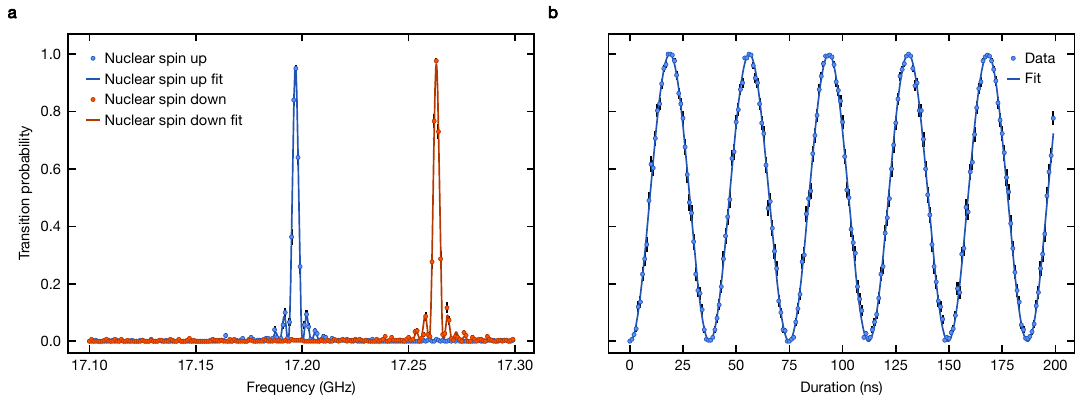}
\caption{\textbf{Microwave properties of the \SiV{} center.} (a) Pulsed ODMR spectrum at 6\,kG. MW pulses are 260\,ns. The two traces show the effect of nuclear spin state initialization. Fits are Rabi lineshapes. (b) Rabi oscillations following nuclear initialization into spin up. The duration of a $\pi$ pulse is 18\,ns, corresponding to a Rabi frequency of approximately $2\pi\times28$\,MHz. Fit is a sinusoid.}
\label{fig:SI:siv_mw}
\end{figure*}

The electron spin is coherently driven by an applied microwave (MW) field. The transition frequencies depend on the nuclear spin state and are measured by optically detected magnetic resonance (ODMR, Fig.~\ref{fig:SI:siv_mw}(a)). During the main experiment sequence, we initialize the nuclear state in the initialization step of each entanglement sequence (main text, Fig.~\ref{fig:sequence}(a)) and drive only the transition at 17.2\,GHz to implement the spin-photon gate. Rabi oscillations of the relevant transition are shown in Fig.~\ref{fig:SI:siv_mw}(b). We note that we observe a MW power-dependent shift of the electron spin resonance frequencies that we must account for in order to achieve maximum pulse fidelity. For a driving field with Rabi frequency $2\pi\times28$\,MHz, we measure a detuning of +3.2\,MHz. We leave the origin of this shift for future work. We characterize the fidelity of the dynamical decoupling sequence applied to the \SiV{} during the main experiment by performing XY8-4 sequences with and without state-preparation and readout $\pi/2$ pulses. By omitting the $\pi/2$ pulses, we isolate the effect of bit-flip errors induced by the pulses that comprise the dynamical decoupling sequence. Without $\pi/2$ pulses we measure a fidelity of 99.36(8)\%, from which we extract an effective MW pulse infidelity 0.019\% per pulse after normalizing to account for SPAM errors. With $\pi/2$ pulses we measure a fidelity of 98.64(12)\% and extract a dephasing infidelity of 0.021\% per pulse after normalizing to account for SPAM errors and MW bit-flip errors.

We rely on nuclear spin state readout to directly verify its successful initialization. This is achieved by reading out the electron spin state, applying a nuclear spin state-selective MW $\pi$ pulse, and then reading out the electron spin state again. If the electron spin flips under the action of the MW pulse, then we conclude that the nuclear spin is in the state targeted by the MW pulse. The nuclear spin state-selective MW pulse is a chirped pulse designed to achieve rapid (1\,$\mu$s) adiabatic passage on the target MW transition. The advantage of the chirped pulse is that it is weakly dependent on small drifts in the MW transition frequency but highly selective of the nuclear spin state. We initialize the nuclear spin state by reading out the nuclear spin state and then applying an RF $\pi$ pulse as necessary until we read out the intended state. We achieve a nuclear spin state initialization fidelity of 99.85(4)\% with this protocol.

Because we only initialize the nuclear spin once at the beginning of each new entanglement generation run, we rely on the nuclear spin to remain in its initial state for the duration of the run. The finite probability that the nuclear spin flips during the run is a source of error for the final experiment fidelity, as the MW pulses are calibrated to target a specific nuclear spin state. Using the results of the nuclear spin readouts performed as part of nuclear spin initialization, we estimate that the nuclear spin flips during 1\% of runs under the experimental conditions of Fig.~\ref{fig:main}(c) of the main text.

\subsection{Mid-circuit SiV electron spin reset}
\label{sec:SI:siv_reset}

\begin{figure*}
\includegraphics[width=\textwidth]{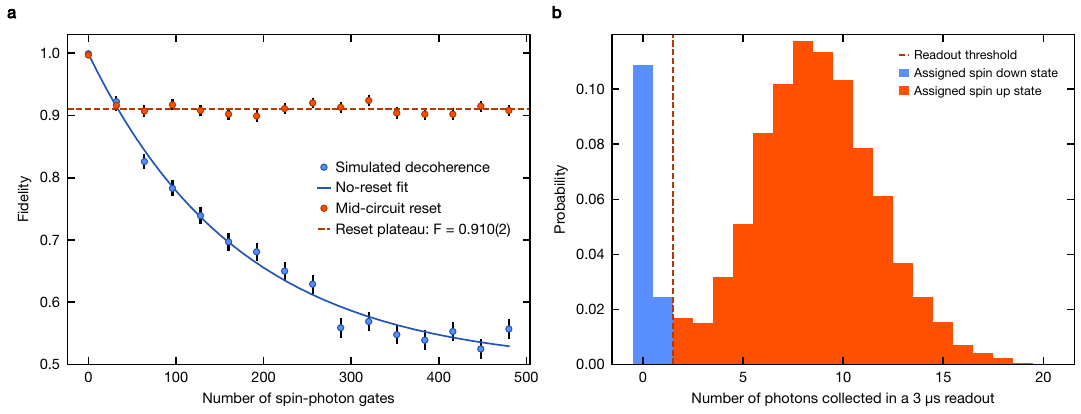}
\caption{\textbf{Mid-circuit reset of the \SiV{} electron spin.} (a) Effect of the reset on measured \SiV{} coherence. Frequency-converted ion photons are simulated using the scheme described in Sec.~\ref{sec:SI:siv_setup}. The experiment consists of an XY8-$N$ sequence with a fixed free precession time matching that used in the main experiment (896\,ns). Without mid-circuit resets (blue) the \SiV{} coherence quickly decays under the effect of the injected photons. With mid-circuit resets applied every XY8 block (32 $\pi$ pulses; orange) the coherence is preserved at 91\%. (b) Fast readout histogram. The readout is performed in 3\,$\mu$s and achieves a fidelity of 99.8\% given a Bayesian prior that the electron spin state is in the bright state with 91\% probability.}
\label{fig:SI:siv_reset}
\end{figure*}

A primary source of infidelity for the quantum memory is electron spin decoherence
caused by photons that become entangled with the \SiV{} but do not register a herald
at the SNSPDs due to losses. Per entanglement attempt, the probability of this effect is
\begin{equation}
  p_\mathrm{d} = p_\mathrm{ph}\left(1-\eta_\mathrm{tot}\right) = 0.0056,
  \label{eq:SI:siv_dropped}
\end{equation}
where $p_\mathrm{ph} = 6.93\times10^{-3}$ is the probability of a photon arriving at
the \SiV{}, and
$\eta_\mathrm{tot} = \eta_\mathrm{R}\,\eta_\mathrm{out}\,\eta_\mathrm{cryo}\,\eta_\mathrm{tp}\,\eta_\mathrm{det} = 0.198$
is the probability that it produces a herald. This is the product of the \SiV{}
reflectivity $\eta_\mathrm{R}=0.44$, the \SiV{}--fiber coupling
$\eta_\mathrm{out}=0.89$, the fiber-to-cryostat transmission
$\eta_\mathrm{cryo}=0.95$, the time-bin-to-polarization conversion efficiency
$\eta_\mathrm{tp}=0.59$, and the SNSPD detection efficiency $\eta_\mathrm{det}=0.90$.
To calculate the impact this effect has on fidelity, we assume that the probability of detecting a herald on any given attempt since performing a mid-circuit reset is uniform. For the experiment shown in Fig.~1(c), we perform a maximum of 15 attempts until performing a mid-circuit reset. The successful attempt is preceded by an average of 7 attempts since the last mid-circuit reset. Failing to detect a photon seen by the SiV$^{-}$, which occurs with probability $p_\mathrm{d}$ per attempt, dephases the SiV$^{-}$. Over the $\bar{n}=7$ attempts this leads to an error for $XX$ and $YY$ measurements with probability
\begin{equation}
  p_\mathrm{err} = \tfrac{1}{2}\left[1-(1-p_\mathrm{d})^{\bar{n}}\right] = 1.92\%.
\end{equation}

To mitigate this effect, we choose a maximum number of entanglement attempts to perform before applying a mid-circuit reset (main text, Fig.~\ref{fig:sequence}(d)). For the reset, we measure the \SiV{} electron spin state in the Y basis and adjust the subsequent MW pulse in the experiment sequence to correct the error in real time. Figure~\ref{fig:SI:siv_reset}(a) shows how the mid-circuit reset effectively preserves the \SiV{} coherence against photons that arrive while the \SiV{} is in superposition.

\subsection{Joint benchmark of the basis converter and SiV memory}
\label{sec:SI:siv_bc}

\begin{figure*}
\includegraphics[width=\textwidth]{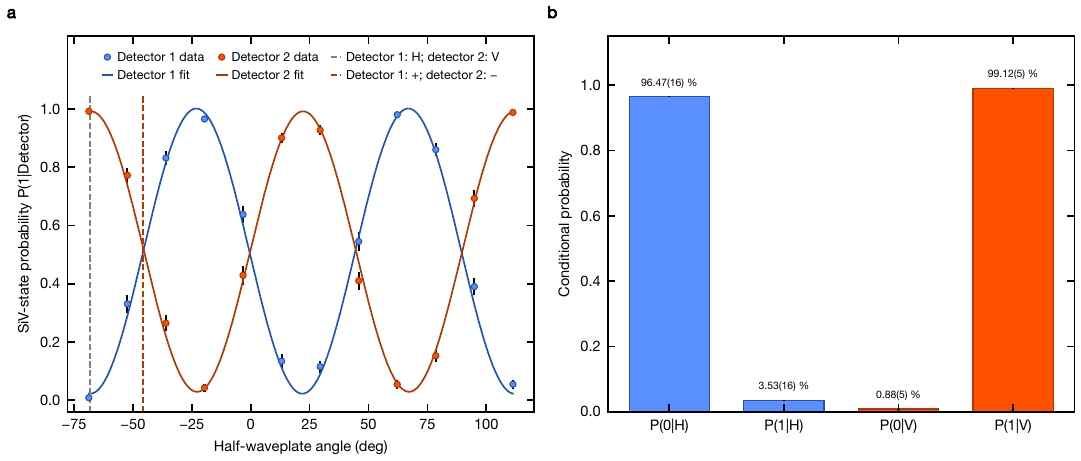}
\caption{\textbf{Correlations between the \SiV{} electron spin state and the polarization of simulated ion photons generated by a spin-photon gate.} (a) Probability of measuring \SiV{} electron spin up conditioned on which detector clicked as a function of analyzer half-wave plate angle. The \SiV{} is measured in the Z basis. The readout basis of the photon is swept by varying the angle of the half-wave plate in front of the analyzer PBS at the detectors. The detector-\SiV{} correlations are maximized (erased) at the gray (red) line. (b) Conditional probabilities at the waveplate angle where correlations are maximized (gray line in panel (a)).}
\label{fig:SI:siv_bc}
\end{figure*}

We benchmark the performance of the basis converter and \SiV{} joint subsystem independently from the rest of the system by running the main experiment sequence using simulated ion photons (see Sec.~\ref{sec:SI:siv_setup}) in place of photons from the ion and QFC. The waveplates at the input to the basis converter are set such that the polarization of the weak coherent pulses is diagonal with respect to the H/V basis. The contrast of the correlations we measure is set primarily by the optical contrast of the \SiV{} cavity QED system, which was $\sim$103 for the measurements shown in Fig.~\ref{fig:SI:siv_bc}. We attribute the asymmetry between the two detectors that can be seen in Fig.~\ref{fig:SI:siv_bc}(b) to a small miscalibration of the microwave pulses used in the experiment, which can bias the \SiV{} electron state towards spin up. In the main experiment, we map the incoming photon state onto the \SiV{} by setting the photon measurement basis to erase the \SiV{}-photon correlations (red line in Fig.~\ref{fig:SI:siv_bc}(a); cf. main text, Fig.~\ref{fig:setup_and_calibration}(b)). We note that for the data shown in Fig.~\ref{fig:setup_and_calibration}(b) of the main text, the \SiV{} featured a reduced contrast of $\sim$60 due to suboptimal cavity--\SiV{} detuning.

\section{Rate and error budget}\label{sec:SI:budget}\label{sec:methods:budget}
Table~\ref{tab:RateBudget} gives the stage-by-stage efficiency budget and Table~\ref{tab:Fid_error_budget} the fidelity error budget; each entry is measured independently where possible. 

\begin{table*}[tb]
  \caption{\textbf{Stage-by-stage efficiency budget.}
  Measured efficiencies of all subsystems per entanglement attempt at a repetition rate of
  $f_{\mathrm{rep}} = 1.116$~MHz. The product of all entries gives the heralded
  entanglement rate through Eq.~(\ref{eq:rate}) of the main text. Uncertainties on subtotals and
  the total are obtained by adding the relative uncertainties of the individual entries in
  quadrature; entries that enter twice (the two basis-converter passes and the two \SiV{}
  fiber-coupling passes) are treated as fully correlated. Entries without an uncertainty are
  taken as exact.}
  \label{tab:RateBudget}
  \centering
  \begin{tabular}{llc}
    \toprule
    Stage & Contribution & Efficiency \\
    \midrule
    Ion & Optical pumping                               & $0.82(2)$  \\
        & $493.5$~nm pulsed excitation                  & $0.97(2)$  \\
        & Branching ratio\footnotemark[1]               & $0.73$     \\
        & Light collection                              & $0.169$    \\
        & Microscope transmission                       & $0.85(3)$  \\
        & Fiber coupling                                & $0.40(5)$  \\
    \cmidrule{2-3}
        & \textbf{Subtotal} $\bm{\etaion}$              & \textbf{\boldmath$0.0334(45)$} \\
    \midrule
    Signal chain
        & $493.5 \rightarrow 737.45$~nm QFC              & $0.54(3)$  \\
        & $737.45$~nm QFC filtering                      & $0.88(1)$  \\
        & Polarization $\rightarrow$ time bin           & $0.59(2)$  \\
        & Time bin $\rightarrow$ polarization (return)  & $0.59(2)$  \\
        & Fiber to cryostat (double pass)               & $0.90(2)$  \\
        & SNSPD detection at $737.45$~nm                 & $0.90(2)$  \\
    \cmidrule{2-3}
        & \textbf{Subtotal} $\bm{\etasig}$              & \textbf{\boldmath$0.134(13)$} \\
    \midrule
    Memory
        & \SiV{} fiber coupling in                      & $0.89(1)$  \\
        & \SiV{} fiber coupling out                     & $0.89(1)$  \\
        & \SiV{} reflection                             & $0.44(1)$  \\
    \cmidrule{2-3}
        & \textbf{Subtotal} $\bm{\etamem}$              & \textbf{\boldmath$0.349(11)$} \\
    \midrule
    Protocol
        & \SiV{} electron init + \Ba{} short cool       & $0.73$     \\
        & \SiV{} nuclear init + \Ba{} long cool         & $0.95$     \\
    \cmidrule{2-3}
        & \textbf{Subtotal} $\bm{\etaprot}$             & \textbf{\boldmath$0.694$} \\
    \midrule
    \multicolumn{2}{l}{\textbf{Total} $\bm{\etaion\,\etasig\,\etamem\,\etaprot}$}
      & \textbf{\boldmath$1.08(18)\times10^{-3}$} \\
    \multicolumn{2}{l}{\textbf{Predicted heralded entanglement rate} $\bm{\Rate}$}
      & \textbf{\boldmath$1.21(20)$~kHz} \\
    \bottomrule
  \end{tabular}
  \footnotetext[1]{Probability of decay from $P_{1/2}$ to $S_{1/2}$ rather than to $D_{3/2}$~\cite{arnold2019}.}
\end{table*}

\begin{table*}[tb]
  \caption{\textbf{Error budget for the ion--\SiV{} Bell-state fidelity.}
  Error probabilities $\epsilon_i$ (in \%) for each two-qubit correlator, for a maximum of
  15 attempts until reset (XY8-2, 16 MW $\pi$ pulses).
  Contributions are assumed independent, so the total success probability is
  $P = \prod_i (1-\epsilon_i)$ and the correlator magnitude is $|\langle \sigma\sigma \rangle| = 2P-1$
  ($\langle YY\rangle$ is negative for $\ket{\Phi^+}$). The fidelity is
  $\mathcal{F} = \tfrac{1}{4}\left(1 + \langle XX\rangle - \langle YY\rangle + \langle ZZ\rangle\right)$.}
  \label{tab:Fid_error_budget}
  \centering
  \begin{tabular}{llccc}
    \toprule
     & & \multicolumn{3}{c}{Error probability $\epsilon_i$ (\%)} \\
    \cmidrule(l){3-5}
    Subsystem & Error source & $XX$ & $YY$ & $ZZ$ \\
    \midrule
    Ion & Pumping leakage after herald (MZ EOM) & 0.87 & 0.87 & 1.45 \\
        & Ion state detection                   & 0.36 & 0.36 & 0.36 \\
        & Double excitation                     & 0.60 & 0.60 & 0.20 \\
        & Other (ion temperature, aberrations)  & 1.26 & 1.26 & 0.57 \\
    \midrule
    QFC & Signal-to-noise                       & 1.85 & 1.85 & 1.85 \\
        & Bidirectional imbalance               & 0.10 & 0.10 & 0.10 \\
    \midrule
    Memory & Electron MW pulse errors           & 0.30 & 0.30 & 0.15 \\
        & Dephasing from environment            & 0.34 & 0.34 & --   \\
        & Electron fast reset                   & 0.13 & 0.13 & --   \\
        & Nuclear spin flip during sequence     & 1.00 & 1.00 & 1.00 \\
        & Electron spin readout                 & 0.04 & 0.04 & 0.04 \\
        & Finite \SiV{} reflection contrast     & 1.00 & 1.00 & 1.00 \\
    \midrule
    Detection & Undetected heralding photons    & 1.92 & 1.92 & --   \\
    \midrule
    \multicolumn{2}{l}{Total error probability $1-P$ (\%)}              & 9.36  & 9.36     & 6.54  \\
    \multicolumn{2}{l}{Correlator $\langle \sigma\sigma\rangle$} & 0.813 & $-0.813$ & 0.869 \\
    \midrule
    \multicolumn{2}{l}{Fidelity $\mathcal{F}$}                   & \multicolumn{3}{c}{0.874} \\
    \bottomrule
  \end{tabular}
\end{table*}

\section{Tomography and Density Matrix Reconstruction}\label{sec:SI:tomography}
We reconstruct the density matrix of the generated ion-\SiV{} entangled state using the nine measurement settings $A,B\in\{X,Y,Z\}$. 
For each setting, we generate the entangled state, apply local rotations to map the selected measurement bases onto the readout basis, and measure both qubits.
We collect 1000 joint ion-\SiV{} measurement outcomes per setting, for a total of 9000 events. We chose this number before the acquisition so that the
statistical uncertainty of the fidelity is below one percentage point. These outcomes form a four-bin histogram with counts $n_{00}^{AB}$, $n_{01}^{AB}$, $n_{10}^{AB}$, and $n_{11}^{AB}$.
Assigning eigenvalues $+1$ and $-1$ to outcomes 0 (dark) and 1 (bright), respectively, where the first index refers to the ion and the second to the \SiV{}, we calculate each correlator as
\begin{equation}
    C_{AB} =
    \frac{n_{00}^{AB}+n_{11}^{AB}-n_{01}^{AB}-n_{10}^{AB}}{\sum_{i,j=0}^{1}n_{ij}^{AB}}.
\end{equation}

We estimate the correlator uncertainties using a multinomial model with a uniform Dirichlet prior on the four outcome probabilities.
For each setting, we draw $10^5$ samples from the posterior distribution
\begin{equation}
    \boldsymbol{p}_{AB}\sim
    \operatorname{Dirichlet}
    \left(n_{00}^{AB}+1,n_{01}^{AB}+1,n_{10}^{AB}+1,n_{11}^{AB}+1 \right),
\end{equation}
and calculate $C_{AB}=p_{00}^{AB}+p_{11}^{AB}-p_{01}^{AB}-p_{10}^{AB}$ for each sample. The lower and upper errors are the distances from the sampled posterior mean to the 15.865th and 84.135th percentiles, respectively. We plot these error magnitudes around the correlators calculated from the measured counts. No readout-error correction is applied.
The results are shown in Fig.~\ref{fig:si:ion-siv-correlators}.

We reconstruct $\rho$ by maximum-likelihood estimation~\cite{James2001_PRA64_052312_MeasurementOfQubits} using the full set of joint outcome counts, rather than the nine correlator values alone. For a candidate density matrix $\rho$, the predicted probability of outcome $ij$ in setting $AB$ is
\begin{equation}
    p_{ij}^{AB}(\rho)=
    \operatorname{Tr}\!\left(\rho\,\frac{I+(-1)^i A}{2}\otimes
    \frac{I+(-1)^j B}{2}\right).
\end{equation}
We maximize the multinomial log-likelihood
\begin{equation}
    \hat{\rho}=
    \underset{\rho}{\operatorname{arg\,max}}\,
    \sum_{A,B}\sum_{i,j=0}^{1}
    n_{ij}^{AB}\log p_{ij}^{AB}(\rho),
\end{equation}
subject to $\rho$ being a valid quantum state: Hermitian ($\rho=\rho^\dagger$), positive semidefinite ($\rho\succeq0$), and of unit trace ($\operatorname{Tr}\rho=1$). We implement this convex optimization in Python using CVXPY~\cite{Diamond2016_JMLR17_83_CVXPY}. Small negative eigenvalues caused by numerical solver precision are set to zero, and the resulting matrix is normalized to unit trace.

We estimate the reconstruction uncertainty by generating 1000 synthetic datasets from the fitted density matrix. For each measurement setting, we draw 1000 outcomes according to the probabilities predicted by $\hat{\rho}$, then repeat the maximum-likelihood reconstruction.
The reported uncertainties are the standard deviations of the real and imaginary parts of each matrix element across these reconstructions. We calculate the fidelity to the target Bell state as $F=\langle\Phi^+|\hat{\rho}|\Phi^+\rangle$, where $|\Phi^+\rangle=(|00\rangle+|11\rangle)/\sqrt{2}$, and obtain its uncertainty from the standard deviation of the fidelities across the reconstructed states.
These uncertainties account for the finite number of measurement outcomes, but not systematic errors in state preparation, analysis rotations, or readout.
In the basis $\ket{00},\ket{01},\ket{10},\ket{11}$ of $\ket{\mathrm{ion},\mathrm{SiV}}$:
\[
\resizebox{\columnwidth}{!}{$
\operatorname{Re}[\rho] = 10^{-3}
\begin{pmatrix}
516(8) & -20(8) & 29(8) & 410(6) \\
-20(8) & 42(6) & -11(5) & -13(8) \\
29(8) & -11(5) & 19(4) & 25(7) \\
410(6) & -13(8) & 25(7) & 423(8) \\
\end{pmatrix}
$}
\]
\[
\resizebox{\columnwidth}{!}{$
\operatorname{Im}[\rho] = 10^{-3}
\begin{pmatrix}
0 & 16(8) & -47(8) & -3(11) \\
-16(8) & 0 & 14(8) & 29(8) \\
47(8) & -14(8) & 0 & 11(8) \\
3(11) & -29(8) & -11(8) & 0 \\
\end{pmatrix}
$}
\]

\begin{figure*}
\includegraphics[width=\columnwidth]{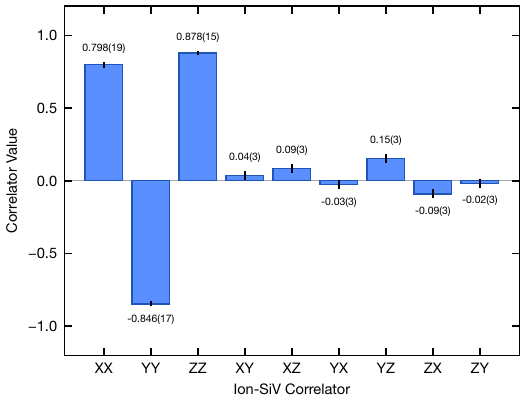}
  \caption{\textbf{Ion--\SiV{} correlators.} Measured two-qubit correlators $C_{AB}$ for all nine settings $A,B\in\{X,Y,Z\}$, with 1000 shots for each setting. For the target state $|\Phi^+\rangle$, the ideal values are $C_{XX}=C_{ZZ}=+1$, $C_{YY}=-1$, and zero for all other correlators. Error bars show the 68.27\% credible interval from the Dirichlet posterior. We apply no readout-error correction.}
  \label{fig:si:ion-siv-correlators}
\end{figure*}

\section{Control System, State Machine and Experimental Sequence}

\subsection{Control hardware}
\label{sec:SI:control-system}
The experiment is controlled by a Quantum Machines OPX1000, equipped with five LF-FEMs (Low-Frequency Front-End Modules), which generates all time-varying control signals, executes the real-time logic and flow control, and serves as the primary time base for the experiment. A Quantum Machines QDAC-II Compact supplies quasistatic control voltages to the Mach--Zehnder EOM bias electrodes and the quadrupolar trap rods, which remain fixed throughout each experimental sequence. A Quantum Machines Octave upconverts OPX1000 outputs to microwave frequencies for \SiV{} spin control, and a Hamamatsu ORCA-Quest 2 camera reads out the ion via fluorescence imaging. Herald photons are detected by ID Quantique SNSPDs. Each SNSPD pulse is converted to a TTL pulse by a Pulse Research Lab comparator, and a Pulse Research Lab 1:4 fan-out buffer distributes the herald signal to three destinations: the OPX1000, which time-tags the photon arrival; a Swabian Instruments timetagger used for calibration and debugging; and a Pulse Research Lab analog OR gate, which combines the two detector channels and whose output is also measured by the OPX1000.

\subsection{State machine}
\label{sec:SI:state-machine}
Each experiment run begins with a 30\,s \textit{warm-up} phase during which the ion is Doppler cooled and the basis-converter Pockels cell is pulsed at the experimental duty cycle to bring it to thermal steady state. After the warm-up, we initialize the \SiV{} nuclear and electron spins, and then the experiment enters the state machine shown in Fig.~\ref{fig:sequence} (a,b), which consists of Initialization, Entanglement Attempt, Reset, and Readout states.

\textit{Initialization.} Each entanglement event begins in the Initialization state. The OPX1000 timestamps the start of this state, beginning the interval used to measure the entanglement rate. The ion is Doppler cooled for 20\,$\mu$s, and the \SiV{} nuclear spin is initialized. The electron spin state is known at this point, either from its initialization after the warm-up or from the final \SiV{} measurement in the preceding Readout state. We therefore apply a nuclear-state-selective microwave $\pi$ pulse, which flips the electron spin only if the nuclear spin is in the target state, and then read out the electron spin again. If it has flipped, the nuclear spin is in the target state; otherwise, we apply an RF $\pi$ pulse on the nuclear spin transition and repeat the process. The final electron spin readout of the nuclear spin initialization sequence sets the sign of a subsequent $\pi/2$ pulse around the $X$ axis, preparing the electron in $\ket{+y} \equiv \left(\ket{\downarrow_\mathrm{SiV}} + i \ket{\uparrow_\mathrm{SiV}}\right)/\sqrt{2}$ regardless of the readout result. This pulse is applied half an attempt period (448\,ns) before the first $\pi$ pulse of the XY8-$N$ sequence.

\subsection{Entanglement attempt sequence}
\label{sec:SI:experimental-sequence}

\begin{figure*}
\includegraphics[width=\textwidth]{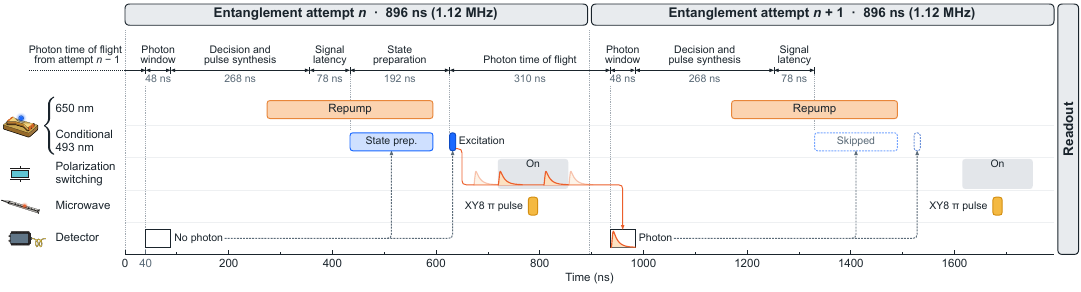}
\caption{
    \textbf{Timing of two consecutive entanglement attempts.} Time is measured from the start of attempt $n$ on the OPX1000; the detection window opens after 40\,ns of on-board latency. Attempt $n$ detects no herald, so the $493.5$\,nm optical pumping and excitation (blue) are applied, whereas attempt $n+1$ detects a herald, so they are skipped (dashed outlines) and the system proceeds to Readout. The 650\,nm repump, the basis-converter Pockels cell (gray window), and the XY8 $\pi$ pulse are applied in every attempt. The photon emitted in one attempt, whose two time bins are shown passing through the Pockels cell (darker traces: switched time bins), is detected in the window at the start of the next.
}
\label{fig:si:detailed_sequence}
\end{figure*}

Figure \ref{fig:si:detailed_sequence} shows the timing of two consecutive attempts, the first without a herald and the second with one. Each entanglement attempt begins with three parallel 48\,ns photon-detection measurements on the OPX1000 analog inputs, which open after 40\,ns of on-board latency that overlaps the photon time of flight. Two of these time-tag the individual SNSPD signals with 0.5\,ns resolution. The third integrates the OR gate output, determining whether either SNSPD detected a herald photon (this is faster than performing on-board logic of the individual time-tagging measurements).

The rest of the attempt is conditioned on this result. While the herald decision is pending, we apply an unconditional 320\,ns pulse of 650\,nm repump light, which depopulates the $D_{3/2}$ manifold after a failed attempt and leaves the entangled state unaffected after a successful attempt, since the ion has then already decayed to $S_{1/2}$. Evaluating the integration result and synthesizing the conditional pulses takes 268\,ns, followed by 78\,ns of signal latency before the Mach--Zehnder EOMs open. If no herald was detected, we then apply 160\,ns of $\sigma^{+}$-polarized $493.5$\,nm optical pumping for state preparation, wait 32\,ns, and trigger the pulsed excitation laser. The timing of the excitation pulse jitters by 12.6\,ns because the laser is not phase-locked to the experimental sequence.

After excitation, as the ion emits a photon, we switch on the quantum frequency conversion (QFC) pump. Keeping the pump off earlier in the attempt prevents light scattered during optical pumping from being converted and reaching the \SiV{}, where it would cause dephasing. We then trigger the basis-converter Pockels cell to rotate the polarization of the late time bin as the photon travels toward the \SiV{} and of the early time bin after reflection, so that each time bin traverses each path of the basis converter exactly once. Between the reflections of the two time bins, we apply one $\pi$ pulse of the XY8 sequence, preserving the electron spin coherence across attempts. The time of flight from the start of the excitation pulse to the opening of the photon-detection windows is 310\,ns, bringing the attempt period to 896\,ns. The photon emitted in one attempt is detected at the start of the next. 

A batch of attempts spans one XY8-$N$ sequence. It opens with an abridged lead-in attempt that contains the $\pi/2$ pulse, an excitation, the QFC pump and basis-converter Pockels cell pulses, and the first XY8 $\pi$ pulse, but no detection window, so that a photon can arrive in the detection window of the next attempt. Up to $8N-1$ full attempts follow, completing the XY8-$N$ sequence. The block ends early if a herald is detected. The AOMs are triggered one attempt in advance so that their latency does not slow the attempt rate. For the lead-in attempt, these triggers are issued during the preceding cooling interval.

For the data in Figs.~\ref{fig:sequence}(d), \ref{fig:setup_and_calibration}(b--d), \ref{fig:SI:ionphoton737} and \ref{fig:SI:siv_bc}, an error in the sequence script prevented the first full attempt of each block from heralding. We found this error after data taking was complete, while we prepared this manuscript. In this attempt, the ion was not excited and a photon was not sent to the \SiV{}. In these data, the first attempt gives fewer than 0.1\,\% of the heralds, compared with 4.9\,\% for the data in Fig.~\ref{fig:main}. A block with $8N-1$ full attempts therefore has $8N-2$ attempts that can herald, and Fig.~\ref{fig:sequence}(d) shows this number on its horizontal axis. The rates in Fig.~\ref{fig:sequence}(d) include the time of the inactive attempt. The data in Fig.~\ref{fig:main} are not affected and use 15 attempts per block.

\textit{Reset.} If the XY8-$N$ sequence completes without a herald, the system enters the Reset state and the ion is Doppler cooled for 3.1\,$\mu$s. During this interval, a $\pi/2$ pulse maps the \SiV{} electron spin to $\ket{\uparrow}$, and a measurement of the electron spin again sets the sign of the subsequent $\pi/2$ pulse, re-preparing $\ket{+y}$. The system then returns to the Entanglement Attempt state for the next block of attempts.

\textit{Readout.} The system enters the Readout state when a photon is detected by either SNSPD. A second OPX1000 timestamp at the start of this state ends the entanglement rate measurement interval. A 1762\,nm $\pi$ pulse first shelves $\ket{\uparrow_\mathrm{ion}}$ into $\ket{D}$ (see Sec.~\ref{sec:methods:ion} for ion state definitions). For X- and Y-basis measurements, a 1762\,nm $\pi/2$ analysis pulse then sets the measurement basis. The phase of this pulse is set in real time based upon the herald time tag to cancel the phase that the ion qubit accumulates after photon emission. This compensates for variations in emission time caused by both the 12.6\,ns excitation jitter and the stochastic timing of the spontaneous decay from the $\ket{P_{1/2},m_J=-1/2}$ state, whose lifetime is 7.9\,ns. The ion is then read out by fluorescence imaging. In parallel, the remaining pulses of the current XY8 cycle are applied to the \SiV{}, followed by an additional $\pi$ pulse that compensates for the resulting bit flip when the herald occurs at an odd position in the cycle.
After these pulses the \SiV{} spin is in the frame of its initial state for every herald position.
Heralds from the two SNSPDs project onto entangled states that differ by a $\pi$ phase, so a conditional $\pi$ frame rotation of the \SiV{} drive corrects for this when the second SNSPD detects the herald. The \SiV{} is then rotated into its measurement basis and read out. The basis-converter Pockels cell continues to be pulsed throughout the Readout state to maintain its thermal steady state. The system then returns to the Initialization state to prepare for the next entanglement event.

\clearpage
\SIbibliography{references_SI,references}

\end{supplement}

\end{document}